\documentclass[11pt,a4paper]{article}

\usepackage[dvips]{graphicx}
\usepackage{amsfonts,amsthm,amsbsy,amssymb}
\usepackage{float}
\usepackage{xspace}
\usepackage[tbtags]{amsmath}
\usepackage{amssymb}
\newcommand{\ar}[2]{
   $^{+#1}_{-#2}$}
\newcommand{\als}{
   \mbox{$\alpha_{\rm{s}}$}}
\newcommand{\PZo}{$\mathrm{Z}$\xspace}
\newcommand{\PMZ}{$M_{\mathrm{Z}}$}
\newcommand{\yc}{\mbox{$y_{\rm{cut}}$}}
\newcommand{\ivpb}{\mbox{pb$^{-1}$}}
\newcommand{\ZZ}{$\rm{Z}\rm{Z}$}
\newcommand{\WW}{$\rm{W}^{+}\rm{W}^{-}$}
\newcommand{\WWx}{$\mathrm{W}^{+}\mathrm{W}^{-}$\xspace}
\newcommand{\alsx}{\mbox{$\alpha_{\rm{s}}$}\xspace}

\newcommand{\alsn}[1]{\mbox{$\alpha_{\rm{s}}^{#1}$}}

\newcommand{\lqqqq}{\mbox{\lag$_{\mathrm{q}\overline{\mathrm{q}}\mathrm{q}\overline{\mathrm{q}}}$}\xspace}
\newcommand{\lqqln}{
   \mbox{\lag$_{\mathrm{q}\overline{\mathrm{q}}\ell\nu}$}\xspace}
\newcommand{\epem}{$\mathrm{e}^{+}\mathrm{e}^{-}$\xspace}
\newcommand{\roots}{$\sqrt{s}$\xspace}
\newcommand{\qqbar}{$\mathrm{q}\overline{\mathrm{q}}$\xspace}

\newcommand{\order}[1]{\mbox{$\mathcal{O}(#1)$}}

\newcommand{\avn}{\langle N \rangle}
\newcommand{\cfm}{centre-of-mass\xspace}

\newcommand{\eeqq}{
   \mbox{$\mathrm{e^{+}e^{-}}\hspace*{-.2cm}\rightarrow\mathrm{q\overline{q}}$}}
\newcommand{\h}{\hspace*{-1mm}}

\newcommand{\lag}{\mbox{$\mathcal{L}$}}

\newcommand{\eeqqqq}{
\mbox{$\mathrm{e^{+}e^{-}}\hspace*{-.2cm}\rightarrow\mathrm{q\overline{q}
q\overline{q}}$}}
\begin{document}
\begin{titlepage}

\begin{center}
   {\large EUROPEAN ORGANISATION FOR NUCLEAR RESEARCH}
\end{center}
\bigskip
\begin{flushright}
      CERN-PH-EP/2005-024 \\
      13 June, 2005
\end{flushright}
\bigskip
\begin{center}
   {\huge\bf\boldmath
      \vspace{.3cm}
      \center{Determination of $\alpha_{\rm{s}}$ Using Jet Rates}
      \center{at LEP with the OPAL Detector}
   }
\end{center}
\bigskip\bigskip\bigskip
\begin{center}
   {\LARGE
      The OPAL Collaboration
   }
   \vspace{0.25cm}
\end{center}
\bigskip\bigskip

\begin{minipage}[t]{.95\textwidth}
\centerline{\bf Abstract:}
Hadronic events produced in \epem collisions
by the LEP collider and recorded by the OPAL detector were used to form
distributions based on the
number of reconstructed jets. The data were collected between 1995 and 2000 and
correspond to
energies of 91 GeV, 130-136 GeV and 161-209 GeV. The jet rates were determined
using four different
jet-finding algorithms (Cone, JADE, Durham and Cambridge). The differential
two-jet rate and the
average jet rate with the Durham and
Cambridge algorithms were used to measure \alsx in the LEP energy range by
fitting an expression in
which \order{\alsn{2}} calculations were matched to a NLLA prediction and fitted
to the data.  Combining the
measurements at different centre-of-mass energies, the value of \als(\PMZ) was
determined to be
         
\begin{center}
\mbox{
\als(\PMZ)=$0.1177\pm0.0006$(stat.)$\pm0.0012$(expt.)$\pm0.0010$(had.)$\pm0.0032
$(theo.)}. \\
\end{center}         
\end{minipage}

\begin{center}
\vspace{.5cm}
\Large
\vspace{.5cm}


(Submitted to European Physical Journal C)

\end{center}

\end{titlepage}

\begin{center}{\Large        The OPAL Collaboration
}\end{center}\bigskip
\begin{center}{
G.\thinspace Abbiendi$^{  2}$,
C.\thinspace Ainsley$^{  5}$,
P.F.\thinspace {\AA}kesson$^{  3,  y}$,
G.\thinspace Alexander$^{ 22}$,
G.\thinspace Anagnostou$^{  1}$,
K.J.\thinspace Anderson$^{  9}$,
S.\thinspace Asai$^{ 23}$,
D.\thinspace Axen$^{ 27}$,
I.\thinspace Bailey$^{ 26}$,
E.\thinspace Barberio$^{  8,   p}$
% ...
}\end{center}\bigskip
\bigskip
$^{  1}$School of Physics and Astronomy, University of Birmingham,
Birmingham B15 2TT, UK
\newline
$^{  2}$Dipartimento di Fisica dell' Universit\`a di Bologna and INFN,
I-40126 Bologna, Italy
\newline
$^{  3}$Physikalisches Institut, Universit\"at Bonn,
D-53115 Bonn, Germany
\newline
% ...
$^{  5}$Cavendish Laboratory, Cambridge CB3 0HE, UK
\newline
% ...
$^{  8}$CERN, European Organisation for Nuclear Research,
CH-1211 Geneva 23, Switzerland
\newline
$^{  9}$Enrico Fermi Institute and Department of Physics,
University of Chicago, Chicago IL 60637, USA
\newline
% ...
$^{ 22}$Department of Physics and Astronomy, Tel Aviv University,
Tel Aviv 69978, Israel
\newline
$^{ 23}$International Centre for Elementary Particle Physics and
Department of Physics, University of Tokyo, Tokyo 113-0033, and
Kobe University, Kobe 657-8501, Japan
\newline
% ...
$^{ 26}$University of Victoria, Department of Physics, P O Box 3055,
Victoria BC V8W 3P6, Canada
\newline
$^{ 27}$University of British Columbia, Department of Physics,
Vancouver BC V6T 1Z1, Canada
\newline
% ...
$^{  p}$ now at The University of Melbourne, Victoria, Australia
\newline
% ...
$^{  s}$ now at University of Antwerpen, Physics Department,B-2610 Antwerpen, 
Belgium; supported by Interuniversity Attraction Poles Programme -- Belgian
Science Policy
\newline
% ...
$^{  y}$ now at CERN
\newline
% ...

\section{Introduction}

In the Standard Model of elementary particle interactions, the strong
interaction is described by the theory of Quantum Chromodynamics
(QCD), and depends on just one fundamental parameter, the strong
coupling \als.  The value of \alsx is expected to depend on the
energy scale of the interaction. It is therefore an important test
of the theory to determine the value of \alsx experimentally at as
many different energies as possible.  It is also important to use
as many different techniques as possible, as different measurements
of \alsx are sensitive to different theoretical and hadronization
variations.

Indeed, many methods have already been employed to evaluate
\als~\cite{als_ref}.  At very low energies the value of \alsx can be
measured using the hadronic decays of the $\tau$ lepton and heavy
quarkonia. Low energy determinations are also available using scaling
violations and    sum rules from deep inelastic scattering
experiments.  Higher energy determinations of \alsx come from collider
experiments (\epem\!\!\!, pp, p$\bar{\rm{p}}$ or ep) using properties
of the created hadron system which are explicitly dependent on the
value of \als($Q$), where $Q$ corresponds to the energy scale at which
the interaction takes place\footnote{For \epem collisions
$Q=$\roots.}.   

During the LEP1.5 (\roots$\sim$133~GeV) and LEP2 (above \WWx
threshold) operational phases of the Large Electron-Positron collider
at CERN, events were recorded with \cfm collision energies ranging
from 91~GeV to 209~GeV.  Events of the form \epem$\rightarrow$hadrons
can  be used to determine distributions based on the ensemble of final
state hadrons (event shapes) or on the ensemble of jets (jet rates).
Previous results by OPAL for an \alsx determination based on event
shapes and jet rates using the \PZo dataset collected during the LEP1
phase can be found in~\cite{als_lep1}.  Determinations of \alsx from
LEP1.5 and LEP2 datasets up to 189~GeV have already been reported by
OPAL based on event shape distributions~\cite{als_lep1.5,
als_lep161,als_lep172} and on jet rates~\cite{peter_pub}.  Another
OPAL paper~\cite{mford_event_shapes} uses the same data that
have been presented here to measure event shapes.

For the analysis presented in this paper we used data collected during
the LEP1.5 and LEP2 phases to construct jet rate distributions using
several jet clustering algorithms.  The differential two-jet rate,
$D_{2}$, and the average jet rate, $\avn$, were used to determine
values of \als(\roots) at the four combined \cfm energies composed
of data within the LEP1.5 and LEP2 datasets. Theoretical predictions
were fitted to these distributions to extract the value of
\als(\roots).

The paper is organized as follows. Section 2 contains a brief
description of the OPAL detector.  A summary of the data and the Monte
Carlo samples used in the analysis is given in Section 3.  In Section
4, we define the jet rate distributions.  The methods used to select
signal events and reject backgrounds are presented in Section 5.  The
variations used for systematic studies are detailed in Section
6.  Finally, the results of this analysis are given in Section 7,
followed by a conclusion and summary in Section 8.

\section{The OPAL Experiment}

A full description of the OPAL detector can be found
in~\cite{opaldetector}.  The critical components of the detector in
the identification of jets were the central tracking chambers, which
were used to reconstruct charged particles, and the electromagnetic
calorimeters, which measured the total energy deposited by electrons
and photons.

The tracking chambers were located inside a solenoidal magnet which
provided a 0.435~T axial magnetic field along the beam axis.  The main
component of the tracking system was a large-volume jet chamber,
which was approximately 4.0 m long with an outer radius of 1.85 m.
The jet chamber was separated into 24 sectors, each with a radial
plane of 159 sense wires separated by 1 cm.  The momenta of tracks in
the $x-y$ plane\footnote{The right-handed OPAL coordinate system is
defined so that $z$ is the coordinate parallel to the e$^{-}$ beam
direction and the $x$ axis points to the centre of the LEP ring, $r$
is the distance normal to the $z$ axis, $\theta$ is the polar angle
with respect to the $z$ axis and $\phi$ is the azimuthal angle with
respect to the $x$ axis.}  were measured with a precision
parametrized by \mbox{$\sigma_{p}/p=\sqrt{0.02^{2}+(0.0015\cdot
p[\rm{GeV}/\mathit{c}])^{2}}$}.

The calorimetry systems were outside the solenoidal magnet.  The electromagnetic
calorimeter was
composed of 11704 lead glass blocks in the barrel and endcap regions,
representing about 25
radiation lengths in the barrel and more than 22 in the endcap.  The iron
sampling hadron calorimeter was located just outside the electromagnetic
calorimeter, and provided
the stopping power to contain most hadronic showers. Luminosity was determined
using small-angle
Bhabha events detected in the forward detectors and silicon-tungsten
calorimeter~\cite{opallumi_sw}.

After an event was
triggered~\cite{op:trig}, data were collected from the subdetectorsand processed
by the OPAL data acquisition system~\cite{op:daq}.  The raw event data were
transferred to a farm of computer processors where the events were fully
reconstructed and
written to tape for offline analysis.

\section{Data and Monte Carlo Samples}

The data used in this analysis were collected by OPAL between 1995 and 2000 and
correspond to integrated luminosities of 14.7~\ivpb\ of data taken with \cfm
energy 91~GeV, 11.3~\ivpb\ of LEP1.5 data with \cfm energies between 130~GeV and
136~GeV and 707.4~\ivpb\ of LEP2 data with \cfm energies ranging from 161 to
209~GeV.  The 91~GeV data, known as \PZo-calibration data, were primarily
collected for calibrating parameters used in the OPAL reconstruction algorithms.
 This \PMZ\ sample had the same detector configuration as the other \cfm energy
points. The exact breakdown of the \cfm energies together with the respective
luminosities and numbers of selected events are given in Table~\ref{tab:lumi}. 
The thirteen points in Table~\ref{tab:lumi} represent the main samples of the
spread of energies in the LEP1.5 and LEP2 data.

The data were combined into four datasets. The LEP1.5 data provided a single
energy point at an event-weighted \cfm energy of 133~GeV, while the LEP2 data
were split into two energy points, one with an event-weighted \cfm energy of
177~GeV using data in the range 161--185~GeV (with a total integrated
luminosity of 78.1 \ivpb) and another at 197~GeV (with a total integrated
luminosity of 628.3 \ivpb) using data in the range 188--209~GeV.  Together
with the \PZo-calibration data this provided for a determination of \alsx at
four \cfm energies.

A number of Monte Carlo samples were created to correct for detector acceptance
and resolution effects, to correct for hadronization effects and to estimate the
contribution of background processes.  These Monte Carlo samples were produced
using a full simulation of the detector~\cite{op:gopal}, followed by the same
reconstruction and selection algorithms applied to the real data, and are
referred to as ``detector-level'' samples.  Other samples without the full
detector simulation are discussed in Section 5.2.

PYTHIA 6.150~\cite{pythia6.1} was used to provide the default Monte Carlo
samples (for the process \mbox{\epem$\rightarrow\rm{Z}/\gamma^{\ast}\rightarrow\
$\qqbar$\rightarrow$ hadrons}) which were used to correct the high energy
datasets.  The \PZo-calibration dataset was corrected using
JETSET~7.408~\cite{jetset7.4}.  The use of JETSET for the lower energy data is a
matter of convenience only, and not due to any inconsistencies in PYTHIA at this
energy.  Any differences between the two generators is expected to be negligible.
Hadronization corrections were evaluated by
comparing results with an alternative Monte Carlo sample,
HERWIG~6.2~\cite{herwig6.2} which uses the cluster model of hadronization. This
was compared with the string model of hadronization in PYTHIA.  The parameters
which were involved in the Monte Carlo simulation, both for JETSET/PYTHIA and
HERWIG, were tuned to OPAL data collected at the \PZo peak, including global
event shapes, particle multiplicities and fragmentation
functions~\cite{opaltune,opaltune_hw62}.  The generation of the initial
quark-antiquark pair for each Monte Carlo sample was implemented at LEP2 using
the $\mathcal{KK}$2f~4.13 event generator~\cite{kk2f}, which has an improved
description of photon production in the initial and final states with respect to
the one currently implemented in the PYTHIA generator.  The available
detector-level Monte Carlo samples are listed in Table~\ref{tab:lumi}.

Above the \WWx production threshold (161 GeV), the main background was expected
to come from four-fermion events (\mbox{\epem$\rightarrow$\WW$\rightarrow$4f}),
in particular those events in which two or all four of the fermions were quarks.
The contribution of these backgrounds in data was estimated using Monte Carlo
samples generated using KORALW 1.42~\cite{koralw} (for
q$\bar{\rm{q}}$q$^{\prime}\bar{\rm{q}}^{\prime}$ and
q$\bar{\rm{q}}\ell\bar{\ell}^{(\prime)}$ where $\ell=\mathrm{e},\mu,\tau,\nu$
but $\ell\bar{\ell}\neq \mathrm{e}^+\mathrm{e}^-$) and grc4f 2.1~\cite{grc4f}
(for eeq$\bar{\rm{q}}$). Grc4f 2.1 was used to generate all the expected
four-fermion background samples for the 161 and 172~GeV data. The background
distributions were normalized to the luminosity of the dataset and subsequently
subtracted from the measured distributions. The LEP1.5 energies were well below
the \WWx and \ZZ\ production thresholds~\cite{wwzzcross} and were therefore
expected to have no significant four-fermion backgrounds.  The total expected
background contribution from 
``four-fermion'' \eeqqqq\
events is 1.2\% of the
combined LEP1.5 data sample and it was neglected in the analysis.   

\section{Jet Rate Distributions}

Jets were formed from the final state objects by applying jet clustering
algorithms.
These algorithms use the kinematic and spatial (geometric) properties of
the individual objects in order to classify them as belonging to a specific jet.
 We used here the
Durham~\cite{durham}, Cambridge~\cite{camb}, JADE~\cite{jadealg} and the $R$ and
$\varepsilon$
variants of the Cone~\cite{cone} jet clustering algorithms. 

The Durham and Cambridge algorithms construct a test variable built from the
energy
and angular separation between two particles,
\begin{eqnarray*}
y_{ij}=\frac{2\min\{E_{i}^{2},E_{j}^{2}\}(1-\cos\theta_{ij})}{E_{\rm{vis}}^{2}}
\end{eqnarray*}
where $E_{i}$ is the energy of particle $i$, $\theta_{ij}$ the angle between the
particle $i$ and
$j$ and $E_{\rm{vis}}$ is the total visible energy in the event. The pair that
produces the smallest
value of $y_{ij}$ is chosen first.  The value of this test variable is compared
to a predefined
parameter, $\yc$, called the jet resolution parameter.  If the test variable is
smaller than
$\yc$ particles $i$ and $j$ are merged into a pseudo-particle.  Merging means
that the momenta
of particles $i$ and $j$ are removed from the set of momenta and the the sum of
their
four-momenta is added to the set of momenta.  After the merging, the clustering
starts again 
using the momentum set and it continues until all test variables become larger
than $\yc$. 
After the clustering stops, all remaining (pseudo-) particles are classified as
jets.

The Cambridge algorithm differs slightly from Durham in its implementation.
In the Cambridge algorithm particles are first paired together by minimizing
the variable $v_{ij} = 2(1-\cos \theta_{ij})$.  The standard test variable is
then
constructed and compared to the jet resolution parameter, \yc.  The procedure
followed is then identical to that of the Durham algorithm, except that
Cambridge freezes out
soft jets by accepting only the lowest energy (pseudo-)particle as the jet when
$y_{ij}>\yc$. 
The number of jets reconstructed in the event, using Durham or Cambridge, is
therefore a function of the jet resolution parameter. The JADE algorithm follows
the same procedure
as the Durham algorithm; however, it uses the scaled invariant mass,
$y_{ij}=2E_{i}E_{j}(1-\cos\theta_{ij})/E_{\rm{vis}}^{2}$, of particles $i$ and
$j$ as the test
variable.

In the Cone jet finding algorithm, a jet is defined as a set of particles whose
three-momentum vectors lie inside a cone of half angle $R$, where the direction
of the sum of their
three-momentum vectors defines the cone axis. In addition, the total energy of
the particles
assigned to a jet is required to exceed some minimum value $\varepsilon$.  
Typical values are $R=0.7$~rad and $\varepsilon=7$~GeV for jets in \epem
annihilation at LEP~1 energies.  When analysing events at the detector level, we
replaced
$\varepsilon$ by $\varepsilon^{\prime}=\varepsilon\cdot E_{\rm{vis}}/$\roots to
compensate for the
incomplete detection of the energy of the event.  In our studies, the jet rate
was computed at
fixed $\varepsilon=7$~GeV as $R$ was varied, and at fixed $R=0.7$ as
$\varepsilon$ was varied. The
former is sensitive to the angular structure of jets, and the latter to their
energy distribution.

The fraction of multihadronic events in a given sample that are classified as
containing $n$ jets for a given value of the jet resolution parameter ($\yc$,
$R$ or $\varepsilon$) is referred to as the $n$-jet rate. This $n$-jet rate is
explicitly defined as
\begin{equation}
R_{n}(\yc)=\frac{\sigma_{n}(\yc)}{\sigma_{\rm{tot}}}\equiv
\frac{N_{n}(\yc)}{N_{\rm{tot}}},
\end{equation}
where $\sigma_{n}$ is the cross-section for the production of a hadronic event
with $n$ jets 
at fixed $\yc$, $\sigma_{tot}$ is the total hadronic cross-section, $N_{n}(\yc)$
is the number of events in a sample with $n$ jets for a given value of $\yc$ and
$N_{tot}$ is the total number of events in that sample.

The differential $n$-jet rate was also determined.  It is the derivative of the 
$n$-jet
rate with respect to $\yc$,
\begin{eqnarray}
   D_{n}(\yc)&=&\frac{\mathrm{d}R_{n}(\yc)}{\rm{d}\yc} .
\end{eqnarray} 
For the case when $n=2$, the differential 2-jet rate reduces to $D_{2}=y_{23}$,
where $y_{23}$ is the value of the jet resolution parameter where the event
flips from a 2- to a 3-jet event.  When the Durham algorithm is used to define
jets the value of $D_{2}$ (denoted $y_{23}^{D}$) is also an event shape
variable.

The average number of jets per event in a given sample, as a function of the jet
resolution
parameter, is defined to be
\begin{eqnarray}
\avn(\yc)&=&\frac{1}{\sigma_{\rm{tot}}}\sum_{n}n\sigma_{n}(\yc) \nonumber\\
&=&\frac{1}{N_{\rm{tot}}}\sum_{n}nN_{n}(\yc).
\end{eqnarray}
A QCD prediction which matches an \order{\alsn{2}} (next-leading order)
prediction, based on the QCD matrix elements~\cite{qcd_matrix_elements}, with a
resummed, next-leading logarithmic approximation (NLLA)~\cite{NLLA} prediction,
such that terms that appear in both predictions are not double counted, was
fitted to data. In this analysis we used the $\ln R$ matched
$D_2$~\cite{improved_d2,coefficients,durham} and $\avn$~\cite{peter_pub,
als_pred} predictions to fit to the distributions of the observables. The
differential and average jet rates were determined using the Durham and
Cambridge clustering algorithms, since resummed predictions only exist for these
algorithms.  This provided four separate observables ($D_{2}^{D}$, $D_{2}^{C}$,
$\avn^{D}$ and $\avn^{C}$) which were used to determine a value of \alsx at the
four different \cfm energy values. 

\section{Analysis Procedure}

\subsection{Selection Method}

\subsubsection{Preselection}

All events within a dataset were required to contain information from both the
central jet chamber
and the electromagnetic calorimeter, meaning both these subdetectors must have
been
flagged as being on and in good operational condition.  In addition events were
required
to be tagged as multihadronic in order to be preselected for analysis. 
Multihadronic
events were identified using the criteria described in~\cite{wwspr} for events
with \roots$>$\PMZ\ 
and in~\cite{tkmh} for \PZo-calibration events. To pass the preselection, an
event was required to
contain at least seven good tracks to reduce potential backgrounds arising from
the production of
$\tau$ leptons (\epem$\rightarrow\tau^+\tau^-$) decaying into hadrons and from
two-photon
interactions producing quarks. Good tracks were defined as those which had
\begin{itemize}
 \item at least 40 hits in the jet chamber
 \item at least 150 MeV/$c$ transverse momentum relative to the beam axis.  
 \item the distance of closest approach to the interaction point in the $r-\phi$
plane satisfying
$d_{0}\leq 2$~cm 
\item  the point of closest approach $\leq 25$~cm from the interaction point in
the $z$-direction
\end{itemize}
Clusters of energy in the calorimeters were also used in the analysis; good
clusters were defined as
those which produced a signal in at least one block in the barrel
electromagnetic calorimeter
corresponding to an uncorrected energy of at least 100~MeV or of 2 blocks in the
endcap
electromagnetic calorimeter corresponding to an uncorrected energy of 250~MeV.
The hadron
calorimeter was not used in this analysis.

All of the good quality tracks and clusters in the event were used to define
``objects''
representing particles using an algorithm (MT) to correct for double counting of
energy.  This MT
algorithm produced a uniquely defined array of track and cluster objects. The
trajectories of the
tracks measured in the central tracking chambers were extrapolated to the
clusters in the
electromagnetic calorimeters. If the energy of the cluster was less than
expected from the track,
then the cluster was omitted to avoid double counting of energy, since the
momentum resolution for
tracks was typically better than the calorimeter energy resolution.  If the
energy of the cluster
was larger than expected the energy of the cluster was reduced by the expected
amount with the
remaining energy interpreted as due to photons or neutral hadrons. These
remaining clusters and
those which were not matched defined the four-vectors of ``neutral'' particles. 
In all cases tracks
were treated as charged pions and neutral particles were treated as being
massless.

\subsubsection{Containment}

We ensured that most particles in the event were well contained in the detector
and
not lost down the beam line by imposing a cut on the direction of the thrust
axis~\cite{thrust1},
\begin{itemize}
\item $|\cos\theta_{\rm{T}}|<0.9$,
\end{itemize} 
where $\theta_{\rm{T}}$ is the angle between the beam axis and the direction of
the thrust axis.  The thrust axis direction was determined from all tracks and
clusters in the event,
without correcting for double counting with the MT algorithm.

\subsubsection{Initial State Radiation (ISR) Cuts}
The events of interest for this analysis were \eeqq\ events where the
final-state \qqbar pair had the full centre-of-mass energy. The effective
centre-of-mass energy of the \epem collision can be reduced by the emission of
one or more ISR photons.  At LEP2, approximately three quarters of the
multihadronic events were such ``radiative return events'', where the invariant
mass of the \qqbar pair was close to the \PZo mass. The effective centre-of-mass
energy of the collision after ISR, $\sqrt{s^{\prime}}$~\cite{sprime}, was
evaluated, and the requirement
\begin{itemize}
\item $\sqrt{s}-\sqrt{s^{\prime}} <$10~GeV 
\end{itemize}
was imposed to select full-energy events.

To calculate $\sqrt{s^{\prime}}$, all isolated photon candidates with energies
greater than 10 GeV were identified.   The Durham jet reconstruction algorithm
[31] was then used to group the remaining tracks and clusters into jets. ISR
photons are often emitted close to the beam direction. Three kinematic fits were
performed, under the assumptions that
\begin{itemize}
 \item there were two undetected photons (in opposite directions along the beam
pipe)
 \item there was one undetected photon
 \item all photons were observed in the detector,
\end{itemize}
respectively. The fit with the most acceptable $\chi^{2}$ was selected, and
$\sqrt{s^{\prime}}$ was calculated
from the invariant mass of the jets, excluding any photons.

The power of this cut can be seen in Figure 1. The efficiency for selecting
non-radiative \qqbar
events is given in Table 2. The purity of non-radiative events was found to be
approximately
73\% in all of the LEP1.5 and LEP2 data samples.  Non-radiative \qqbar events
are defined as those in which $\sqrt{s}-\sqrt{s^{\prime}_{\rm{true}}}<1$~GeV,
where $s^{\prime}_{\rm{true}}$ was determined from generator-level information
in the PYTHIA samples.  This ISR cut was applied to all analyzed datasets with
the exception of the \PZo calibration data.

\subsubsection{Final Cuts}

The dominant background to the process \mbox{ \epem$\rightarrow
\rm{Z}/\gamma^{\ast}\rightarrow$\ \qqbar$\rightarrow$\ hadrons} at LEP2 came
from the four-fermion process \mbox{\epem$\rightarrow$\WW} in which one or both
of the bosons decayed hadronically, producing two or four quarks in the final
state. This background was expected to make up approximately 30\% of all
observed events which pass the first stage of cuts in each of the LEP2 datasets.
These backgrounds were addressed by placing a cut~\cite{joost} on two likelihood
values which indicate how likely an event is to be a non-QCD four-quark or a
semi-leptonic event:
\begin{itemize}
\item \lqqqq $< 0.25$
\item \lqqln $< 0.50$
\end{itemize} 
The effect of these cuts in each of the LEP2 datasets and the expected
backgrounds can be seen on Figure~\ref{fig:qqqq}.

The four-quark likelihood value~\cite{wwlikelihood}, \lqqqq, was estimated from
four kinematic variables describing characteristics of hadronic \WW\ decays like
their four-jet nature and angular structure.  These variables were used to
construct event probabilities based on two hypotheses: first, that the event was
due to a hadronically decaying \WW\ pair
(\mbox{\WW$\rightarrow$\qqbar\h\qqbar}) and, second, that the event was due to a
hadronically decaying $\rm{Z}/\gamma^{\ast}$
(\epem$\rightarrow\rm{Z}/\gamma^{\ast}\rightarrow$\qqbar).  The probabilities
were combined to produce the discriminating likelihood, \lqqqq.   This cut
reduced the expected background by approximately 80\% so that it constituted
only 9\% of the observed number of events.

The semi-leptonic likelihood~\cite{wwlikelihood}, \lqqln, was based upon three
separate likelihoods, one for each lepton species ($\ell=$e, $\mu$, $\tau$). 
Each of these likelihoods was based on ten variables describing the properties
of the lepton, the jets produced by the \qqbar pair and the missing energy
carried away by the neutrino.  This cut in conjunction with the cut on the
four-quark likelihood removed almost 90\% of the background expected in the
observed LEP2 dataset. The effect of these cuts can be seen in
Figure~\ref{fig:qqqq}.  

These likelihood cuts also reduced the backgrounds arising from
\mbox{\epem$\!\rightarrow$\ZZ} in which one or both of the \PZo bosons decay
hadronically. \ZZ\ production contributed only a small fraction of the 
background due to its 
lower cross-section compared to \WW\ production in the energy
ranges used in this experiment.  The likelihood
cuts were applied only to those datasets with \roots$\geq$161~GeV.

The expected size of the total background contribution to each dataset was
determined by Monte Carlo predictions after scaling to the luminosity of the
dataset.  The effect of the final cuts  and the expected four-fermion
backgrounds for each \cfm energy dataset can be seen in Table~\ref{tab:cuts}. 
As seen in the table the likelihood cuts greatly increased the purity of
selected non-radiative \qqbar events.  The LEP2 datasets data with \roots $\geq$
183~GeV were typically $\sim$70\% pure following the ISR cuts; however, after
the final cuts this increased to a purity of 94--95\%.

\subsection{Monte Carlo Corrections}

The values of the variables $R_n$, $D_n$ and $\avn$ were determined for each
accepted event using the MT-corrected tracks and clusters. These values were
then compiled into histograms as a function of \yc\ with bins of varying size. 
The background rejection cut did not completely remove all of the expected
background events from \WW\ and \ZZ\ production (referred to as four-fermion
background in this paper).  The remaining backgrounds, taken from the Monte
Carlo, were subtracted from the corresponding measured distributions on a
bin-by-bin basis.  Systematic uncertainties in this procedure will be discussed in
\mbox{Section 6}.

Corrections to the distributions were also made for effects arising from finite
detector resolution and a limited detector acceptance (recall that the cut on
$|\cos\theta_{\rm{T}}|$ 
reduced the fiducial volume) and for residual ISR events which were not removed
by the
$\sqrt{s^{\prime}}$ cut. These corrections were done separately for each
variable and were
accomplished by comparing distributions from two separate Monte Carlo samples,
one of which had gone through a full detector simulation, including effects of
detector resolution and acceptance and initial state radiation, called the
``detector'' level.  The other sample used only the generator-level hadrons and
had not gone through the detailed
detector simulation.  In this sample all short-lived particles ($\tau\leq3\times
10^{-10}$s) had decayed and a requirement that
$\sqrt{s}-\sqrt{s^{\prime}_{\rm{true}}}<1$~GeV was imposed. This ``hadron
level'' sample was thus expected to produce distributions arising solely from
the properties of the underlying hadrons, free of any detector biases determined
over the full acceptance without any limitations arising from limited
resolution. 

Correction factors for each bin of the distributions were determined from the
ratio of the two Monte Carlo distributions. Thus, any bin, $i$, of the measured
distributions was corrected via
\begin{equation}
      \mathcal{H}_{i}^{\rm{data}}=\left(\frac{\mathcal{H}_{i}^{\rm{MC}}}
      {\mathcal{D}_{i}^{\rm{MC}}}\right)(\mathcal{D}_{i}^{\rm{data}}-
      \mathcal{D}_{i}^{\rm{bkgd}})
\end{equation}
where $\mathcal{H}$ and $\mathcal{D}$ represent distributions at the hadron
level and
the detector level respectively, and $\mathcal{D}_{i}^{\rm{bkgd}}\!\!$
corresponds to the expected size of the total background in bin $i$.  The hadron
level was used in this analysis when determining jet rate distributions;
theoretical predictions which were fitted to these distributions were obtained
from computations valid at the ``parton level''. The parton level corresponds to
distributions that would be produced if only the partons created immediately
following the \epem annihilation and before the hadronization phase were used in
the analysis.  The parton level Monte Carlo sample was built from quarks and
gluons that were produced during the parton shower simulated by the generator
before the hadronization phase began.  As in the case of the hadron level, the
parton level sample gave rise to distributions that were free of initial state
radiation without any detector simulation applied.   The correction factor
determined from the ratio of the hadron level to the parton level was applied to
the theoretical predictions before the fitting procedure. This factor corrected
the prediction to the hadron level so that it could be compared to the corrected
hadron level distribution determined from the data,
\begin{equation}
\mathcal{H}_{i}^{\rm{pred}}=\left(\frac{\mathcal{H}_{i}^{\rm{MC}}}
      {\mathcal{P}_{i}^{\rm{MC}}}\right)\mathcal{P}_{i}^{\rm{pred}}
\end{equation}
where $\mathcal{H}$ and $\mathcal{P}$ represent distributions at the hadron
level and
the parton level respectively. Statistical
errors on the data distributions included the effects of Monte Carlo statistics.

\section{Systematic Uncertainties}

\subsection{Experimental Systematic Variations}

Several selection algorithms and selection cuts were varied to determine their
impact on the results of the analysis.  In all cases the result from the
variation was compared to the result from the standard selection, and the
difference was taken as a contribution to the total systematic error.  The
systematic variations that were used (given in descending order of their
contributions to the overall experimental uncertainty) are:

\begin{description}
\item[Track-Cluster Matching] Systematic errors relevant to the definition of
objects defined
within the tracking chambers and calorimeters, and hence used for jet
clustering, were estimated by comparing the results using the MT package to a
method using all selected tracks and clusters without taking into account the
possibility of double counting.

\item[Detector Correction] The uncertainty in modelling the detector was
investigated by using HERWIG Monte Carlo datasets in place of PYTHIA to correct
from the detector level to the hadron level.  The hadronization correction was
still performed using the PYTHIA Monte Carlo samples.

\item[Containment] The constraint on the direction of the thrust axis was
tightened to
$|\cos \theta_{T}|<0.7$, restricting events to the barrel region.

\item[ISR]A possible systematic effect introduced through the selection of
events with
little or no initial state radiation was estimated by repeating the selection
using a second method of ISR determination~\cite{wwspri}. This alternative
method assumed while performing the kinematic fit that there was always a single
photon which either escaped undetected down the beam line or was detected in the
electromagnetic calorimeter.  

\item[\boldmath \lqqqq and \lqqln]  To account for the systematic uncertainty
that arises from the value of the cut placed on the \WW\ hadronic and
semi-leptonic likelihoods, the values of the likelihood cuts were changed to 0.1
and 0.4 for \lqqqq and to 0.25 and 0.75 for \lqqln.  In each case the largest
deviation was taken as the systematic uncertainty.

\item[Backgrounds]To account for uncertainties introduced during background
subtraction, arising from imprecise knowledge of the four-fermion
cross-sections, these cross-sections were conservatively varied by $\pm$5\% and
the largest deviation from the standard value
was used to determine an overall systematic error.

\end{description} 

It should be noted that there was no single dominant contribution to the overall
experimental
systematic error.  In general, the largest contributions occurred for the
track-cluster matching and detector correction variations, while the least
significant contributions came from the variations of the background
cross-section and the semi-leptonic likelihood.

\subsection{Hadronization Systematic Variations}

Systematic uncertainties arise from the modelling of the hadronization. These
were estimated by using HERWIG Monte Carlo samples, which employ a different
hadronization model, in the determination of the theoretical prediction
correction factor (Eqn 5). The PYTHIA-HERWIG differences were taken to be the
systematic uncertainties from hadronization. The statistical component due to
limited Monte Carlo statistics was included in the total  determination of the
hadronization uncertianty.  


\subsection{Theoretical Systematic Variations}

Three further systematic variations were considered when fitting the predictions
to the data to determine \alsx.  
\begin{description}
\item[Fit Range]We investigated the choice of the range of bins used in
   performing the fit
   of the theoretical prediction to the data.  The fit range was first
   increased by two bins, by adding one bin to each endpoint of the fit range.
   In the case where one of the endpoints was
   already at the maximum allowed bin value, only the other point was
   extended by one bin.  A second variation decreased the fit range by
   two bins, by removing a bin from each endpoint of the fit range.  The
   largest deviation from the standard fit value was taken as the
   contribution to the systematic error.

\item[Renormalization Scale]The second fit-related systematic variation
accounted for the  
   uncertainty due to the  dependence of the
fixed order and resummed predictions on the renormalization scale, $x_{\mu}$,
where $x_{\mu}=\mu/\sqrt{s}$.  The value
 of $x_{\mu}$, which   was set to 1.0 in the standard fits, was varied to 0.5
and 2.0 respectively.  The largest deviation from the standard fit value was
taken as the contribution to the systematic error representing renormalization
scale uncertainty.

\item[Logarithm rescaling, \boldmath $x_{L}$] In the resummation process for
event shape  
   variables, like $D_{2}$, there is an arbitrariness due to the definition of
the logarithms which were resummed. In this analysis we used $\als\ln(1/\yc)$;
however, this could be generalized to powers of
$\als\ln[1/(x_{L}\,\yc)]$~\cite{xl}.  The standard value for this rescaling
($x_{L}$=1) was varied to $\frac{4}{9}$ and $\frac{9}{4}$, to investigate the
systematic effect of this arbitrariness.  No analoguous rescaling prescription has yet been developed for the case of the average jet rates, so the $x_{L}$ variations are
only shown for the $D_{2}$ distributions for comparison to the $x_{\mu}$
variation.  Hence this variation was not used in the determination of the total
theoretical systematic error and was included only as a cross-check and for
comparison with other analyses.  A comprehensive study of the combination of
various theoretical variations to produce a global theoretical uncertainty for
event shape distributions is given in~\cite{evsh_syst}.

\end{description}

The differences between the standard result and those due to the above
variations were
separated into three categories: experimental, hadronization and theory (see
Table 3).  The
variations in each of these categories were added in quadrature and the result
taken as the
systematic uncertainty for that category.  In the case of asymmetric errors, the
error was
symmetrized by taking the largest systematic variation and applying it as the
full systematic
contribution. A summary of the systematic variations is provided in
Table~\ref{tab:systerr}.

\section{Results}

Data from thirteen datasets were used in this analysis: one \PZo-calibration
dataset,
two LEP1.5 datasets and ten datasets from LEP2.  These thirteen datasets were
combined to
produce four higher statistics datasets (\roots=\PMZ, 133, 177 and 197~GeV)
which were analysed separately. The raw distributions ($n$-jet fractions,
$D_{2}$ and $\avn$) for each of the datasets were determined as functions of the
jet resolution parameters defined using four different jet clustering
algorithms. The distributions underwent a bin-by-bin correction which included
subtraction of expected backgrounds and correction for detector and residual ISR
effects. Systematic effects were examined by varying the parameters used in
selecting events (see Table 3). The difference between the corrected
distributions using these variations and those from the standard selection then
determined the size of the systematic error on each bin of the distribution. 

Matched predictions of NLO and resummed calculations were fitted to the
corrected Durham and Cambridge $D_{2}$ and $\avn$ distributions over a
predetermined fit range (see Section 7.2.1), taking into account bin-to-bin
correlations in $\avn$ (see Section 7.2.2). These fits provided four values of
\alsx with statistical and systematic errors at each of the four \cfm energies. 
Taking into account the statistical and systematic correlations between the four
measurements, they were combined into a single \alsx result at each energy.  

\subsection{\boldmath $n$-Jet Fractions}
The $n$-jet fractions for the \PZo-calibration sample and those for the LEP1.5
and the two LEP2 samples are shown in Figures~\ref{fig:cn91} to~\ref{fig:jr103}.
 Each plot shows the fraction of events in a given sample that were determined
to be $n$-jets for a given value of the jet resolution parameter at the hadron
level\footnote{Further details of the data will be made available in the HEPDATA
database, http://durpdg.dur.ac.uk/HEPDATA.}. The jet fractions were calculated
using four different clustering algorithms.  For the Cone algorithm, results
from using both the $R$ and $\varepsilon$ resolution parameters are plotted,
showing the individual $n$-jet fractions for $n\leq 2$, $n=3$ and $n\geq 4$ in
Figures~\ref{fig:cn91} to~\ref{fig:cn103}. Similarly the $n$-jet fractions for
the JADE algorithm are shown in Figure~\ref{fig:jd91} for $n=2,3,4,5$.
Figures~\ref{fig:jr91} to~\ref{fig:jr103} show the individual $n$-jet fractions
for $n=2,3,4,5$ for the Durham and Cambridge algorithms respectively. The error
bars on the points represent the total statistical and systematic errors added
in quadrature. The Monte Carlo expectations corresponding to PYTHIA and HERWIG
are also displayed on each plot, for each algorithm and energy.  The Monte Carlo
expectations match the measured $n$-jet fractions reasonably well.

The differential two-, three- and four-jet rates, $D_{n}$, are plotted as a
function of \yc\ for the Durham and Cambridge algorithms for the
\PZo-calibration, the LEP1.5 and the two LEP2 data samples in
Figures~\ref{fig:dn91} and \ref{fig:dn186} respectively.   Similarly, the
average jet rates are plotted as a function of \yc\ for the Durham and Cambridge
algorithms for the \PZo-calibration, the LEP1.5 and the two LEP2 data samples
in Figures~\ref{fig:avn91} and \ref{fig:avn186} respectively.  The curves on all
the plots represent the expected Monte Carlo distributions.  There is good
agreement between the data and the expectations from both PYTHIA and HERWIG.   

\subsection{Fits to Determine the Value of \boldmath\als}
\subsubsection{Differential Two-jet Rates}

The range over which the $D_{2}$ and $\avn$ distributions were fitted was
determined by splitting the 91~GeV (and 189~GeV) PYTHIA Monte Carlo samples into
100 statistically independent subsamples. 
The QCD predictions for each distribution were fitted to each of these
subsamples, with $\alpha_\mathrm{s}$ as a free parameter, for a number of  
possible end-points of the fit range (requiring
that the fit range be at least six bins).  A $\chi^{2}$ per degree of freedom
was determined for each fit range.  The $\chi^{2}$ values were averaged over the
100 subsamples. The fit range that produced the smallest average $\chi^{2}$ per
degree of freedom was chosen to be the default fit range for the
\PZo-calibration (and high energy) datasets.  Where no clear $\chi^{2}$ minimum
was found, the largest reasonable range was chosen. The size of the fit range
was then
adjusted to ensure the range did not extend into a region where the
hadronization corrections
exceeded 10\%, in particular at smaller \yc\ values.  A further adjustment was
made to exclude fit ranges where one of the endpoints produced a contribution of
more than 30\% to the overall $\chi^{2}$ value. Potential correlations
introduced in the correction of the distributions were accounted by including a
covariance matrix in the $\chi^2$ fit.  The covariance matrices were determined
in the manner detailed in Section 7.2.2.

The fits of the $\ln R$ matching prediction to the $D_{2}$ distribution are
shown in
Figure~\ref{d2cfits} for the Cambridge algorithm and in Figure~\ref{d2dfits} for
the Durham
algorithm.  The values of \alsx determined from the $\ln R$ fits for the four
datasets, together with a complete breakdown of statistical and systematic
uncertainties, are given
in Tables~\ref{d2_camb_tab} and~\ref{d2_durh_tab} for the Cambridge and
Durham\footnote{Note
that~\cite{mford_event_shapes} also determines \alsx using $D_{2}^{D}$, denoted
there as
$y_{23}^{D}$.  The small differences between the results have been investigated
in detail, and are not significant.  They may be attributed to differences in
fit regions, the use of statistically different Monte Carlo samples, and the
adoption of slightly different strategies for the assessment of theoretical
errors.} algorithms respectively.

\subsubsection{Average Jet Rates}

For the average jet rates, $\avn$, the statistical errors were strongly
correlated between points, since the same events were used to determine $\avn$
at each value of $\yc$. The fit was performed using a correlated $\chi^{2}$ fit
in which the covariance matrix was determined from the PYTHIA Monte Carlo sample
divided into many detector level subsamples.  Each subsample was then corrected
to the hadron level using a second, statistically independent, PYTHIA sample.
There were 1500 subsamples created for the \PZo-calibration dataset and 1000
subsamples created for the high energy datasets.  The number of subsamples was
chosen so that the elements of the covariance matrix would be stable and would have 
negligible fluctuations.  These subsamples were used to build a standard covariance matrix,
which was then converted into a correlation matrix. The statistical errors on
the bins of the data distributions were then applied to this correlation matrix
to produce the covariance matrix used in the fits.

An example of the fit of the $\ln R$ matching prediction to the $\avn$
distribution is seen in Figure~\ref{ncfits} for the Cambridge algorithm and in
Figure~\ref{ndfits} for the Durham
algorithm. The values of \alsx determined from the $\ln R$ fits for the four
datasets, together with a complete breakdown of statistical and systematic
uncertainties, are given in
Tables~\ref{avn_camb_tab} and~\ref{avn_durh_tab} for the Cambridge and Durham
algorithms
respectively. 

\subsubsection{Running of \boldmath\als}

The four values of \alsx determined from the $D_{2}$ and $\avn$ distributions
were combined into a single value at each \cfm energy.  The large statistical
correlations between the four
values were handled in a manner similar to that for the bin-to-bin correlations
of the average jet rates. For each of the four distributions, 1000 Monte Carlo
samples were used to determine the statistical correlations between the \alsx
values.  The correlation matrices determined for the four separate energy points
are given in Table~\ref{als_corr}. Using the statistical error for the \alsx
value from each observable, the statistical covariance matrix was then
determined. The full covariance matrix also included contributions from
experimental, hadronization and theoretical uncertainties
\begin{equation}
V=V_{\rm{stat}}+V_{\rm{expt}}+V_{\rm{hadr}}+V_{\rm{theo}}.
\end{equation}
A weight was determined for each of the \alsx values from the inverse of the
covariance matrix,
\mbox{$w_{i}=\sum_{j}(V^{-1})_{ij}/\sum_{ij}(V^{-1})_{ij}$}. The combined \alsx
value was then determined from the weighted sum of the \alsx values. 
Only statistical and experimental systematic uncertainties were allowed to
contribute to the off diagonal elements of the covariance matrix $V$, to ensure
undesirable features such as negative weights were avoided.  Hadronization and
theoretical systematics were added only to the diagonal elements of the
covariance matrix.

The statistical and experimental uncertainties on the combined \alsx value were
determined from the product of the weights with the individual covariance
matrices~\cite{blue_method},
\begin{equation}
\sigma_{\rm{err}}^{2}=w^{T}V_{\rm{err}}w \qquad \rm{where\ err=stat,expt}.
\end{equation}
The hadronization and theoretical systematic uncertainties were determined by
repeating the combination for each systematic variation separately using the same
weights. The difference between these systematic combinations and the central
value is taken as the systematic contribution to the error on the central value.
For the experimental systematic covariance matrix, off diagonal elements were
determined using a ``minimum overlap" method,
\begin{equation}
{(V_\mathrm{expt})}_{ij}=\max[{(V_\mathrm{expt})}_{ii},{(V_\mathrm{expt})}_{jj}].
\end{equation}
  The values of \alsx determined for each
\cfm energy are given in Table~\ref{tab_blue} along with the breakdown of the
uncertainties, both statistical and systematic.  A comparison of the combined
\alsx values in Table~\ref{tab_blue} with those determined from the individual
$D_{2}$ and $\avn$ distributions is seen in Figure~\ref{asrun_comp}. Taking the
combined \alsx values at each \cfm energy as an input, the value of \alsx can be
run back to the \PZo pole using an \order{\alsn{3}} prediction.  These
\als(\PMZ) values are also shown in Table~\ref{tab_blue} and plotted against the
world average value of \als(\PMZ)=0.1187$\pm$0.0020~\cite{pdg} in
Fig~\ref{as_comp_mz}.  Using these values and taking into account proper
statistical and systematic correlations, a weighted mean of
\als(\PMZ)=$0.1177\pm0.0006$(stat.)$\pm0.0012$(expt.)$\pm0.0010$(had.)$\pm0.0032
$(theo.) is
determined. The four combined \alsx values are plotted in Figure~\ref{fit:asrun}
as a function of the \cfm energy, compared to the \order{\alsn{2}} energy
evolution of \alsx based on the determined value of \als(\PMZ).

\section{Summary and Conclusion}

Data from twelve LEP~1.5 and  LEP~2 datasets, with \cfm energies ranging from
130~GeV to 209~GeV, were combined into three higher statistics datasets.  These
datasets and one combining several \PZo-calibration runs at 91~GeV were used to
determine the $n$-jet fractions, the differential $n$-jet rates and the average
jet rates for each of the energies.  The different jet multiplicity
distributions were compared to both PYTHIA and HERWIG Monte Carlo expectations.

Hadron-level $n$-jet fractions were determined using four jet-clustering
algorithms, Cone, JADE, Durham and Cambridge.  For the Cone algorithm,
measurements of the fraction of events with $n\leq 2$, 3, $\geq$4 jets were
presented as functions of $R$ and $\varepsilon$. In the case of JADE, Durham and
Cambridge, measurements of the 2-, 3-, 4-, and 5-jet fractions were presented as
functions of the jet resolution parameter, \yc.  In all cases there was
generally good agreement between the measured jet fractions and the Monte Carlo
expectations.

Hadron-level determinations of the differential $n$-jet and average jet rates
were performed for the Durham and Cambridge algorithms.  The differential
two-jet rate, $D_{2}$, and the average jet rate, $\avn$ were used to determine
the value of \als(\roots) for each of the four combined energy points.  The
determinations were carried out by fitting the ln$R$ matching predictions,
appropriately corrected to the hadron level, to the hadron level data
distributions over an appropriate range of \yc, with \als(\roots) as the
variable parameter. 
The running of \als(\roots) was demonstrated by comparing the four values of
\alsx as determined from the combined datasets as a function of their \cfm
energies. 

Using the measured values of \als(\roots) a value of \als(\PMZ) was determined
for each of the four datasets.  A weighted mean taking account of correlations
determined a final value of the strong coupling at \roots=\PMZ\ of 
\begin{equation*}
\als(M_{\mathrm{Z}})=0.1177\pm0.0006(\mathrm{stat.})\pm0.0012(\mathrm{expt.}
)\pm0.0010(\mathrm{had.}
)\pm0.0032 (\mathrm{theo.})
\end{equation*}
where the error contains contributions from statistical, experimental,
hadronization and theoretical uncertainties. The error on the determined value
is slightly larger than that for the previous OPAL publication~\cite{peter_pub}
which also used resummed predictions for $D_{2}$ and average jet rate
distributions, but explored slightly different energy ranges, including 35 and
45~GeV, and all LEP energies up to only 189~GeV.  There is good agreement
between these four values of \als(\PMZ) measured in this analysis and previous
determinations of \alsx summarized in~\cite{als_ref} and with the world average
value of 0.1187$\pm$0.0020~\cite{pdg}. 

\section*{Acknowledgements}

We particularly wish to thank the SL Division for the efficient operation
of the LEP accelerator at all energies
 and for their close cooperation with
our experimental group.  In addition to the support staff at our own
institutions we are pleased to acknowledge the  \\
Department of Energy, USA, \\
National Science Foundation, USA, \\
Particle Physics and Astronomy Research Council, UK, \\
Natural Sciences and Engineering Research Council, Canada, \\
Israel Science Foundation, administered by the Israel
Academy of Science and Humanities, \\
Benoziyo Center for High Energy Physics,\\
Japanese Ministry of Education, Culture, Sports, Science and
Technology (MEXT) and a grant under the MEXT International
Science Research Program,\\
Japanese Society for the Promotion of Science (JSPS),\\
German Israeli Bi-national Science Foundation (GIF), \\
Bundesministerium f\"ur Bildung und Forschung, Germany, \\
National Research Council of Canada, \\
Hungarian Foundation for Scientific Research, OTKA T-038240, 
and T-042864,\\
The NWO/NATO Fund for Scientific Research, the Netherlands.\\

\bibliography{pr408}

\providecommand{\href}[2]{#2}\begingroup\raggedright\begin{thebibliography}{10}

\bibitem{als_ref}
S.~Bethke, {\em Nucl. Phys. Proc. Suppl.} {\bf 121} (2003)
74.
\newblock

\bibitem{als_lep1}
{ OPAL} Collaboration, P.~D. Acton {\em et al.}, {\em Z. Phys.} {\bf C59}
  (1993)
1.
\newblock

\bibitem{als_lep1.5}
{ OPAL} Collaboration, G.~Alexander {\em et al.}, {\em Z. Phys.} {\bf C72}
  (1996)
191.
\newblock

\bibitem{als_lep161}
{ OPAL} Collaboration, K.~Ackerstaff {\em et al.}, {\em Z. Phys.} {\bf C75}
  (1997)
193.
\newblock

\bibitem{als_lep172}
{ OPAL} Collaboration, G.~Abbiendi {\em et al.}, {\em Eur. Phys. J.} {\bf C16}
  (2000)
185.
\newblock

\bibitem{peter_pub}
{ JADE and OPAL} Collaborations, P.~Pfeifenschneider {\em et al.}, {\em Eur.
  Phys. J.} {\bf C17} (2000)
19.
\newblock

\bibitem{mford_event_shapes}
{ OPAL} Collaboration, G.~Abbiendi\etal, {\em Eur. Phys. J.} {\bf C40} (2005)
  287.

\bibitem{opaldetector}
{ OPAL} Collaboration, K.~Ahmet {\em et al.}, {\em Nucl. Instrum. Meth.} {\bf
  A305} (1991) 275.

\bibitem{opallumi_sw}
{ OPAL} Collaboration, G.~Abbiendi {\em et al.}, {\em Eur. Phys. J.} {\bf C14}
  (2000)
373.
\newblock

\bibitem{op:trig}
{ OPAL} Collaboration, M.~Arignon {\em et al.}, {\em Nucl. Instrum. Meth.} {\bf
  A313} (1992)
103.
\newblock

\bibitem{op:daq}
{ OPAL} Collaboration, J.~T.~M. Baines {\em et al.}, {\em Nucl. Instrum. Meth.}
  {\bf A325} (1993)
271.
\newblock

\bibitem{op:gopal}
{ OPAL} Collaboration, J.~Allison {\em et al.}, {\em Nucl. Instrum. Meth.} {\bf
  A317} (1992)
47.
\newblock

\bibitem{pythia6.1}
T.~Sj\"{o}strand {\em et al.}, {\em Comput. Phys. Commun.} {\bf 135} (2001)
238.
\newblock

\bibitem{jetset7.4}
T.~Sj\"{o}strand, {\em Comput. Phys. Commun.} {\bf 82} (1994)
74.
\newblock

\bibitem{herwig6.2}
G.~Corcella {\em et al.}, {\em JHEP} {\bf 01} (2001)
010.
\newblock

\bibitem{opaltune}
{ OPAL} Collaboration, G.~Alexander {\em et al.}, {\em Z. Phys.} {\bf C69}
  (1996)
534.
\newblock

\bibitem{opaltune_hw62}
{ OPAL} Collaboration, G.~Abbiendi {\em et al.}, {\em Eur. Phys. J.} {\bf C35}
  (2004)
293.
\newblock

\bibitem{kk2f}
S.~Jadach, B.~Ward, and Z.~W\c{a}s, {\em Comput. Phys. Commun.} {\bf 130}
  (2000)
260.
\newblock

\bibitem{koralw}
M.~Skrzypek, S.~Jadach, W.~Placzek, and Z.~W\c{a}s, {\em Comput. Phys. Commun.}
  {\bf 94} (1996)
216.
\newblock

\bibitem{grc4f}
J.~Fujimoto\etal, {\em Comput. Phys. Commun.} {\bf 100} (1997) 128.

\bibitem{wwzzcross}
{ ALEPH, DELPHI, L3, OPAL and SLD} Collaborations,
{A combination of preliminary electroweak measurements and constraints on the
Standard Model, 2003,}
\href{http://www.arXiv.org/abs/hep-ex/0312023}{{\tt hep-ex/0312023}}.
\newblock

\bibitem{durham}
S.~Catani, Y.~Dokshitzer, M.~Olsson, G.~Turnock, and B.~Webber, {\em Phys.
  Lett.} {\bf B269} (1991) 432.

\bibitem{camb}
Y.~L. Dokshitzer, G.~D. Leder, S.~Moretti, and B.~R. Webber, {\em JHEP} {\bf
  08} (1997)
001.
\newblock

\bibitem{jadealg}
{ JADE} Collaboration, W.~Bartel {\em et al.}, {\em Z. Phys.} {\bf C33} (1986)
23.
\newblock

\bibitem{cone}
F.~Abe\emph{ et al.}, {\em Phys. Lett.} {\bf D45} (1992) 1448.

\bibitem{qcd_matrix_elements}
R.~K. Ellis, D.~A. Ross, and A.~E. Terrano, {\em Nucl. Phys.} {\bf B178} (1981)
421.
\newblock

\bibitem{NLLA}
S.~Catani, L.~Trentadue, G.~Turnock, and B.~Webber, {\em Nucl. Phys.} {\bf
  B407} (1992) 3.

\bibitem{improved_d2}
A.~Banfi, G.~P. Salam, and G.~Zanderighi, {\em JHEP} {\bf 01} (2002)
018.
\newblock

\bibitem{coefficients}
G.~Dissertori and M.~Schmelling, {\em Phys. Lett.} {\bf B361} (1995) 167.

\bibitem{als_pred}
{ OPAL} Collaboration, P.~D. Acton {\em et al.}, {\em Z. Phys.} {\bf C59}
  (1993)
1.
\newblock

\bibitem{wwspr}
{ OPAL} Collaboration, K.~Ackerstaff {\em et al.}, {\em Eur. Phys. J.} {\bf C2}
  (1998)
441.
\newblock

\bibitem{tkmh}
{ OPAL} Collaboration, G.~Alexander {\em et al.}, {\em Z. Phys.} {\bf C52}
  (1991)
175.
\newblock

\bibitem{thrust1}
S.~Brandt, C.~Peyrou, R.~Sosnowski, and A.~Wroblewski, {\em Phys. Lett.} {\bf
  12} (1964)
57.
\newblock

\bibitem{sprime}
{ \OPAL} Collaboration, K.~Ackerstaff {\em et al.}, {\em Phys. Lett.} {\bf
  B391} (1997)
221.
\newblock

\bibitem{joost}
{ OPAL} Collaboration, G.~Abbiendi {\em et al.}, {\em Eur. Phys. J.} {\bf C16}
  (2000)
185.
\newblock

\bibitem{wwlikelihood}
{ OPAL} Collaboration, G.~Abbiendi {\em et al.}, {\em Phys. Lett.} {\bf B493}
  (2000)
249.
\newblock

\bibitem{wwspri}
{ OPAL} Collaboration, K.~Ackerstaff {\em et al.}, {\em Z. Phys.} {\bf C75}
  (1997)
193.
\newblock

\bibitem{xl}
M.~Dasgupta and G.~P. Salam, {\em JHEP} {\bf 08} (2002)
032.
\newblock

\bibitem{evsh_syst}
R.~W.~L. Jones, M.~Ford, G.~P. Salam, H.~Stenzel, and D.~Wicke, {\em JHEP} {\bf
  12} (2003)
007.
\newblock

\bibitem{blue_method}
L.~Lyons, D.~Gibaut, and P.~Clifford, {\em Nucl. Instrum. Meth.} {\bf A270}
  (1988)
110.
\newblock

\bibitem{pdg}
S.~Eidelman {\em et al.}, {\em Phys. Lett.} {\bf B592} (2004) 1.

\end{thebibliography}\endgroup


\newpage
\vspace{1.cm}
\begin{table}[H]
   \centering
   \begin{tabular}{|c|c|c|r|c||r|r|r|}
 \hline \multicolumn{1}{|c|}{} & \multicolumn{3}{|c|}{Energy (GeV)} &
  \multicolumn{1}{|c||}{Integrated} &  \multicolumn{3}{c|}{Number of
  Events} \\ \cline{2-4}\cline{6-8} Year & Nominal & Range & Mean &
   Luminosity & Data & JETSET/ & HERWIG \\  & &  & & (pb$^{-1}$)& &
  PYTHIA & \\ \hline \hline 1996--2000 & 91 &91.0--91.5  &  91.3 &
  14.7 & 426194 &  459k & 443k \\ \hline 1995, 1997 &130&130.0--130.3 
  & 130.1 &  5.3 &   1628 &   50k &  50k \\ 1995, 1997
  &136&135.7--136.2 & 136.1 &  6.0 &   1504 &   50k &  50k \\ \hline
  1996 & 161&161.2--161.6 & 161.3 &  10.1 &   1369 &  100k &  80k
  \\ 1996 & 172&170.2--172.5 & 172.1 &  10.4 &   1285 &  100k &  80k
  \\ 1997 & 183&180.8--184.2 & 182.7 &  57.7 &   6027 &  100k & 100k
  \\ \hline 1998 & 189&188.3--189.1 & 188.6 & 185.2 &  18559 &  500k
  & 100k \\ 1999 & 192&191.4--192.1 & 191.6 &  29.5 &   2866 &  200k
  & 100k \\ 1999 & 196&195.4--196.1 & 195.5 &  76.7 &   7076 &  200k
  & 100k \\ 1999, 2000 & 200&199.1--200.2 & 199.5 &  79.3 &   6676 &
   200k & 100k \\ 1999, 2000 & 202&201.3--202.1 & 201.6 &  37.8 &
  3225 &  200k & 100k \\ 2000 & 205&202.5--205.5 & 204.9 &  82.0 &
  6721 &  200k & 100k \\ 2000 & 207&205.5--208.9 & 206.6 & 138.8 &
                    10987 &  375k & 100k \\ \hline
   \end{tabular}
                            \vspace{.3cm}
  \caption{Integrated luminosity and the total number of preselected
  events for all samples. Included in this table are the Monte Carlo
    samples used to correct and make comparisons withthe data. The
     91~GeV dataset corresponds to the data collected during the
                       \PZo-calibration runs.}
   \label{tab:lumi}
\end{table}
                            \vspace{1.cm}
\begin{table}[H]
                                \small
                           \hspace*{-.2cm}
   \begin{minipage}[t]{.2\textwidth}
      \begin{tabular}{|c|l|c|c|}
     \hline       \multicolumn{2}{|c|}{Dataset} & ISR Fit & Final
  \\  \hline \hline 91 GeV &Data &   395695 & 395695 \\ \hline\hline 
130 GeV &Data &  318 & 318 \\ \cline{2-4}  &MC Expected &   368 $\pm$
 3 & 368 $\pm$ 3 \\ \cline{2-4}  &non-rad eff(\%)  & 85.4 $\pm$ 1.4 &
85.4 $\pm$ 1.4 \\ \hline\hline  136 GeV &Data  & 312 & 312
      \\ \cline{2-4}  &MC Expected  & 329 $\pm$ 3 & 329 $\pm$ 3
 \\ \cline{2-4}  &non-rad eff(\%)  & 85.5 $\pm$ 1.4 & 85.5 $\pm$ 1.4
    \\ \hline\hline  161 GeV &Data & 304 & 281 \\ \cline{2-4}  &MC
Expected & 299 $\pm$ 4 & 274 $\pm$ 3 \\ \cline{2-4}  &non-rad eff(\%)
& 83.4 $\pm$ 1.0 & 80.0 $\pm$ 1.0 \\ \cline{2-4}  &4f bkg frac (\%) &
6.2 $\pm$ 1.7 & 2.1 $\pm$ 1.4 \\ \hline\hline  172 GeV &Data & 280 &
     218 \\ \cline{2-4}  &MC Expected & 288 $\pm$ 7 & 232 $\pm$ 3
  \\ \cline{2-4}  &non-rad eff(\%) & 83.3 $\pm$ 1.0 & 79.6 $\pm$ 1.0
 \\ \cline{2-4}  &4f bkg frac (\%)  & 19.7 $\pm$ 2.2 & 4.4 $\pm$ 1.7
   \\ \hline\hline  183 GeV &Data & 1456 & 1077 \\ \cline{2-4}  &MC
  Expected  & 1478 $\pm$ 22 & 1083 $\pm$ 11 \\ \cline{2-4}  &non-rad
eff(\%) & 83.0 $\pm$ 1.0 & 79.0 $\pm$ 1.0 \\ \cline{2-4}  &4f bkg frac
(\%) & 27.3 $\pm$ 1.2 & 5.5 $\pm$ 1.2 \\ \hline\hline  189 GeV &Data &
   4448 & 3086 \\ \cline{2-4}  &MC Expected & 4497 $\pm$ 37 & 3130
$\pm$16 \\ \cline{2-4}  &non-rad eff(\%)& 83.0 $\pm$ 0.4 & 78.1 $\pm$
   0.4 \\ \cline{2-4}  &4f bkg frac (\%) & 30.0 $\pm$ 0.6 & 5.5 $\pm$
   0.6 \\ \hline
      \end{tabular}
   \end{minipage}
           \hspace*{5.25cm}\begin{minipage}[t]{.2\textwidth}
      \begin{tabular}{|c|l|c|c|}
  \hline \multicolumn{2}{|c|}{Dataset} & ISR Fit & Final  \\  \hline
   \hline 192 GeV &Data & 717 & 514 \\ \cline{2-4}  &MC Expected &
696$\pm$14 & 471$\pm$4 \\ \cline{2-4}  &non-rad eff(\%) & 82.8$\pm$0.7
  & 77.4$\pm$0.7 \\ \cline{2-4}  &4f bkg frac (\%) & 31.3$\pm$1.5 &
5.3$\pm$1.2 \\ \hline\hline  196 GeV &Data & 1732 & 1137
\\ \cline{2-4}  &MC Expected & 1746$\pm$24 & 1162$\pm$9 \\ \cline{2-4}
&non-rad eff(\%) & 82.7$\pm$0.7 & 77.1$\pm$0.6 \\ \cline{2-4}  &4f bkg
frac (\%) & 32.6$\pm$1.0 & 5.8$\pm$1.0 \\ \hline\hline  200 GeV &Data
     & 1636 & 1090 \\ \cline{2-4}  &MC Expected & 1717 $\pm$25 &
    1130$\pm$10 \\ \cline{2-4}  &non-rad eff(\%) & 82.6$\pm$0.7 &
76.9$\pm$0.7 \\ \cline{2-4}  &4f bkg frac (\%) & 33.6$\pm$1.0 &
6.0$\pm$1.0 \\ \hline\hline  202 GeV &Data & 806 & 519 \\ \cline{2-4}
&MC Expected & 804$\pm$16 & 527$\pm$5 \\ \cline{2-4}  &non-rad eff(\%)
  & 85.4$\pm$0.7 & 76.9$\pm$0.7 \\ \cline{2-4}  &4f bkg frac (\%) &
34.1$\pm$1.4 & 6.2$\pm$1.2 \\ \hline\hline  205 GeV &Data & 1687 &
     1130 \\ \cline{2-4}  &MC Expected & 1693$\pm$25 & 1089$\pm$9
    \\ \cline{2-4}  &non-rad eff(\%) & 82.4$\pm$0.7 & 76.3$\pm$0.7
    \\ \cline{2-4}  &4f bkg frac (\%) & 34.8$\pm$1.0 & 6.2$\pm$1.0
   \\ \hline\hline  207 GeV &Data & 2713 & 1717 \\ \cline{2-4}  &MC
Expected & 2807$\pm$32 & 1804$\pm$12 \\ \cline{2-4}  &non-rad eff(\%)
  & 82.7$\pm$0.5 & 76.5$\pm$0.5 \\ \cline{2-4}  &4f bkg frac (\%) &
34.7$\pm$0.8 & 6.2$\pm$0.6 \\ \hline
      \end{tabular}
   \end{minipage}
\caption{Effect of selection cuts (see text) on the number of observed
   events and the expectation from Monte Carlo simulations for the
  \qqbar signal plus the four-fermion background. The quoted errors
    are purely statistical and do not reflect any experimental or
  hadronization systematics. The efficiency to select non-radiative
  \qqbar events with $\sqrt{s}-\sqrt{s^{\prime}}<1$~GeV and the expected
  fraction of four-fermion background are also shown. The four-fermion
               background is negligible below 161 GeV.}
   \label{tab:cuts}
\end{table}
\begin{table}[H]
                          \centering \small
   \begin{tabular}{|c|l|c|c|c|}
\hline Category & Error Source & Standard & Variation 1 & Variation 2
\\ \hline  \hline &Track-Cluster Matching$^\ast$ & MT objects & Tracks
 + Clusters &  \\ \cline{2-5} & Detector Correction$^\ast$ & PYTHIA &
HERWIG & \\  \cline{2-5} &Containment ($|\cos\theta_{\rm{T}}|$)$^\ast$
  & $<0.9$ & $<0.7$& \\ 
\cline{2-5}(exp.) &qqqq ($\lqqqq$) & $<0.25$
          &$<0.1$ & $<0.4$ \\  
\cline{2-5} &qql$\nu$($\lqqln$) & $<0.5$ & $<0.75$ & $<0.25$ \\
\cline{2-5} &ISR Algorithm &  Ref.~\cite{wwspr}& Ref.~\cite{wwspri}& \\ \cline{2-5}
       &Backgrounds ($\sigma_{\rm{bkgd}}$) & 1.0 & +5\% & $-$5\%
\\  \hline \hline (had.)&Hadronization$^\ast$ & PYTHIA & HERWIG &
   \\ \hline \hline (theo.)&Fit Range$^\ast$ & min 6 bins & +2(1) bins
   & $-$2 bins\\ \cline{2-5} &Renorm. Scale Dep. ($x_{\mu}$)$^\ast$ & 1
                        & 0.5 & 2.0 \\ \hline 
   \end{tabular}
   \caption{Summary of systematic variations applied to the datasets. The $\ast$
indicates 
       which of the systematic variations were used for the 91~GeV dataset.} 
   \label{tab:systerr}
\end{table}
\newpage
\begin{table}[H]
   \centering
   \begin{tabular}{|l|c|c|c|c|}
      \hline
		 &	 91~GeV &	   133~GeV &	 177~GeV &	 197~GeV
\\ \hline 
\alsx(\roots) 	 &	 0.1147 &	   0.1071 &	  0.0991 &	
0.0996\\ \hline 
Fit Range  ($-\log_{10}\yc$)   & 2.06--0.81 & 2.75--0.75 & 2.75--0.75 & 2.75--0.75 \\
$\chi^{2}$ /dof  &        7.98/10 &       7.74/8 &      10.33/8 &       8.63/8
\\ \hline
Experimental	 &	 $\pm$0.0017 &	 $\pm$0.0032 &	 $\pm$0.0019 &	
$\pm$0.0009 \\  \hline
Hadronization	 &	 $\pm$0.0026 &	 $\pm$0.0007 &	 $\pm$0.0005 &	
$\pm$0.0005 \\  \hline
Fit Range Variation &	 $\pm$0.0005 &	 $\pm$0.0035 & $\pm$0.0008&	
$\pm$0.0004 \\[.1cm] 
$x_{\mu}\ ^{x_{\mu}=2.0}_{x_{\mu}=0.5}$   &       \ar{0.0030}{0.0010} &  
\ar{0.0030}{0.0014} & \ar{0.0023}{0.0011} & \ar{0.0020}{0.0010} \\[.1cm]  
\hline
$x_{L}\ ^{x_{L}=9/4}_{x_{L}=4/9}$          &       (\ar{0.0040}{0.0026}) &  
(\ar{0.0033}{0.0022}) & (\ar{0.0026}{0.0018}) & (\ar{0.0025}{0.0019}) \\[.1cm]  
\hline
Theoretical	 &	 $\pm$0.0030 &	 $\pm$0.0046 &	 $\pm$0.0024 &	
$\pm$0.0020 \\  \hline
Total Stat.	 &	 $\pm$0.0004 &	 $\pm$0.0026 &	 $\pm$0.0019 &	
$\pm$0.0008 \\ 
Total Syst.	 &	 $\pm$0.0043 &	 $\pm$0.0056 &	 $\pm$0.0031 &	
$\pm$0.0022 \\  \hline
Total Error	 &	 $\pm$0.0043 &	 $\pm$0.0062 &	 $\pm$0.0037 &	
$\pm$0.0024 \\  \hline
   \end{tabular}
   \caption{Determination of \alsx and the breakdown of statistical and
systematic errors
         from the fit to the Cambridge differential two-jet rate distribution
($D_{2}^{\rm{C}}$) for all \cfm energy values.  The quality of the fit is
characterized by the chi-square ($\chi^{2}$) and the number of degrees of
freedom (dof).  The theoretical error includes contributions from the fit range
and $x_\mu$ variations and excludes the $x_{L}$ variation, which is used only
for comparison.}
   \label{d2_camb_tab}
   \vspace{.5cm}
   \begin{tabular}{|l|c|c|c|c|}
      \hline
		 &	 91~GeV &	 133~GeV &	 177~GeV &	 197~GeV
\\ \hline 
\alsx(\roots) 	 &	 0.1199 &	0.1129 &	0.1060 &	0.1064\\
\hline 
Fit Range ($-\log_{10}\yc$)  & 1.81--0.68 & 2.75--0.75 & 2.75--0.75 & 2.50--0.75\\
$\chi^{2}$/dof   &        4.02/9 &        7.02/8  &      1.59/8 &         9.10/7
\\ \hline
Experimental     &       $\pm$0.0025 &   $\pm$0.0026 &   $\pm$0.0013 &  
$\pm$0.0007 \\  \hline
Hadronization	 &	 $\pm$0.0017 &	 $\pm$0.0005 &	 $\pm$0.0005 &	
$\pm$0.0007 \\  \hline
Fit Range Variation &	 $\pm$0.0005 &	 $\pm$0.0020 &	 $\pm$0.0035 &	
$\pm$0.0015 \\[.1cm]
$x_{\mu}\ ^{x_{\mu}=2.0}_{x_{\mu}=0.5}$        &       \ar{0.0037}{0.0013} &
\ar{0.0028}{0.0020} &
 \ar{0.0031}{0.0017} & \ar{0.0030}{0.0014} \\[.1cm] \hline
$x_{L}\ ^{x_{L}=9/4}_{x_{L}=4/9}$          &    (\ar{0.0044}{0.0027}) &
(\ar{0.0028}{0.0026}) &
(\ar{0.0031}{0.0022}) &
(\ar{0.0032}{0.0020}) \\[.1cm]  \hline
Theoretical	 &	 $\pm$0.0037 &	 $\pm$0.0034 &	 $\pm$0.0047 &	
$\pm$0.0034 \\  \hline
Total Stat.	 &	 $\pm$0.0004 &	 $\pm$0.0026 &	 $\pm$0.0019 &	
$\pm$0.0009 \\ 
Total Syst.	 &	 $\pm$0.0048 &	 $\pm$0.0043 &	 $\pm$0.0049 &	
$\pm$0.0035 \\  \hline
Total Error	 &	 $\pm$0.0048 &	 $\pm$0.0050 &	 $\pm$0.0052 &	
$\pm$0.0037 \\  \hline
   \end{tabular}
   \caption{Determination of \alsx and the breakdown of statistical and
systematic errors
      from the fit to the Durham differential two-jet rate distribution
($D_{2}^{\rm{D}}$) for all
      \cfm energy values.  The quality of the fit is characterized by the
chi-square
      ($\chi^{2}$) and the number of degrees of freedom (dof). The theoretical
error
         includes contributions from the fit range and $x_\mu$ variations and
excludes the
         $x_{L}$ variation, which is used only for comparison. }
   \label{d2_durh_tab}
\end{table}
\begin{table}[H]
   \centering
   \begin{tabular}{|l|c|c|c|c|}
      \hline
		 &	 91~GeV &	 133~GeV &	  177~GeV &	 197~GeV
\\ \hline 
\alsx(\roots) 	 &	 0.1254 &	0.1158   &	  0.1064  &	0.1066\\
\hline 
Fit Range ($-\log_{10}\yc$)&   2.56--0.56 & 2.75--0.50 & 2.75--0.50 & 2.75--0.50 \\
$\chi^{2}$/dof   &       20.70/16       & 7.38/9     &    2.57/9 &      
5.16/9\\ \hline
Experimental	 &	 $\pm$0.0013 &	 $\pm$0.0040 &	 $\pm$0.0030 &	
$\pm$0.0015 \\  \hline
Hadronization	 &	 $\pm$0.0033 &	 $\pm$0.0034 &	 $\pm$0.0018 &	
$\pm$0.0017 \\  \hline
Fit Range Variation &	 $\pm$0.0008 &	 $\pm$0.0042 &	 $\pm$0.0020 &	
$\pm$0.0010 \\ 
$x_{\mu}\ ^{x_{\mu}=2.0}_{x_{\mu}=0.5}$&       \ar{0.0036}{0.0001}
&\ar{0.0021}{0.0001}&\ar{0.0016}{0.0001} &\ar{0.0016}{0.0001}
\\ \hline
Theoretical	 &	 $\pm$0.0037 &	 $\pm$0.0047 &	 $\pm$0.0026 &	
$\pm$0.0019 \\  \hline
Total Stat.	 &	 $\pm$0.0005 &	 $\pm$0.0033 &	 $\pm$0.0023 &	
$\pm$0.0009 \\ 
Total Syst.	 &	 $\pm$0.0051 &	 $\pm$0.0070 &	 $\pm$0.0042 &	
$\pm$0.0030 \\  \hline
Total Error	 &	 $\pm$0.0051 &	 $\pm$0.0077 &	 $\pm$0.0048 &	
$\pm$0.0031 \\  \hline
   \end{tabular}
   \caption{Determination of \alsx and the breakdown of statistical and
systematic errors
      from the fit to the Cambridge average jet rate distribution
($\avn^{\rm{C}}$) for all
      \cfm energy values.  The quality of the fit is characterized by the
chi-square
      ($\chi^{2}$) and the number of degrees of freedom (dof).}
   \label{avn_camb_tab}
   \vspace{.5cm}
   \begin{tabular}{|l|c|c|c|c|}
      \hline
		 &	 91~GeV &	 133~GeV &	 177~GeV &	 197~GeV
\\ \hline 
\alsx(\roots) 	 &	 0.1272 &	0.1193 &	0.1103 &	0.1106\\
\hline 
Fit Range ($-\log_{10}\yc$)& 2.56--0.56 & 2.75--0.50 & 2.75--0.50 & 2.75--0.50\\
$\chi^{2}$/dof &         7.90/16 &      4.73/9 &      0.81/9&       4.14/9 \\
\hline
Experimental	 &	 $\pm$0.0006 &	 $\pm$0.0039 &	 $\pm$0.0026 &	
$\pm$0.0013 \\  \hline
Hadronization	 &	 $\pm$0.0039 &	 $\pm$0.0037 &	 $\pm$0.0009 &	
$\pm$0.0007 \\  \hline
Fit Range Variation &	 $\pm$0.0003 &	 $\pm$0.0029 &	 $\pm$0.0003 &	
$\pm$0.0008 \\ 
$x_{\mu}\ ^{x_{\mu}=2.0}_{x_{\mu}=0.5}$	 &\ar{0.0034}{0.0013}
&\ar{0.0028}{0.0010}
&\ar{0.0023}{0.0008}&\ar{0.0023}{0.0008} \\ \hline
Theoretical	 &	 $\pm$0.0034 &	 $\pm$0.0040 &	 $\pm$0.0023 &	
$\pm$0.0024 \\  \hline
Total Stat.	 &	 $\pm$0.0004 &	 $\pm$0.0030 &	 $\pm$0.0021 &	
$\pm$0.0008 \\ 
Total Syst.	 &	 $\pm$0.0052 &	 $\pm$0.0067 &	 $\pm$0.0035 &	
$\pm$0.0028 \\  \hline
Total Error	 &	 $\pm$0.0052 &	 $\pm$0.0073 &	 $\pm$0.0041 &	
$\pm$0.0029 \\  \hline
   \end{tabular}
   \caption{Determination of \alsx and the breakdown of statistical and
systematic errors
      from the fit to the Durham average jet rate distribution ($\avn^{\rm{D}}$)
for all
      \cfm energy values.  The quality of the fit is characterized by the
chi-square
      ($\chi^{2}$) and the number of degrees of freedom (dof).}
   \label{avn_durh_tab}
\end{table}

\begin{table}[H]
   \centering
\begin{tabular}{|l||c|c|c|c|}
\hline
 91~GeV     & $D_{2}^{C}$ & $D_{2}^{D}$ & $\avn^{C}$ & $\avn^{D}$ \\ \hline
\hline
$D_{2}^{C}$ & 100      &  85         &  53         &  46\\ \hline
$D_{2}^{D}$ &  85      & 100         &  56         &  47\\ \hline
$\avn^{C}$  &  53      &  56        & 100        &  75\\ \hline
$\avn^{D}$  &  46      &  47         &  75         & 100\\ \hline
\end{tabular}
\vspace{.3cm}
\begin{tabular}{|l||c|c|c|c|}
\hline
 133~GeV     & $D_{2}^{C}$ & $D_{2}^{D}$ & $\avn^{C}$ & $\avn^{D}$ \\ \hline
\hline
$D_{2}^{C}$ & 100      &  83         &  70         &  64\\ \hline
$D_{2}^{D}$ &  83      & 100        &  64        &  63\\ \hline
$\avn^{C}$  &  70      &  64        & 100         &  81\\ \hline
$\avn^{D}$  &  64      &  63         &  81         & 100\\ \hline
\end{tabular}
\vspace{.3cm}
\begin{tabular}{|l||c|c|c|c|}
\hline
177~GeV     & $D_{2}^{C}$ & $D_{2}^{D}$ & $\avn^{C}$ & $\avn^{D}$ \\ \hline
\hline
$D_{2}^{C}$ & 100      &  84         &  73         &  66\\ \hline
$D_{2}^{D}$ &  84      & 100         &  71         &  72\\ \hline
$\avn^{C}$  &  73      &  71         & 100         &  82\\ \hline
$\avn^{D}$  &  66      &  72         &  82         & 100\\ \hline
\end{tabular}
\vspace{.3cm}
\begin{tabular}{|l||c|c|c|c|}
\hline
197~GeV     & $D_{2}^{C}$ & $D_{2}^{D}$ & $\avn^{C}$ & $\avn^{D}$ \\ \hline
\hline
$D_{2}^{C}$ & 100      &  90         &  77         &  68\\ \hline
$D_{2}^{D}$ &  90      & 100         &  72         &  73\\ \hline
$\avn^{C}$  &  77      &  72        & 100         &  82\\ \hline
$\avn^{D}$  &  68      &  73         &  82         & 100\\ \hline
\end{tabular}
   \caption{Statistical correlation matrix determined for the combination of
\alsx     
      measurements from the four different observables for each of the four
different datasets.  Each matrix element is presented as a percentage (\%).}
   \label{als_corr}
\end{table}

\vspace{1.5cm}
\begin{table}[H]
   \centering
   \small
   \begin{tabular}{|c||c|c|c|c|c||c|c|c|c|c|}
      \hline
            & \multicolumn{5}{|c||}{Value for \roots}
&\multicolumn{5}{|c|}{Value at \PMZ} \\
            \cline{2-11} 
      \roots~(GeV)&    \als   & $\sigma_{\rm{stat.}}$ &$\sigma_{\rm{exp.}}$
            &$\sigma_{\rm{hadr.}}$ & $\sigma_{\rm{theory}}$ & \als   &
$\sigma_{\rm{stat.}}$
            &$\sigma_{\rm{exp.}}$
            &$\sigma_{\rm{hadr.}}$ & $\sigma_{\rm{theory}}$\\
         \hline
         \PMZ & 0.1213 & 0.0004 & 0.0013 & 0.0029 & 0.0034 &
                0.1213 & 0.0004 & 0.0013 & 0.0029 & 0.0034 \\
          133 & 0.1126 & 0.0025 & 0.0028 & 0.0007 & 0.0039 & 
                0.1191 & 0.0028 & 0.0031 & 0.0008 & 0.0044 \\
          177 & 0.1039 & 0.0018 & 0.0018 & 0.0001 & 0.0028 & 
                0.1140 & 0.0021 & 0.0021 & 0.0001 & 0.0033 \\
          197 & 0.1046 & 0.0008 & 0.0009 & 0.0002 & 0.0023 & 
                0.1163 & 0.0010 & 0.0012 & 0.0003 & 0.0029 \\
      \hline
   \end{tabular}
   \caption{Values of \alsx determined from the weighted average of the
individual
      \alsx results at each \cfm energy, along with statistical and systematic
errors.  The value \alsx for the dataset when run back to the \PZo pole are
given on the right half of the table with full statistical and systematic
errors.}
   \label{tab_blue}
\end{table}
\normalsize

\begin{figure}[H]
\vspace{1.cm}
\centering
\hspace*{.6cm}
   \hspace*{-1.2cm}\includegraphics[width=1.1\textwidth]{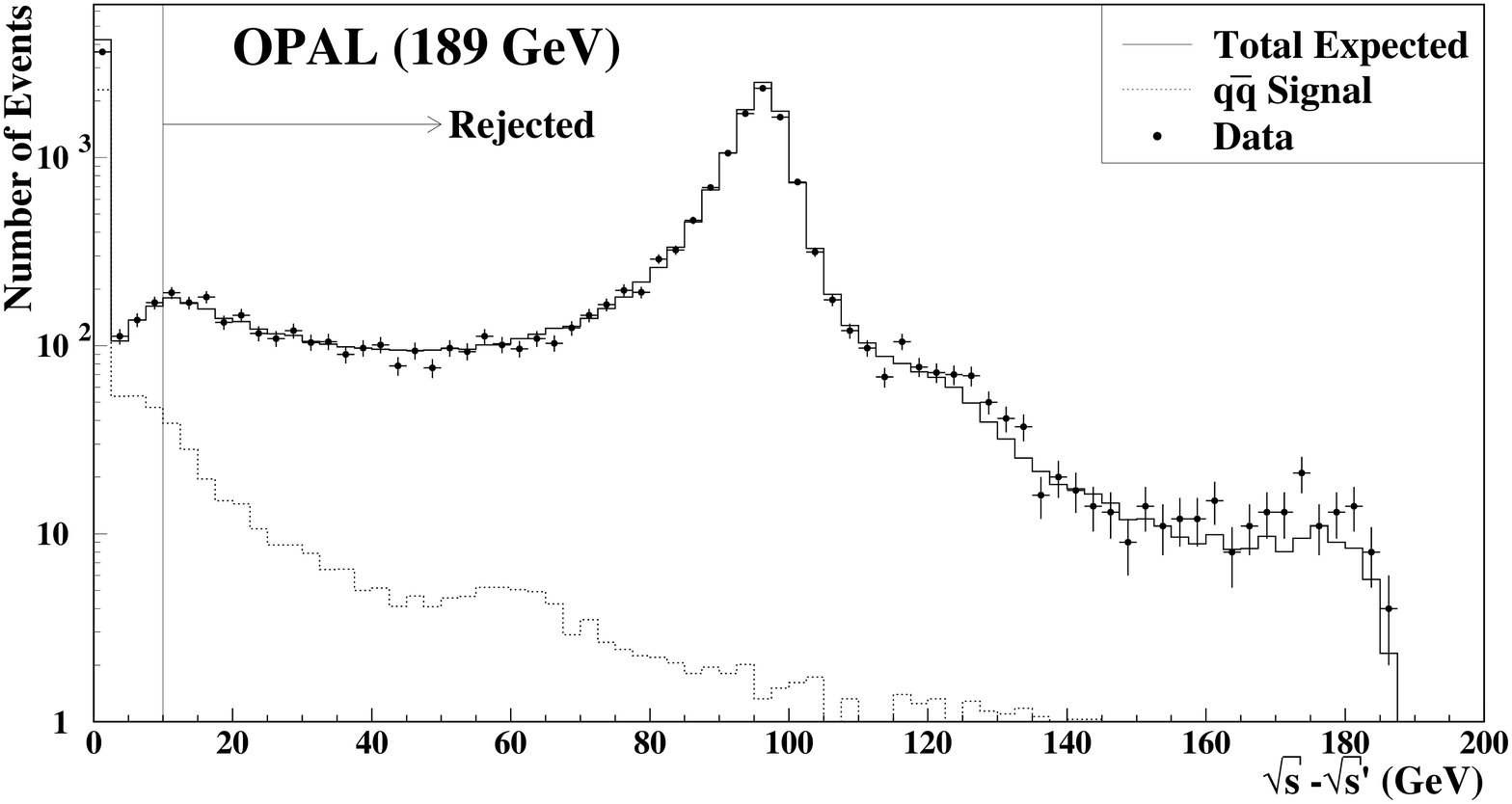}
   \caption{The $\sqrt{s} - \sqrt{s^{\prime}}$ distribution for the 189~GeV
dataset. 
       The vertical line indicates where the cut was applied.  The expected
non-radiative \qqbar signal ($\sqrt{s}-\sqrt{s^{\prime}_{\rm{true}}}<1$~GeV)
was determined from Monte Carlo and normalized to the luminosity of the measured
sample.  Vertical error bars indicate the size of statistical errors, while
horizontal error bars correspond to the bin width.}
   \label{fig_sprime}
\end{figure}
\begin{figure}[H]
\vspace{-.5cm}
\centering
\includegraphics[width=\textwidth]{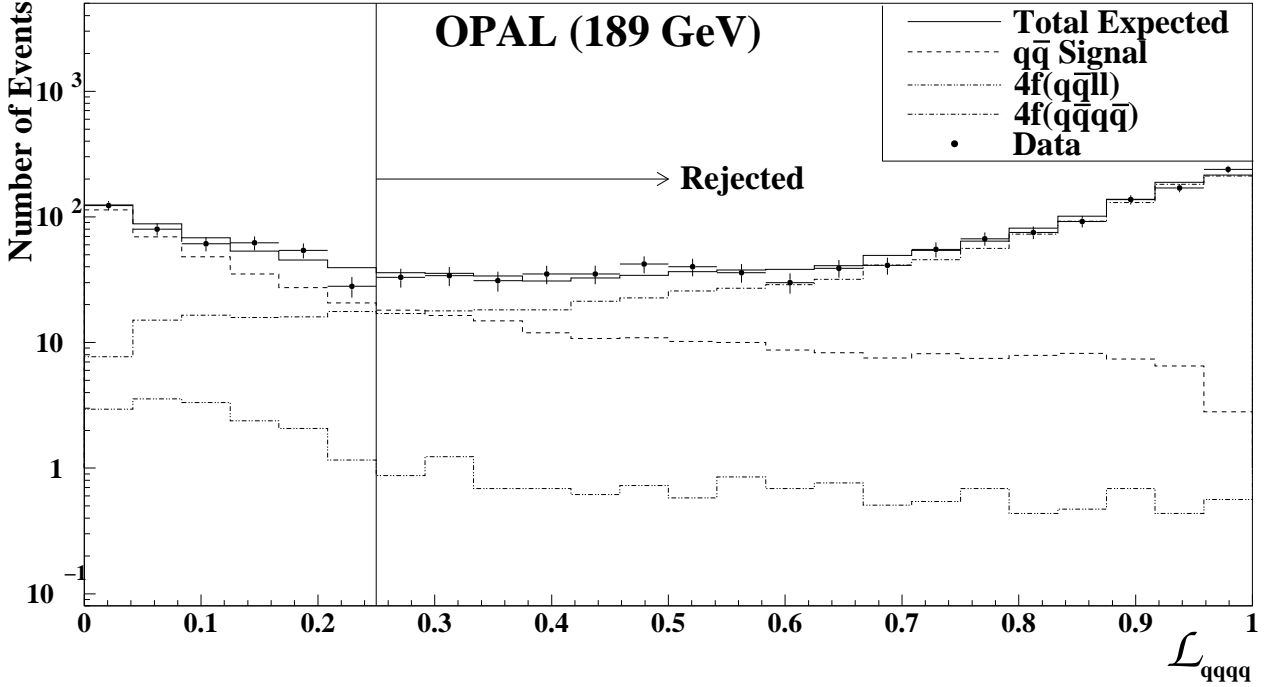}
\vspace{-.3cm} 
\includegraphics[width=\textwidth]{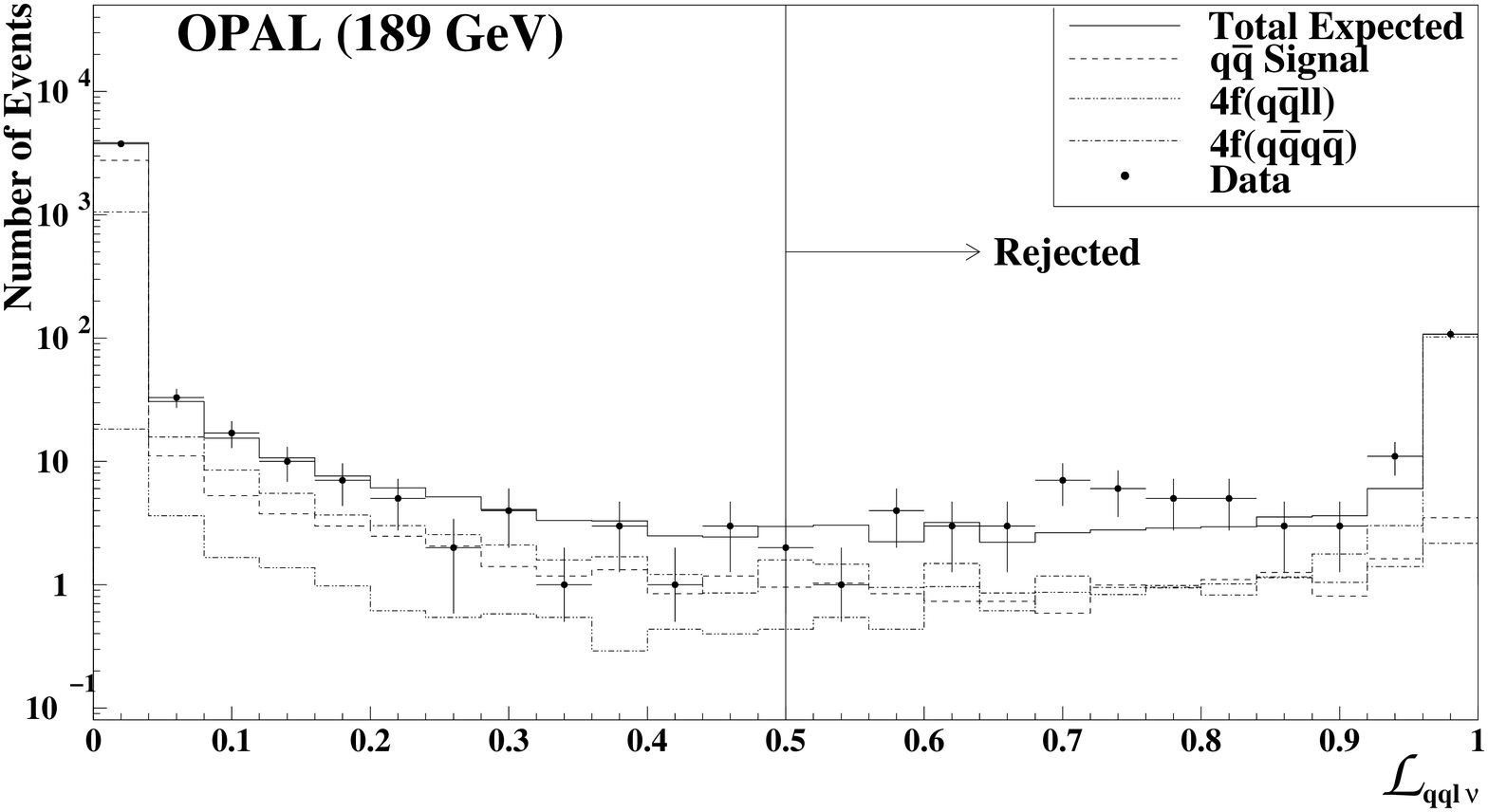}
\caption{Distributions of the four-quark likelihood \lqqqq (top) and
semi-leptonic 
   likelihood \lqqln (bottom) for the 189~GeV dataset.  The expected
contribution of the four-quark non-QCD background to the total sample is shown
by the dotted line, and the expected semi-leptonic four-fermion background is
represented by the dashed line. The vertical lines indicate the values of the
parameters where the cuts were applied.  The  expected \qqbar signal was
determined from Monte Carlo and normalized to the luminosity of the measured
sample.  Vertical error bars indicate the size of statistical errors, while the
horizontal bars correspond to the bin width.}
\label{fig:qqqq}
\end{figure}

\begin{figure}[H]
\vspace{-1cm}
\centering
   \hspace*{-1cm}\mbox{
         \includegraphics[width=.59\textwidth]{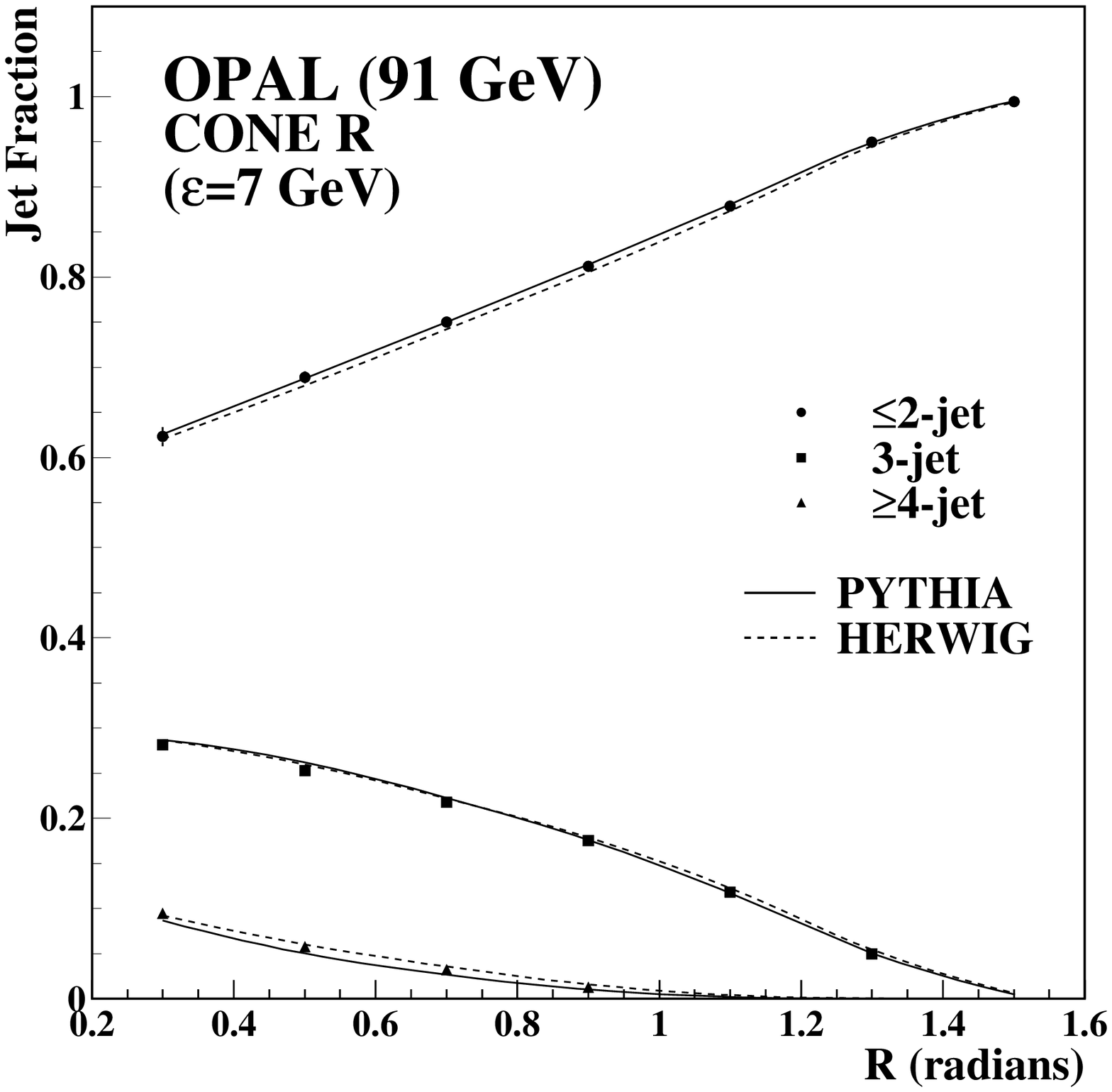}
         \hspace*{-.9cm}
         \includegraphics[width=.59\textwidth]{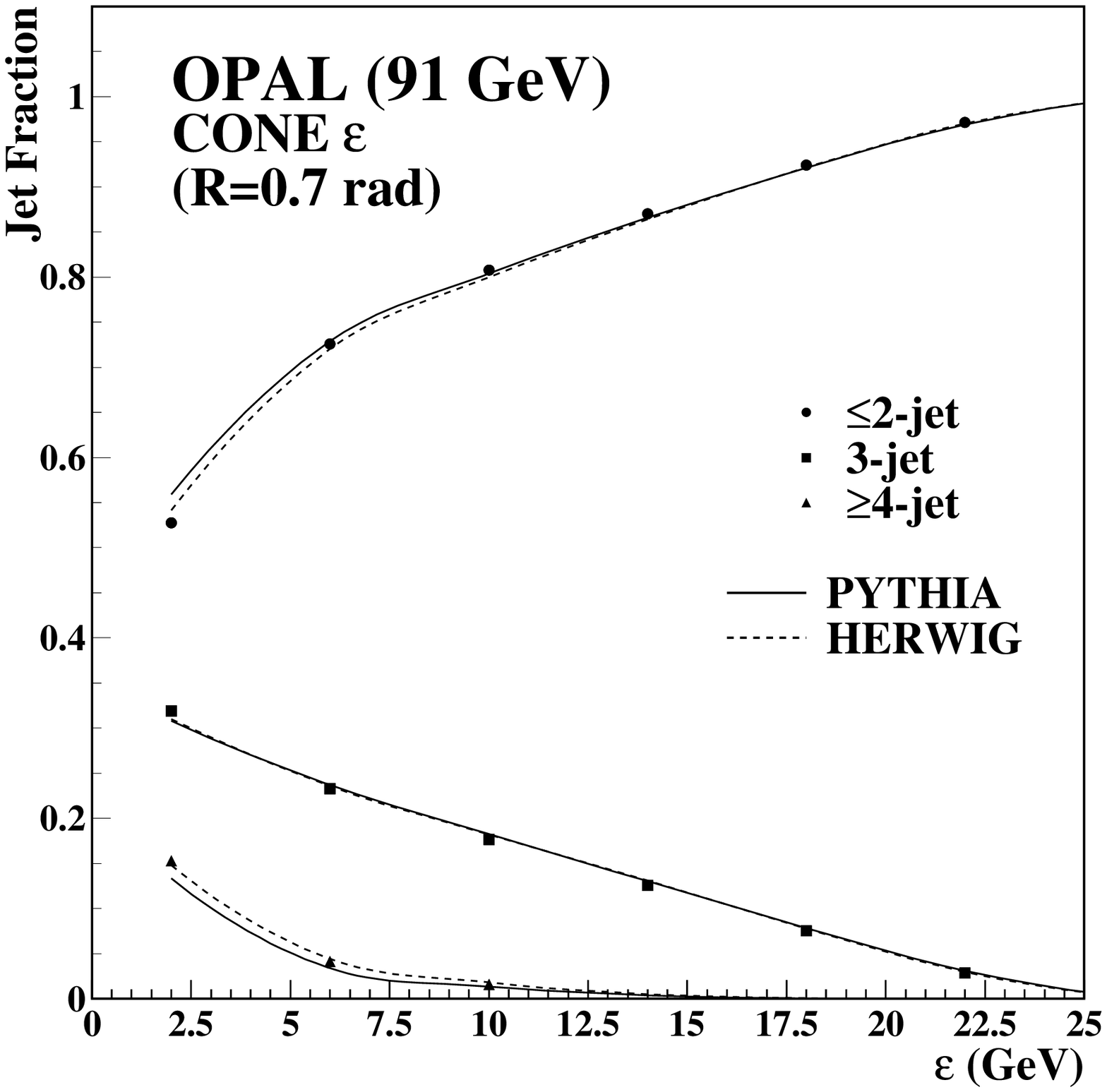}}
   \hspace*{-1cm}\mbox{
         \includegraphics[width=.59\textwidth]{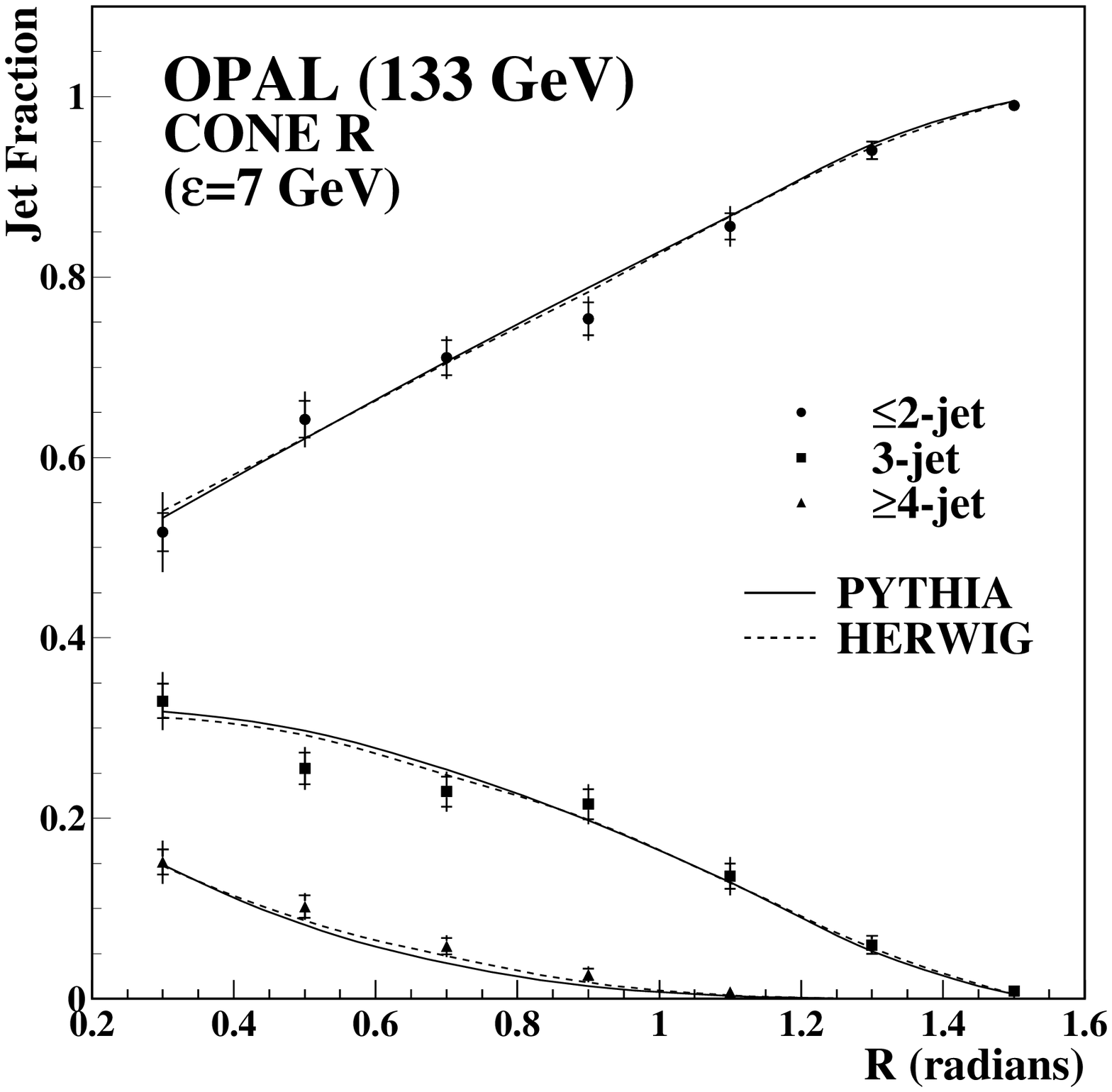}
         \hspace*{-.9cm}
         \includegraphics[width=.59\textwidth]{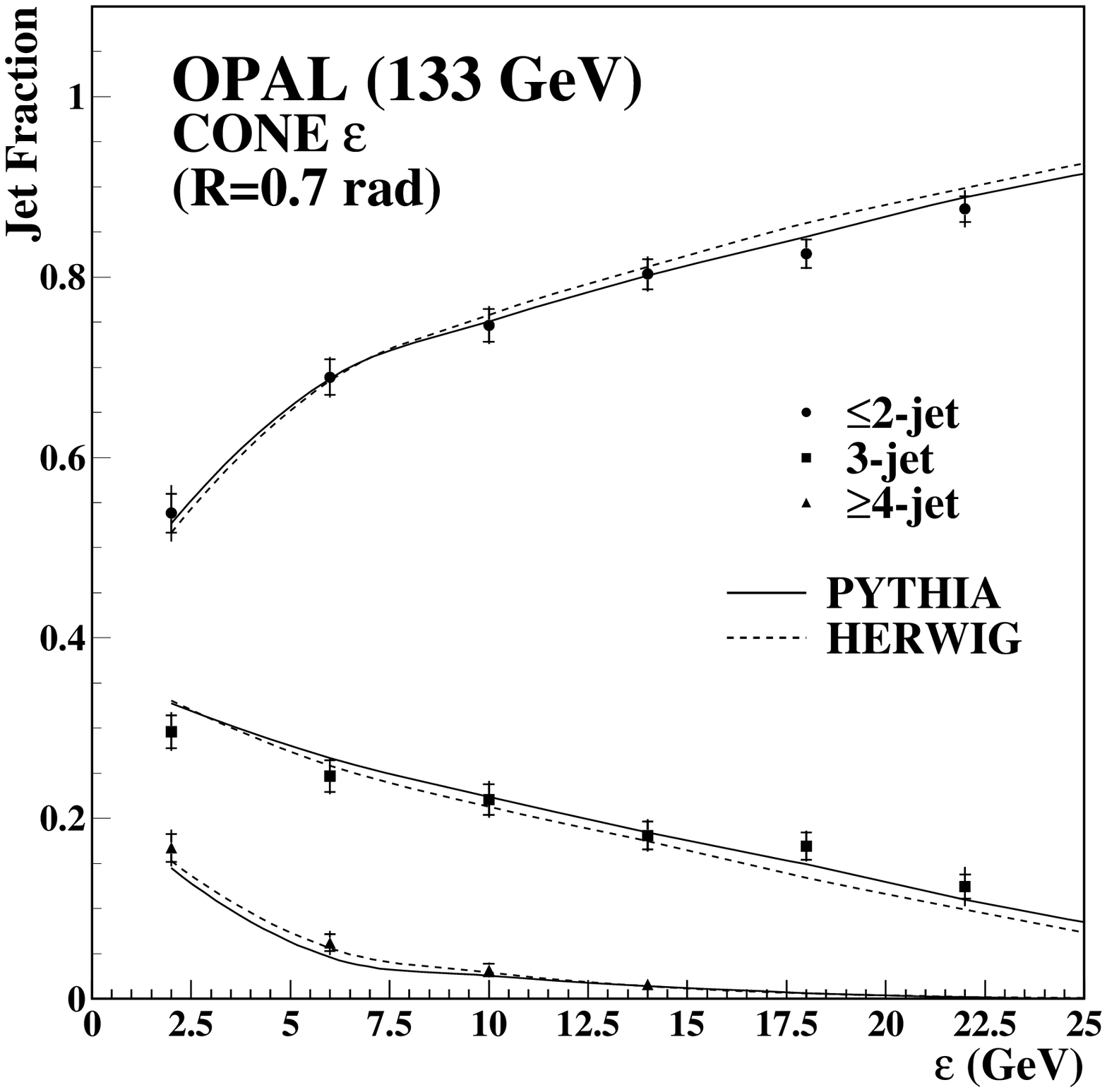}}
\vspace{-.7cm}
\caption{The hadron level $n$-jet rates for the $R$ and $\varepsilon$ variants
of the Cone algorithm
   for the data with $\sqrt{s}=91$~GeV (top) and $\sqrt{s}=133$~GeV (bottom). In
all
   plots the JETSET/PYTHIA and HERWIG Monte Carlo expectations
   are represented by the curves. Outer error bars indicate total
   errors while the inner bars indicate statistical errors.} 
\label{fig:cn91}
\end{figure}

\begin{figure}[H]
\centering
   \hspace*{-1cm}\mbox{
         \includegraphics[width=.59\textwidth]{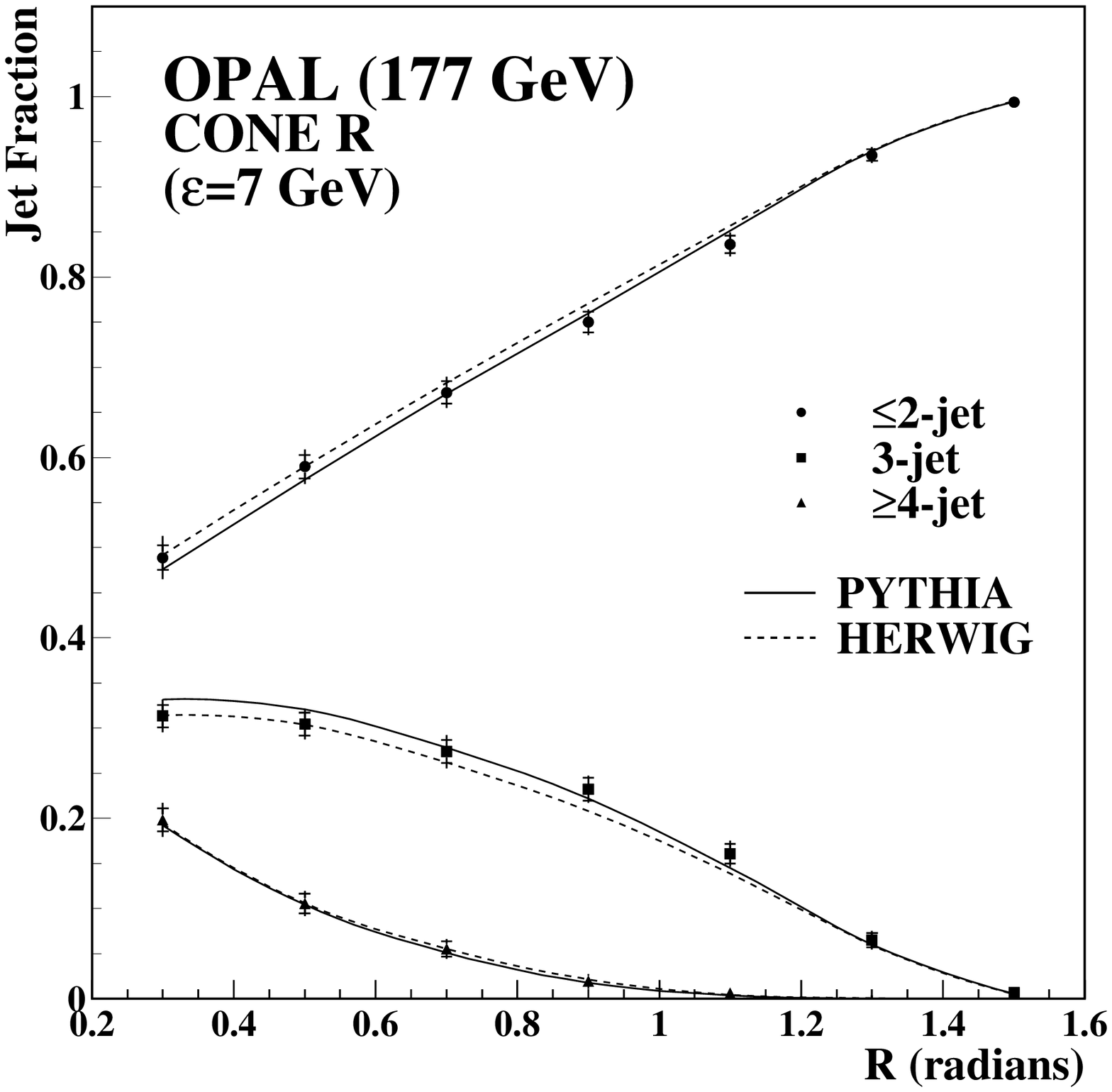}
         \hspace*{-.9cm}
         \includegraphics[width=.59\textwidth]{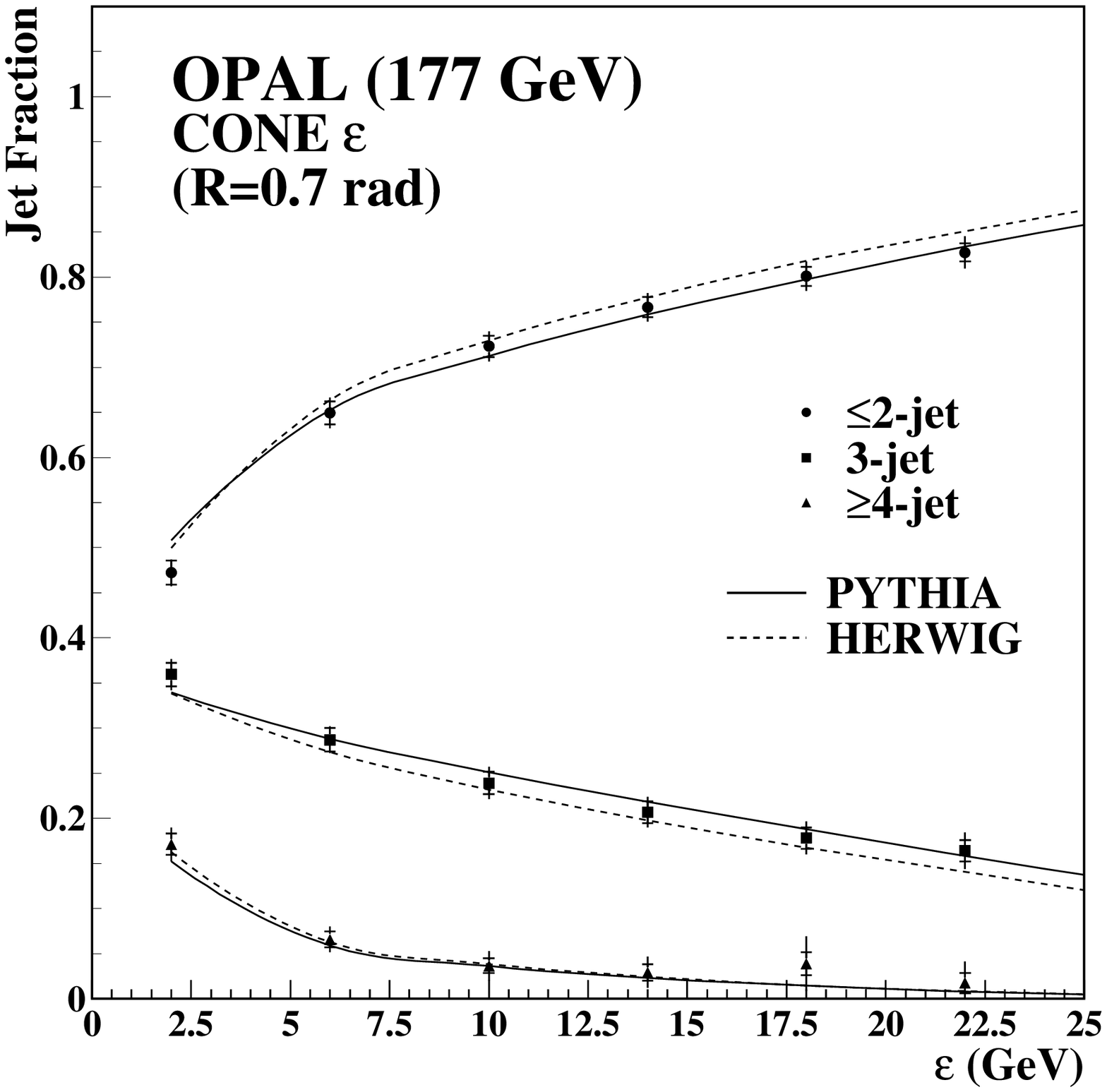}}
   \hspace*{-1cm}\mbox{
         \includegraphics[width=.59\textwidth]{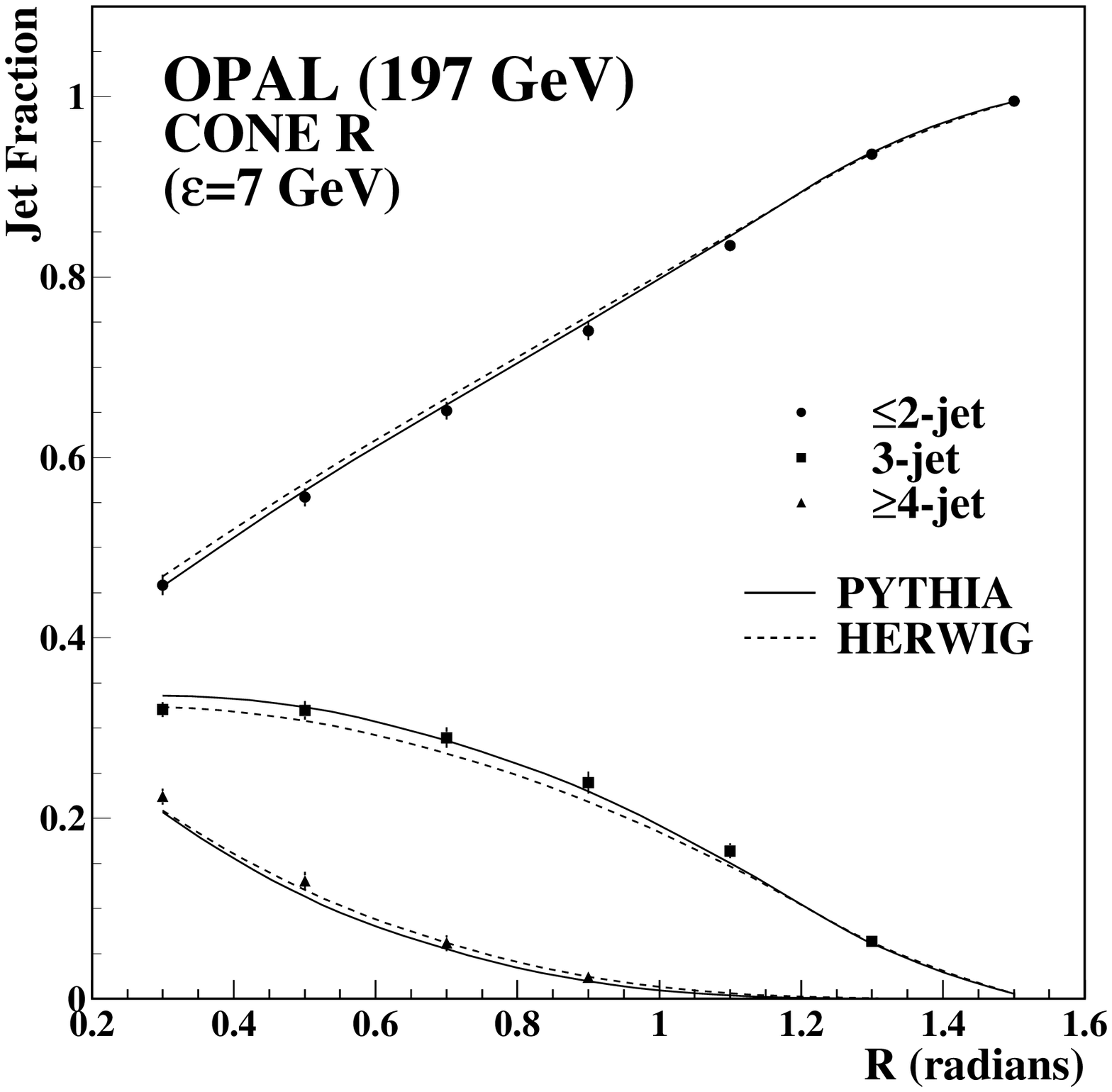}
         \hspace*{-.9cm}
         \includegraphics[width=.59\textwidth]{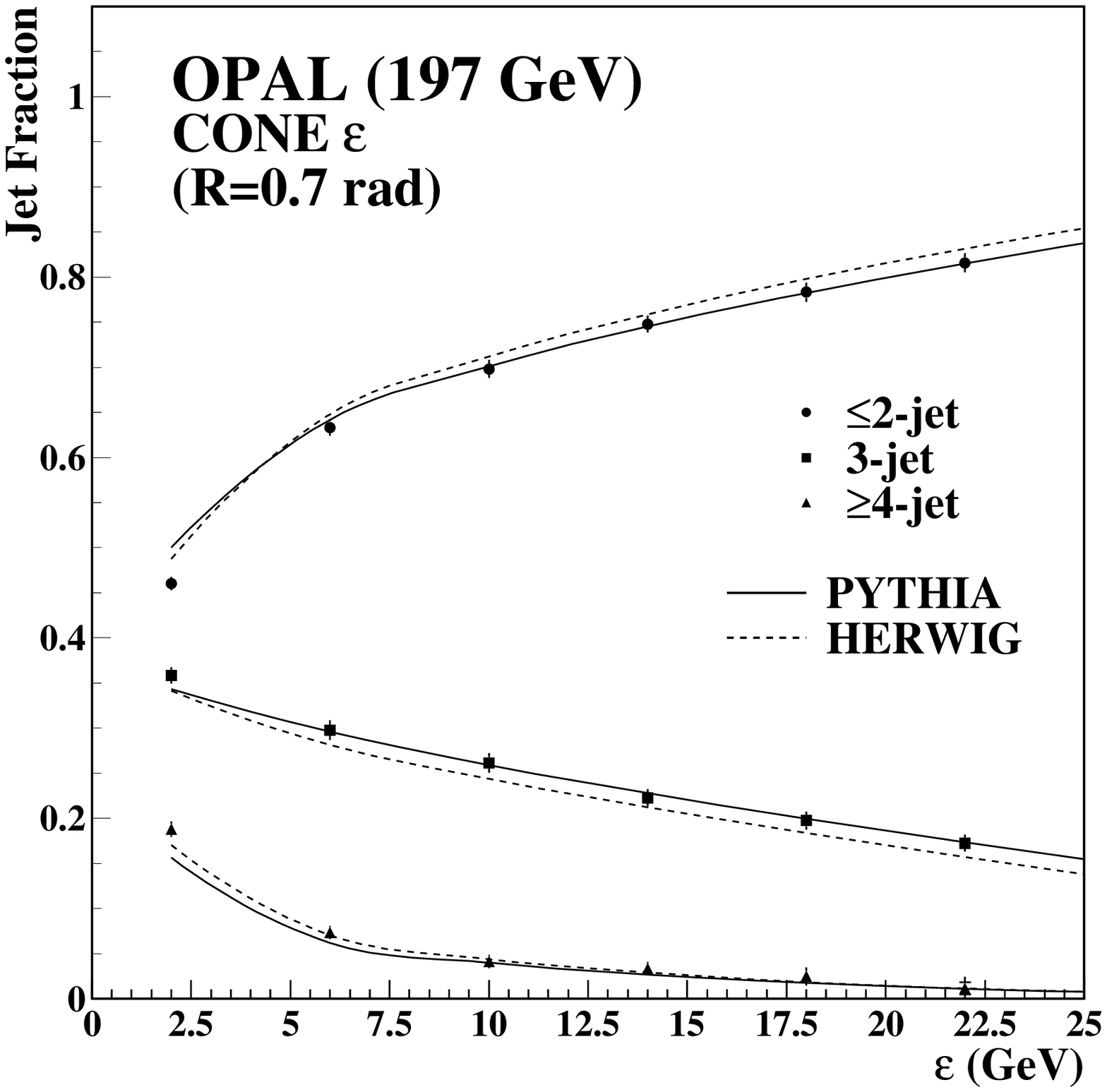}}
\vspace{-.7cm}
\caption{The hadron level $n$-jet rates for the $R$ and $\varepsilon$ variants
of the Cone algorithm
   for the data with $\sqrt{s}=179$~GeV (top) and $\sqrt{s}=198$~GeV (bottom).
In all
   plots the PYTHIA and HERWIG Monte Carlo expectations
   are represented by the curves. Outer error bars indicate total
   errors while the inner bars indicate statistical errors.} 
\label{fig:cn103}
\end{figure}

\begin{figure}[H]
   \centering   
   \hspace*{-1cm}\mbox{
         \includegraphics[width=.59\textwidth]{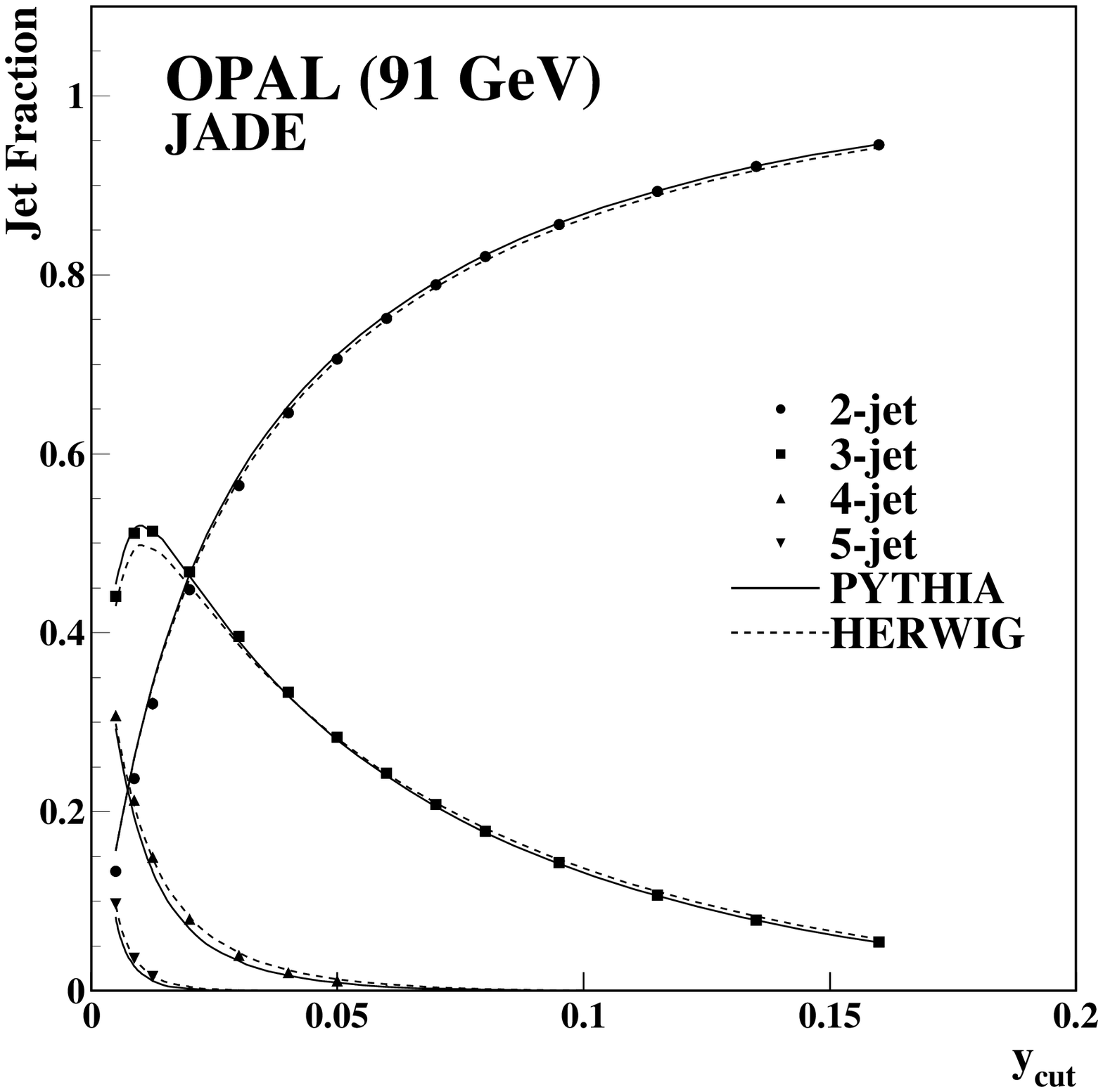}
         \hspace*{-.9cm}
         \includegraphics[width=.59\textwidth]{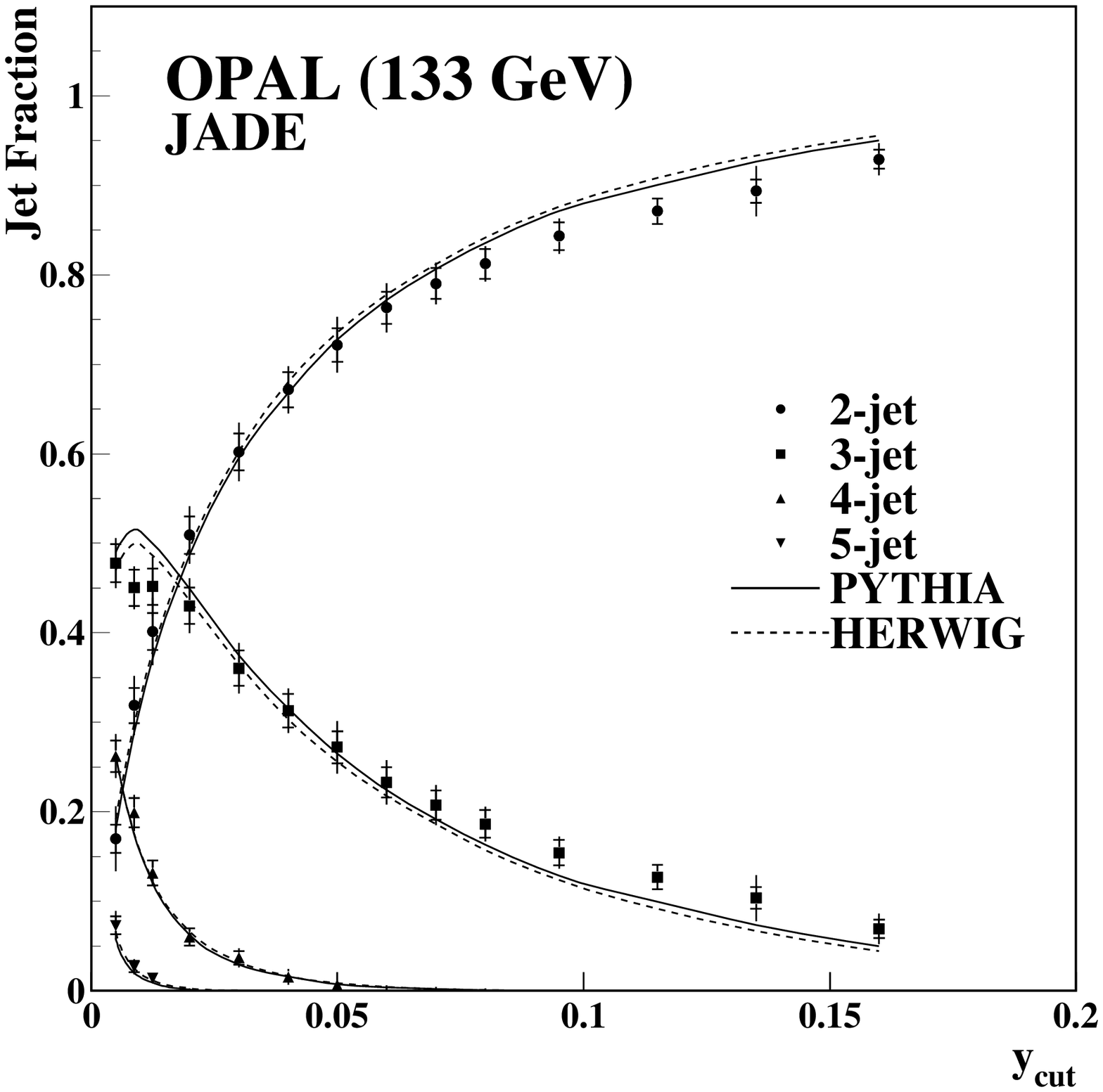}}
   \hspace*{-1cm}\mbox{
         \includegraphics[width=.59\textwidth]{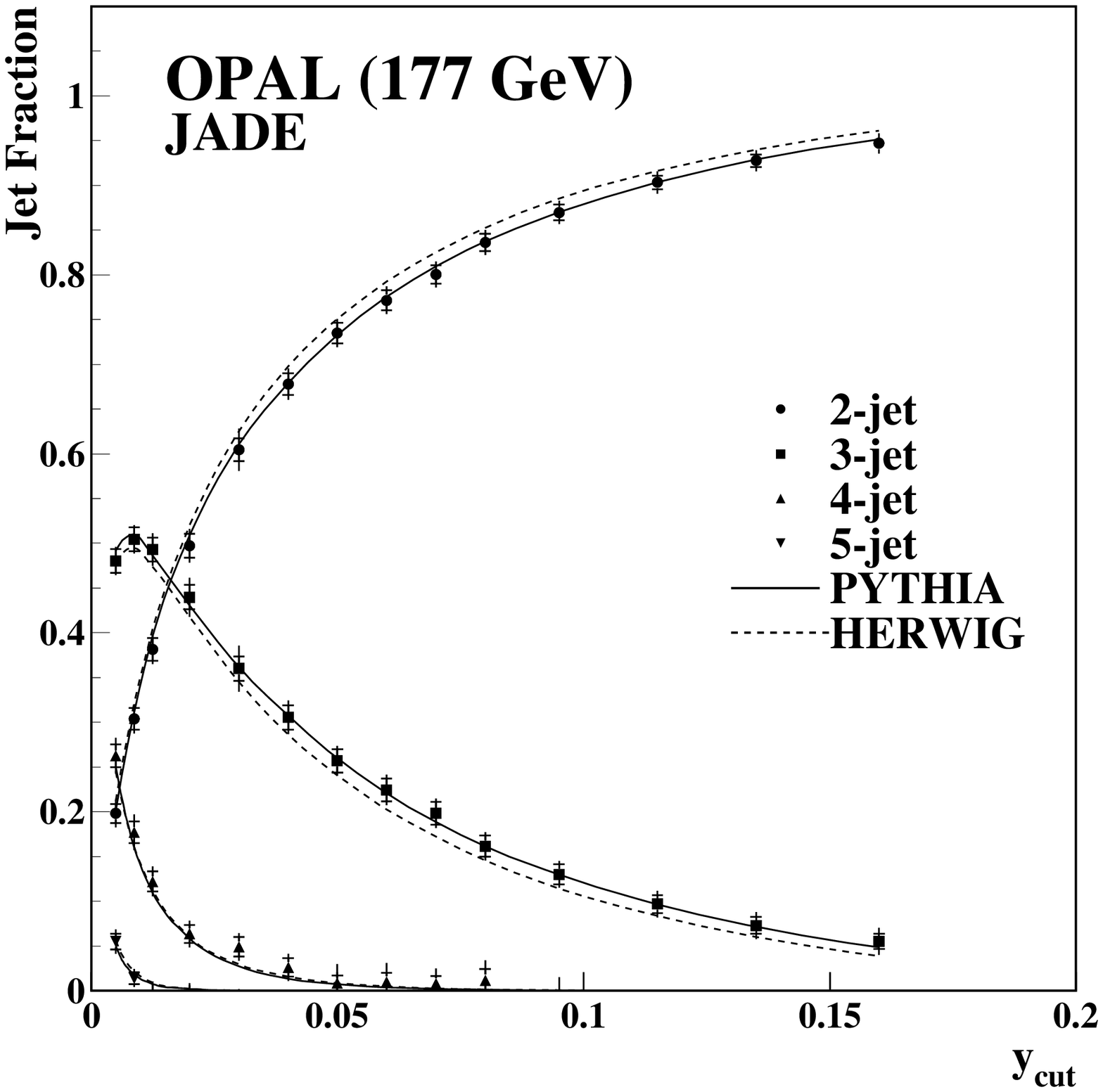}
         \hspace*{-.9cm}
         \includegraphics[width=.59\textwidth]{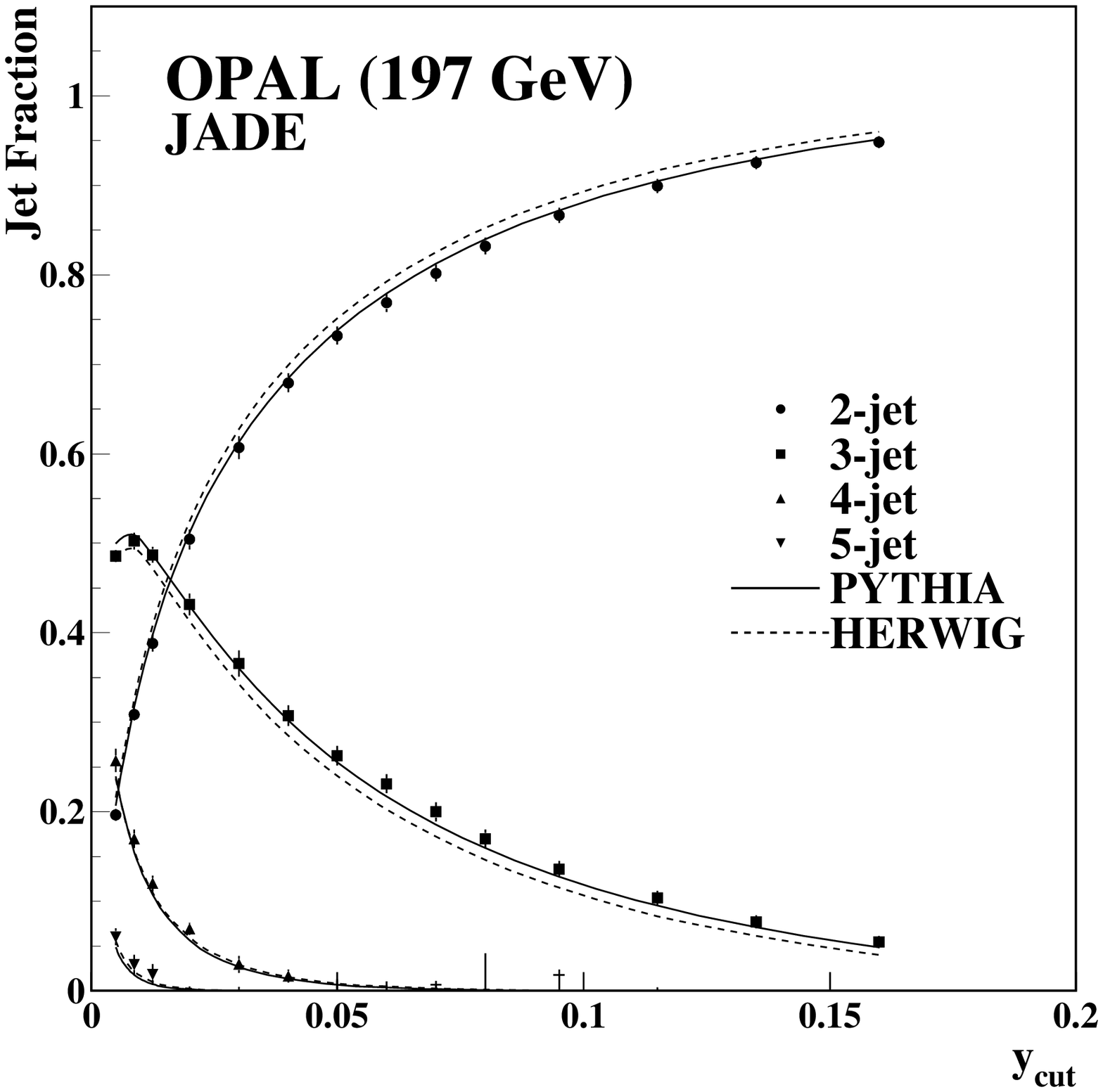}}
\vspace{-.7cm}
\caption{The hadron level $n$-jet rates for the JADE algorithm
   for the data with $\sqrt{s}=91$~GeV (top left), $\sqrt{s}=133$~GeV (top
right),
   $\sqrt{s}=179$~GeV (bottom left) and $\sqrt{s}=198$~GeV (bottom right). In
all
   plots the JETSET/PYTHIA and HERWIG Monte Carlo expectations
   are represented by the curves. Outer error bars indicate total
   errors while the inner bars indicate statistical errors.} 
\label{fig:jd91}
\end{figure}

\begin{figure}[H]
\centering
   \hspace*{-1cm}\mbox{
         \includegraphics[width=.59\textwidth]{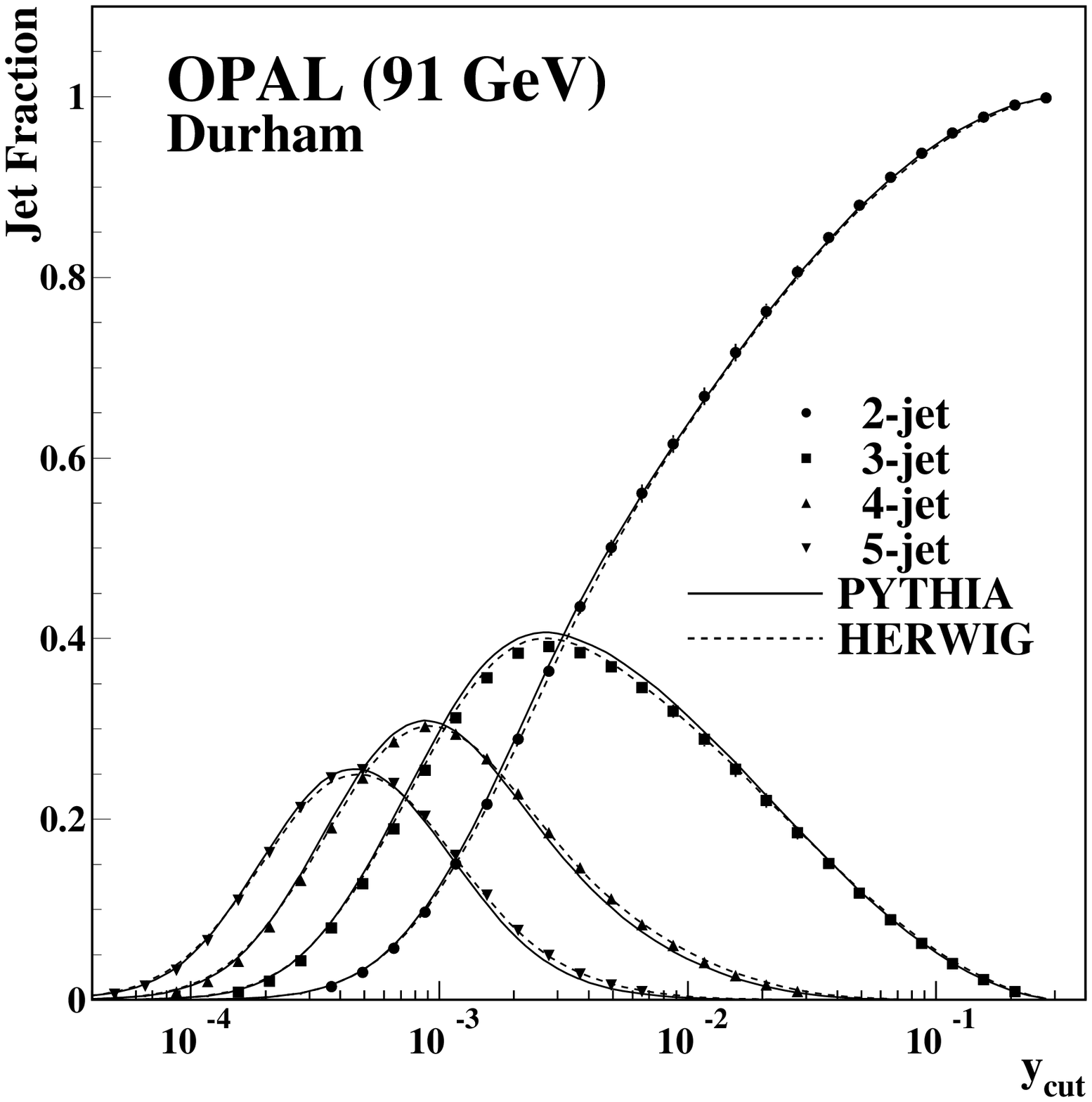}
         \hspace*{-.9cm}
         \includegraphics[width=.59\textwidth]{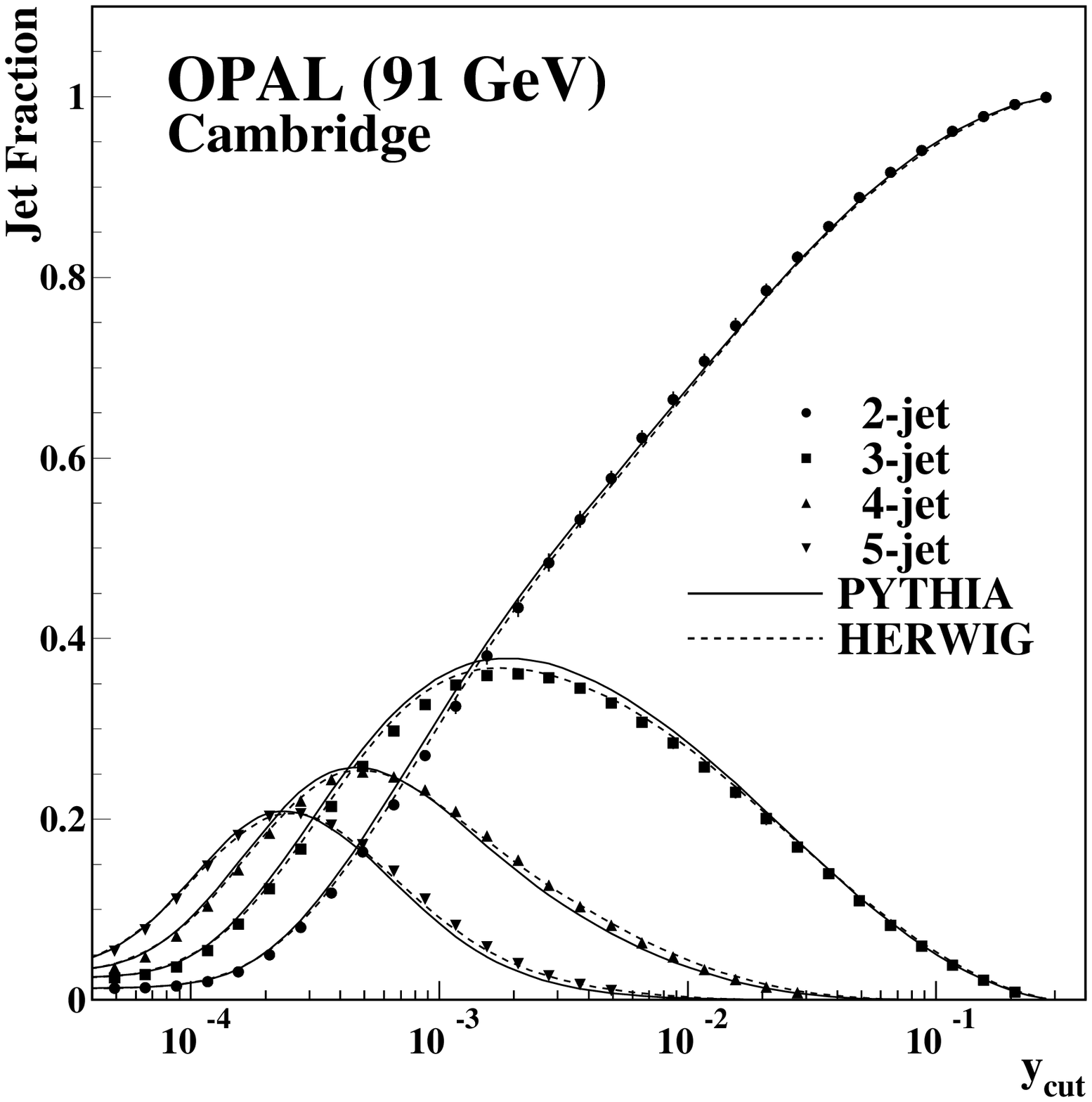}}
   \hspace*{-1cm}\mbox{
         \includegraphics[width=.59\textwidth]{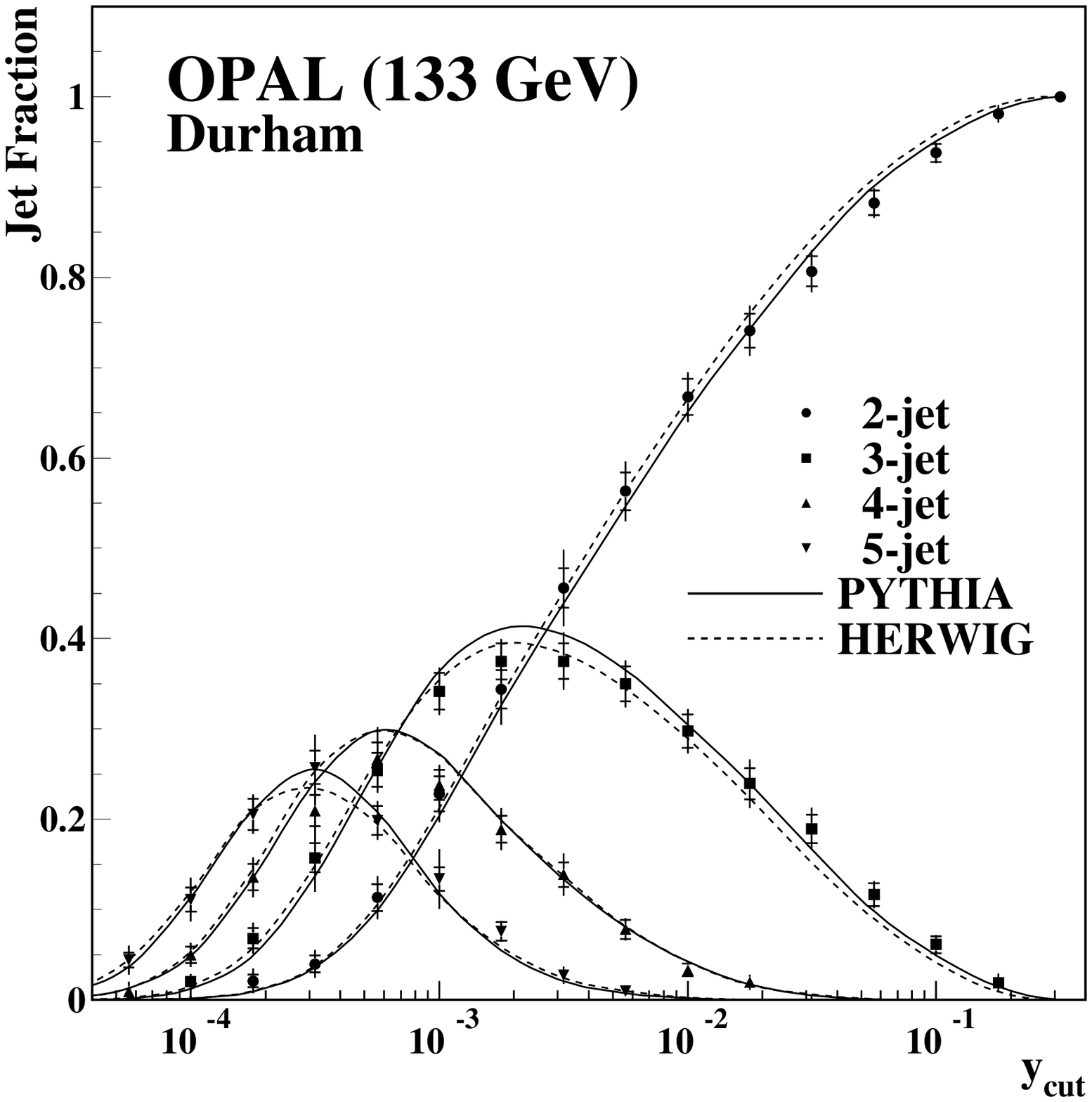}
         \hspace*{-.9cm}
         \includegraphics[width=.59\textwidth]{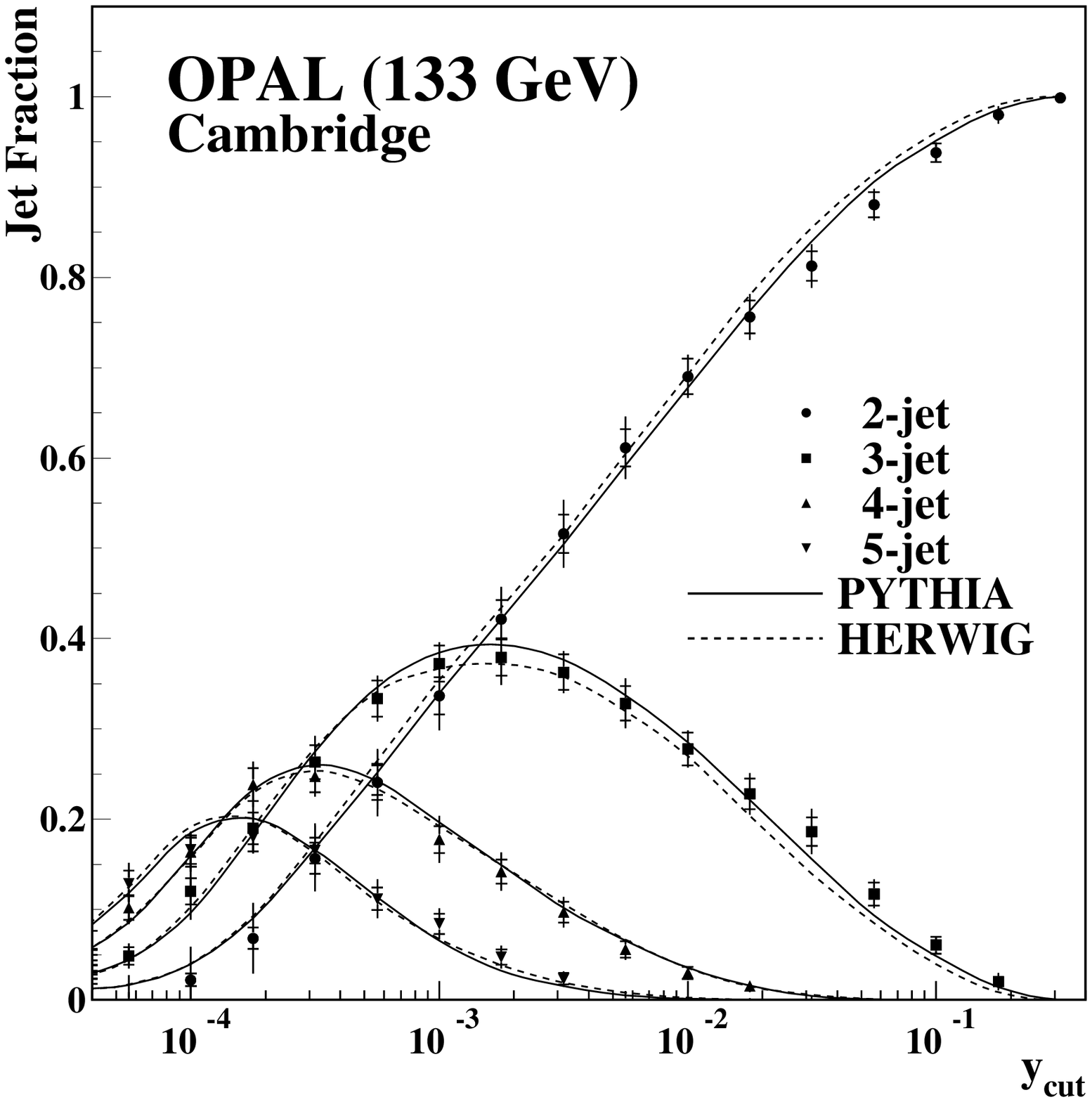}}
\vspace{-.7cm}
\caption{The hadron level $n$-jet rates for the Durham and Cambridge algorithms
   for the data with $\sqrt{s}=91$~GeV (top) and $\sqrt{s}=133$~GeV (bottom). In
all
   plots the JETSET/PYTHIA and HERWIG Monte Carlo expectations
   are represented by the curves. Outer error bars indicate total
   errors while the inner bars indicate statistical errors.} 
\label{fig:jr91}
\end{figure}

\begin{figure}[H]
\centering
   \hspace*{-1cm}\mbox{
         \includegraphics[width=.59\textwidth]{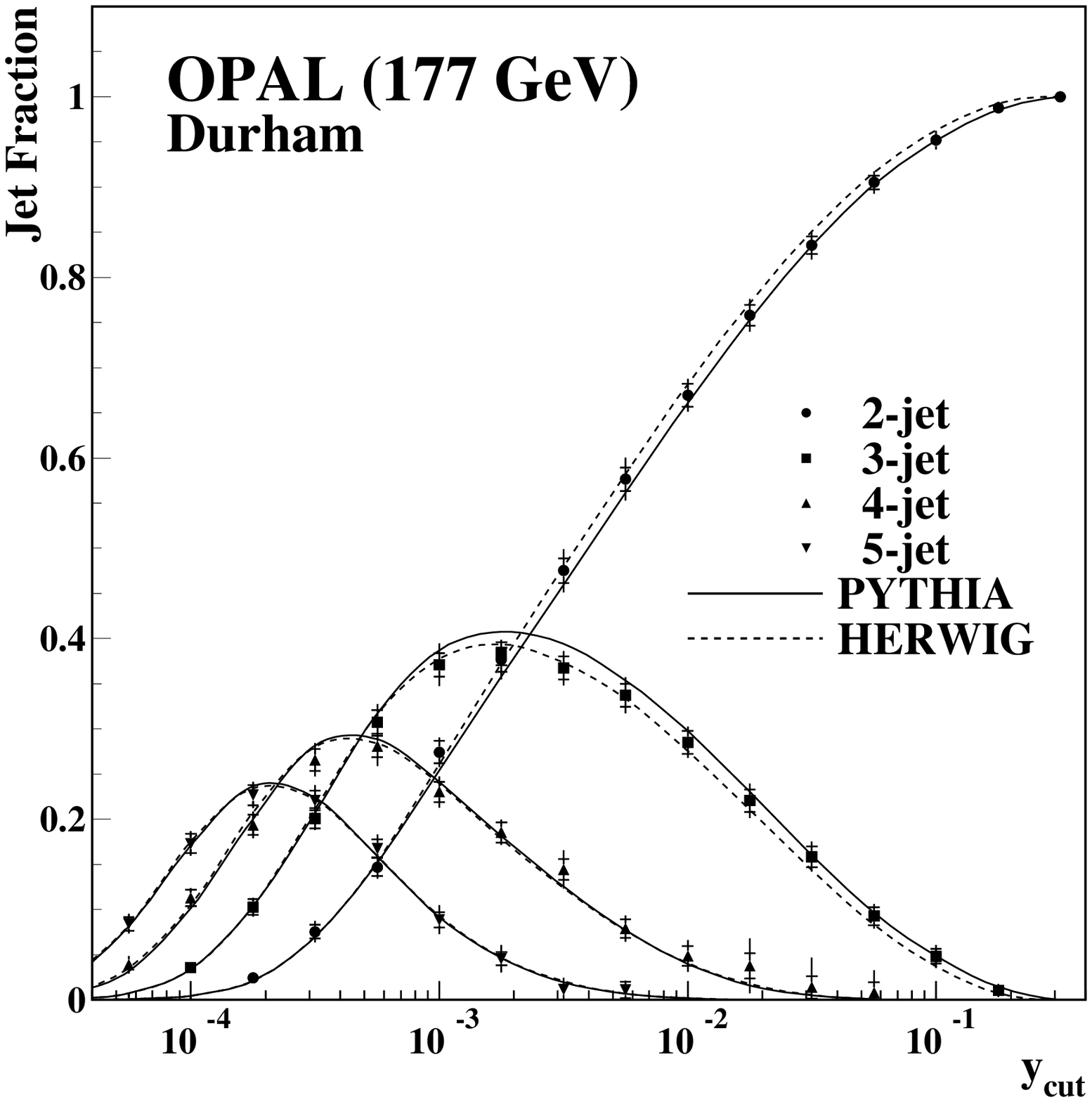}
         \hspace*{-.9cm}
         \includegraphics[width=.59\textwidth]{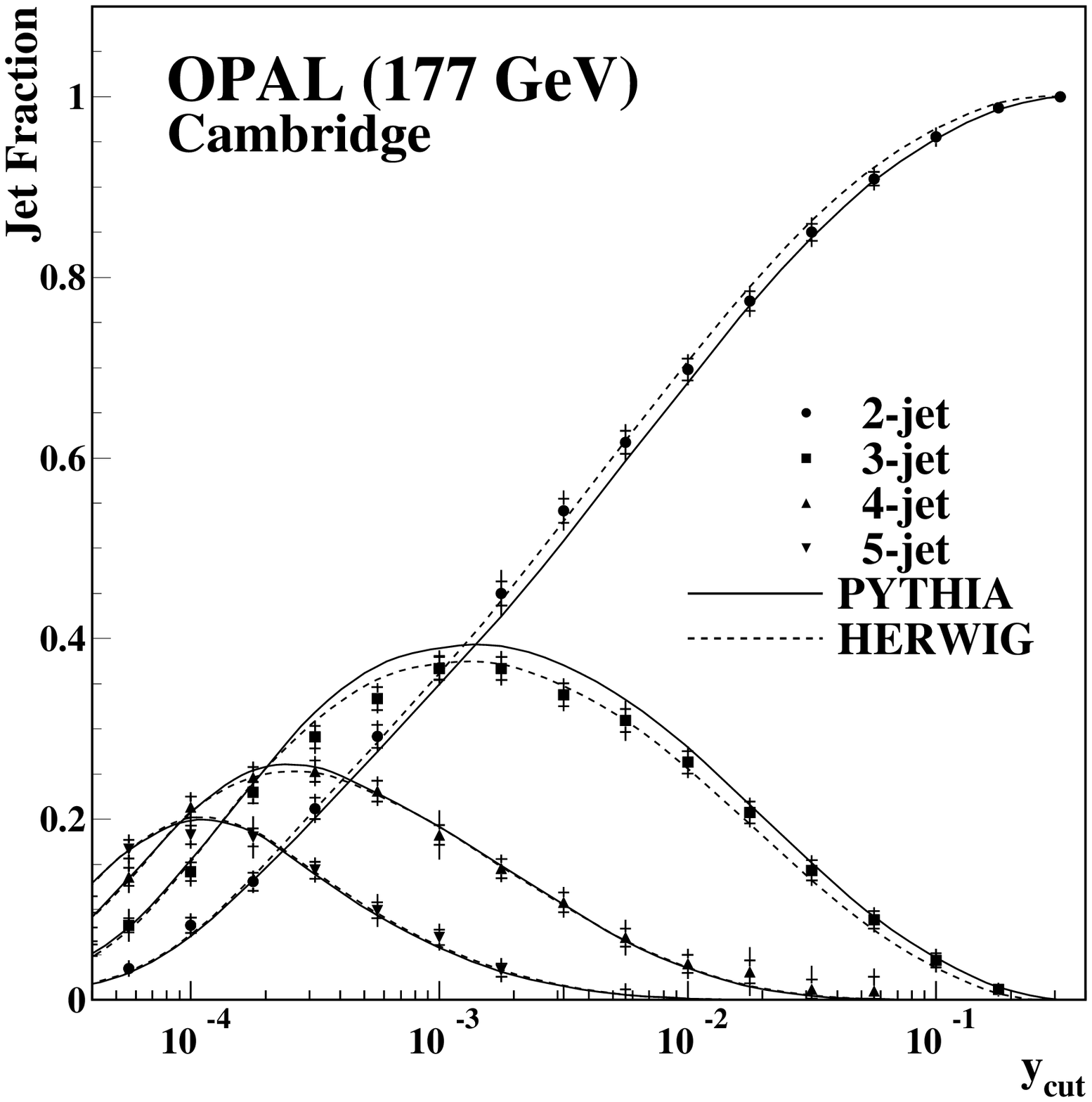}}
   \hspace*{-1cm}\mbox{
         \includegraphics[width=.59\textwidth]{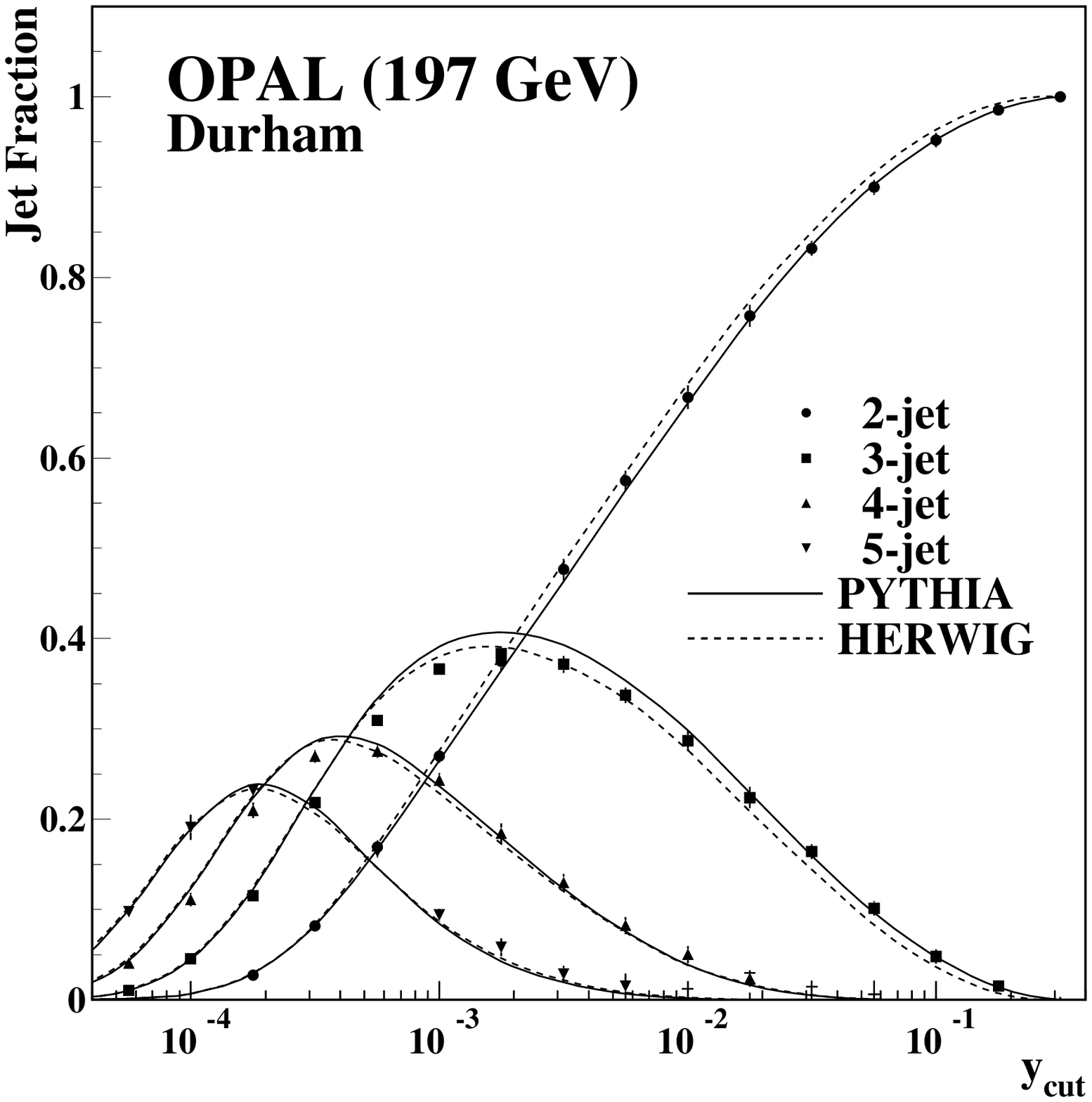}
         \hspace*{-.9cm}
         \includegraphics[width=.59\textwidth]{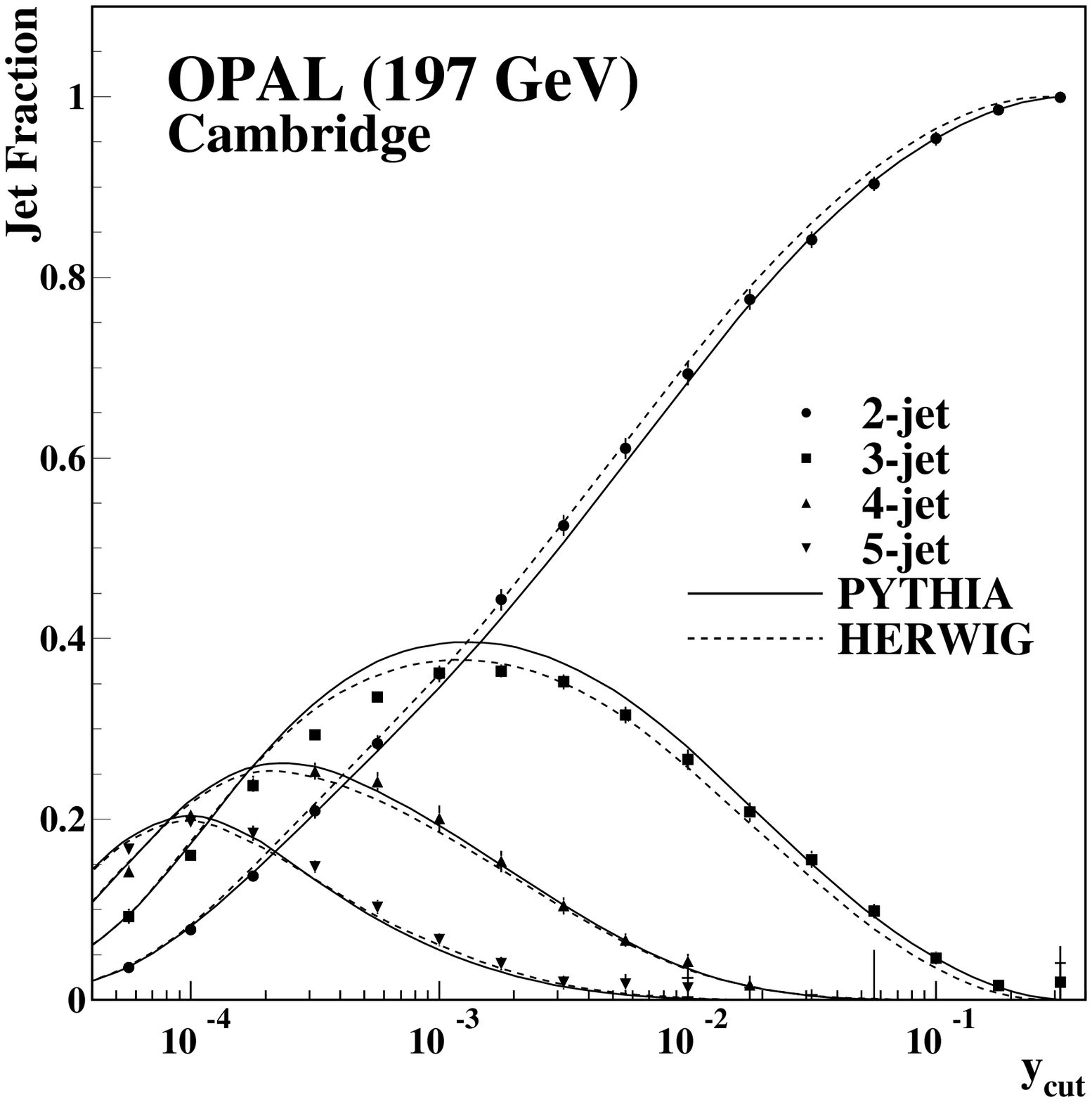}}
\vspace{-.7cm}
\caption[$n$-jet rates for the Cone, JADE, Durham and Cambridge
algorithms with the 91~GeV dataset.]
   {The hadron level $n$-jet rates for the Durham and Cambridge algorithms
   for the data with $\sqrt{s}=179$~GeV (top) and $\sqrt{s}=198$~GeV (bottom).
In all
   plots the PYTHIA and HERWIG Monte Carlo expectations
   are represented by the curves. Outer error bars indicate total
   errors while the inner bars indicate statistical errors.} 
\label{fig:jr103}
\end{figure}

\begin{figure}[H]
\centering
\vspace{-1.cm}
   \hspace*{-1cm}\mbox{
      \includegraphics[width=.59\textwidth]{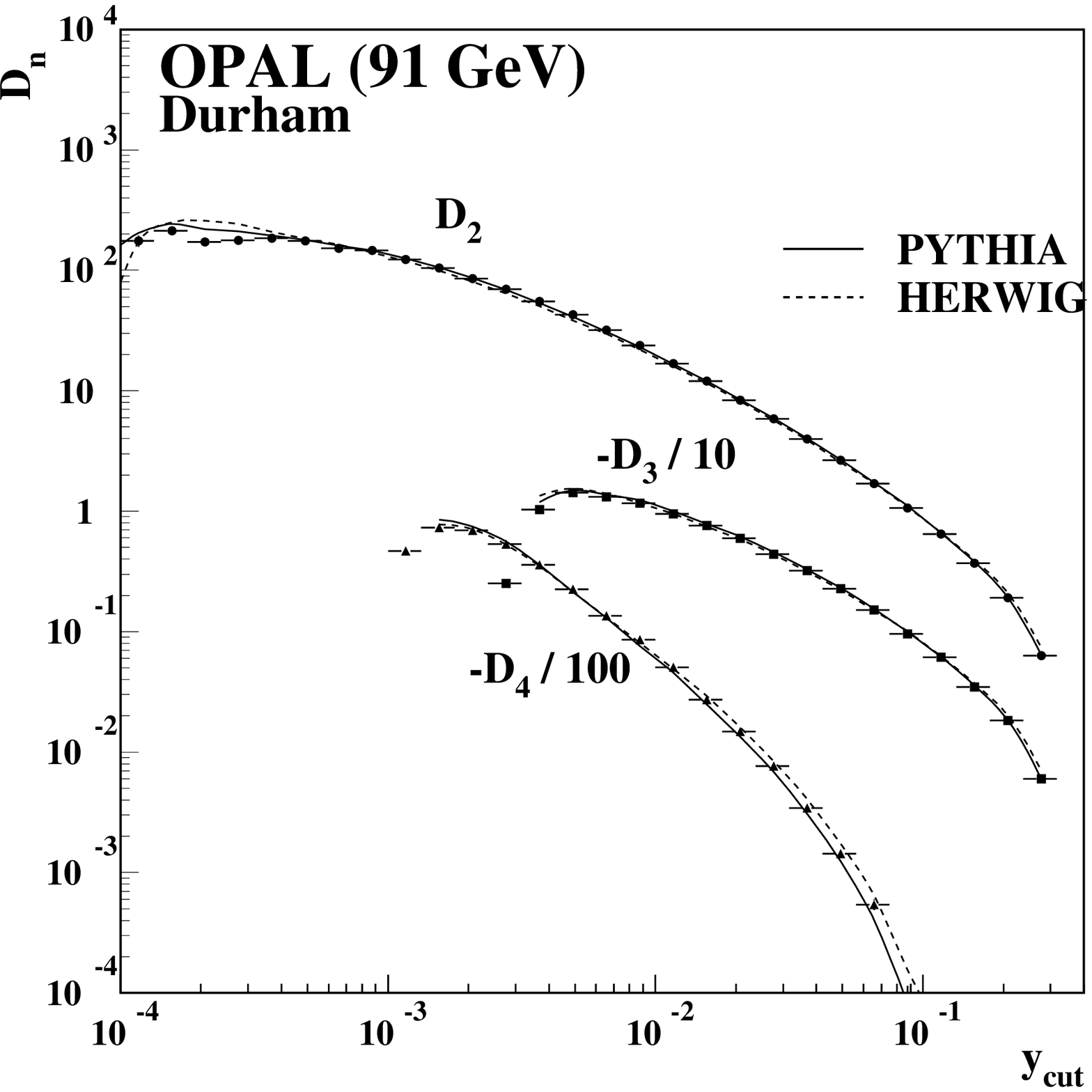}
      \hspace*{-.9cm}
      \includegraphics[width=.59\textwidth]{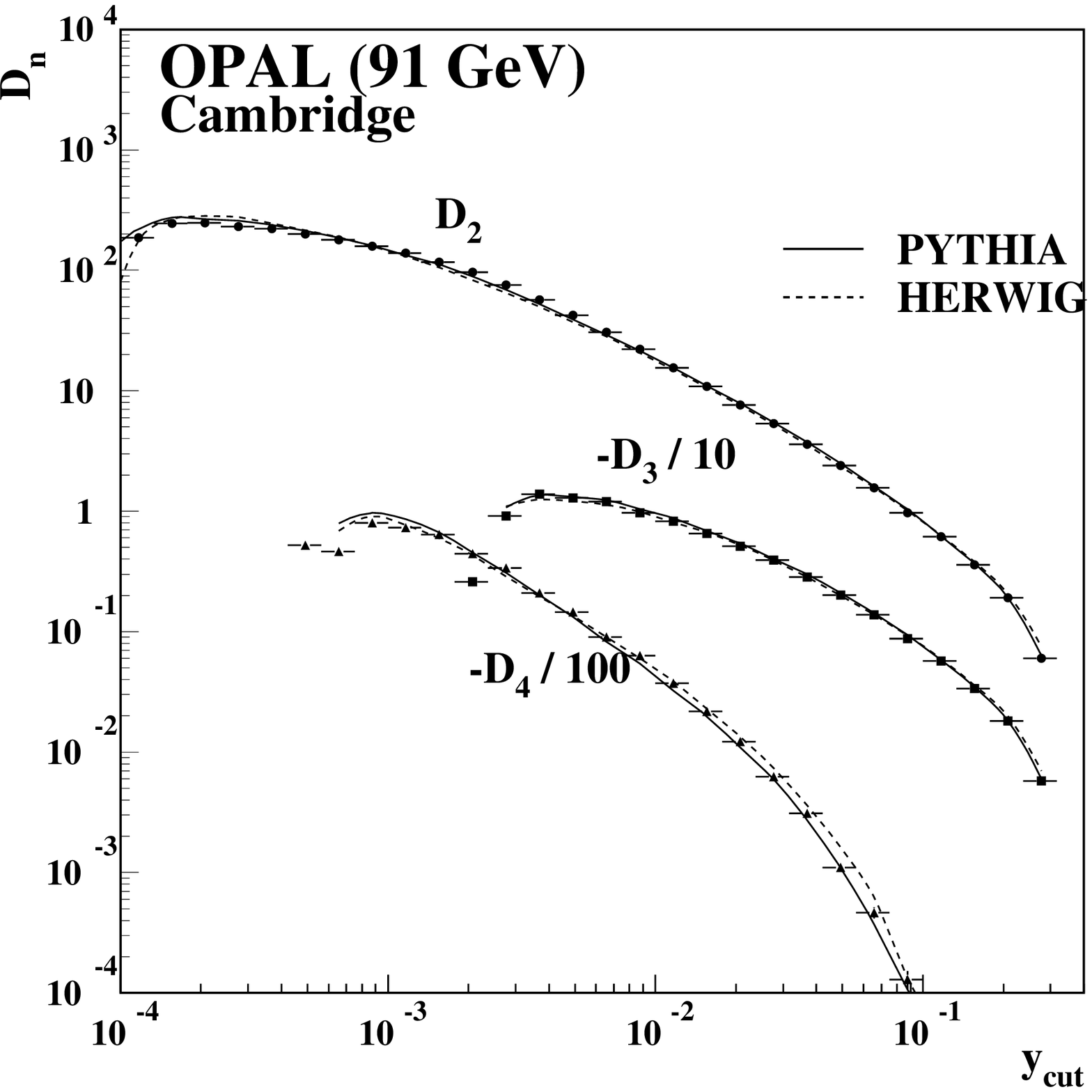}}
\vspace{-1.2cm}
\phantom{test}
   \hspace*{-1cm}\mbox{
      \includegraphics[width=.59\textwidth]{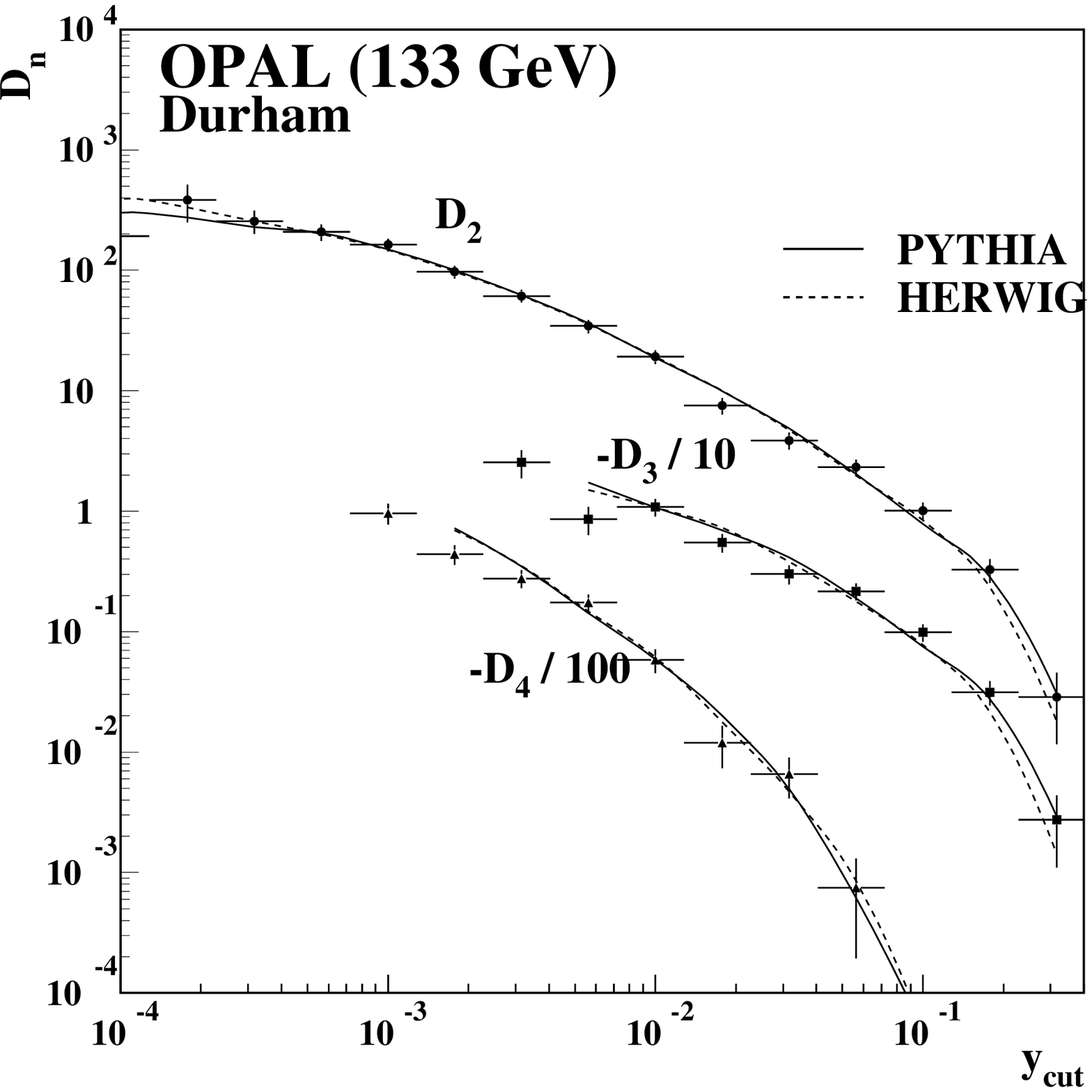}
      \hspace*{-.9cm}
      \includegraphics[width=.59\textwidth]{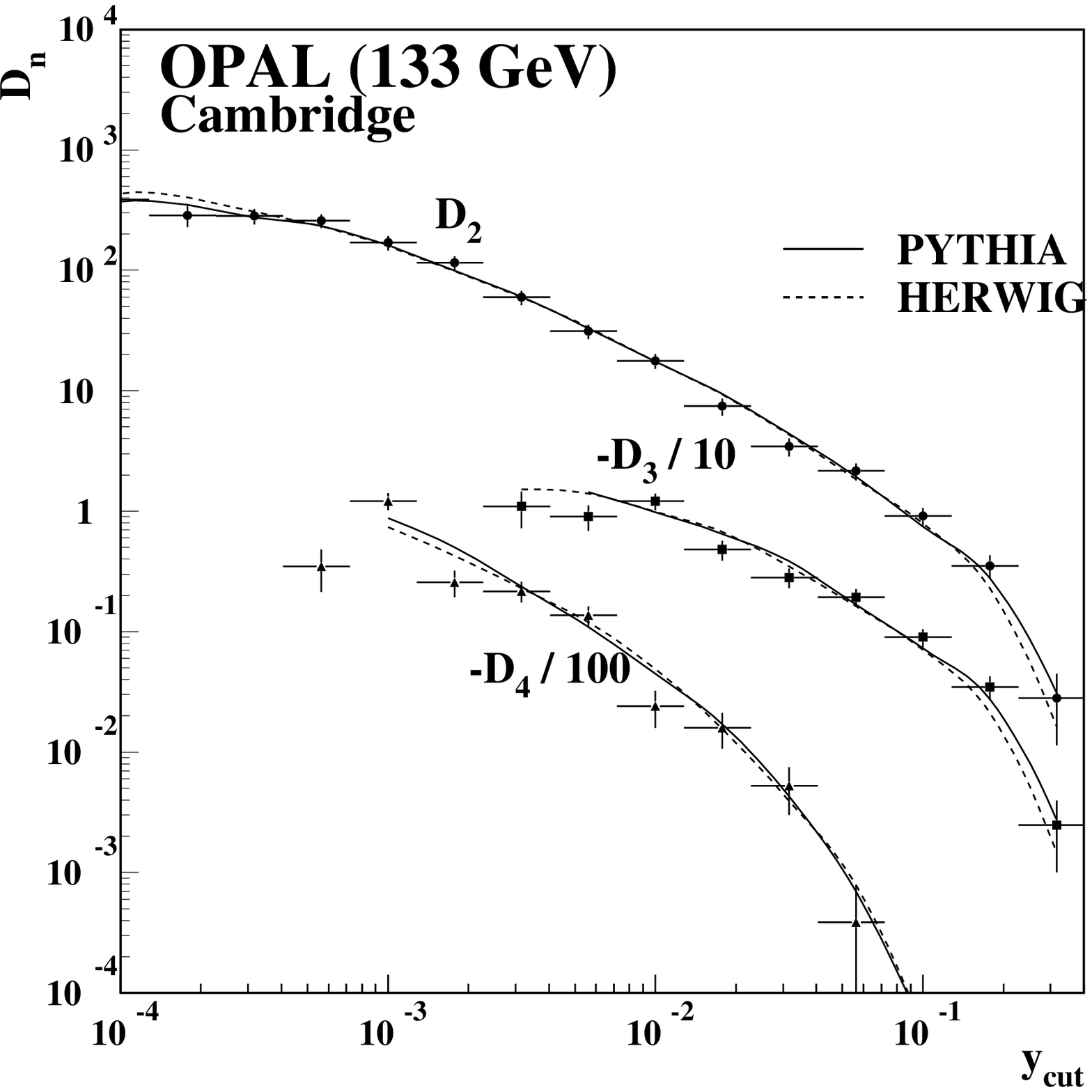}}
\vspace{-.7cm}
\caption{The differential 2-, 3- and 4-jet rates at the hadron level as a
function of 
   $y_{cut}$ for the Cambridge and Durham algorithms for data with
$\sqrt{s}=91$~GeV (top) and for the LEP1.5 combined dataset at
$\sqrt{s}=133$~GeV (bottom).  PYTHIA and HERWIG Monte Carlo expectations are
represented by the curves.  The differential 3- and 4-jet rates have a negative
slope in the region of large \yc\ and therefore the negative values of $D_{3}$
and $D_{4}$ are plotted (the positive values are not seen on the curves).  Note that $D_{3}$ and $D_{4}$ are scaled down by one
and two orders of magnitude, respectively, for clarity.  Error bars indicate
total (statistical + systematic) errors.} 
\label{fig:dn91}
\end{figure}
\begin{figure}[H]
\centering
\vspace{-1.cm}
   \hspace*{-1cm}\mbox{
      \includegraphics[width=.59\textwidth]{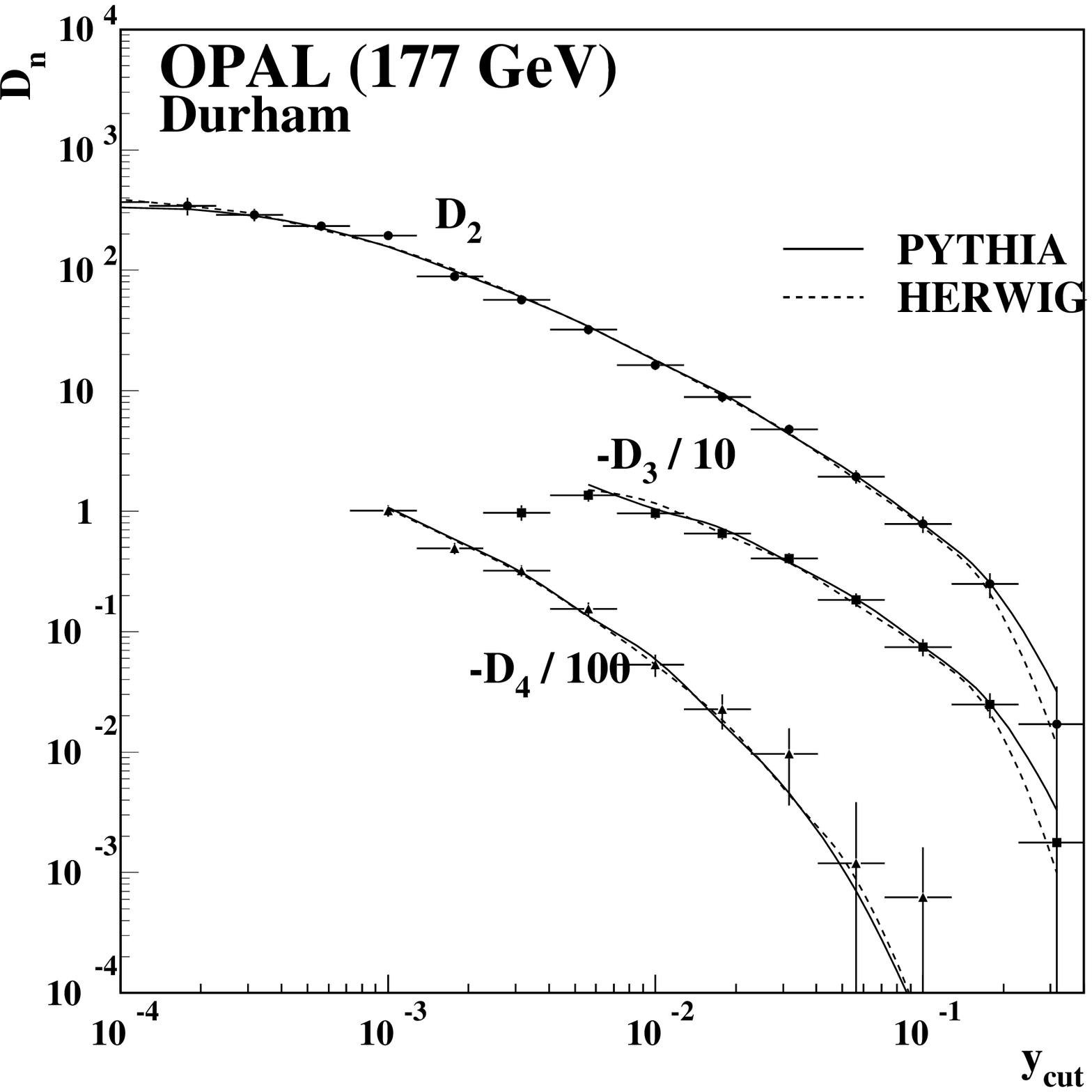}
      \hspace*{-.9cm}
      \includegraphics[width=.59\textwidth]{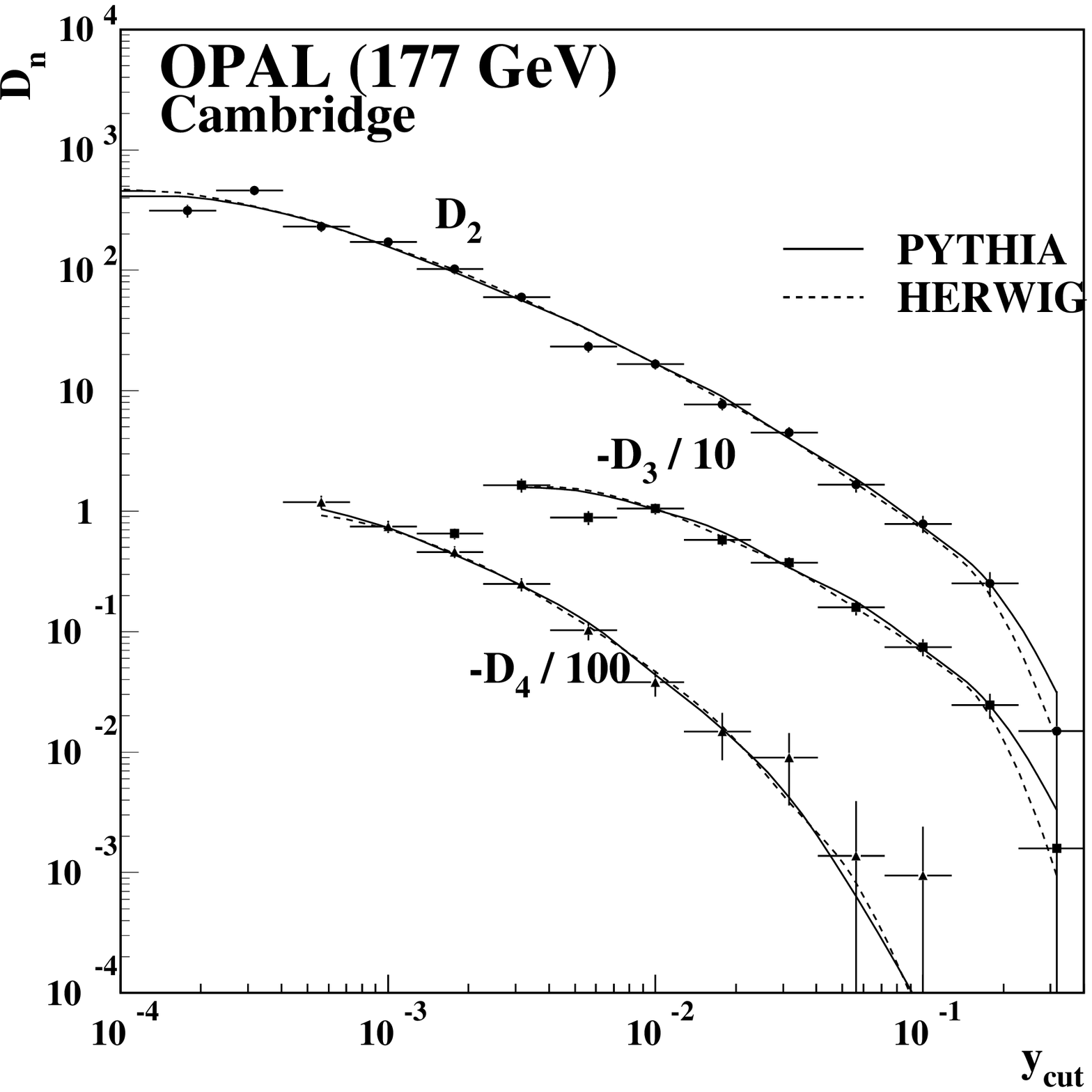}}
\vspace{-1.2cm}
\phantom{test}
   \hspace*{-1cm}\mbox{
      \includegraphics[width=.59\textwidth]{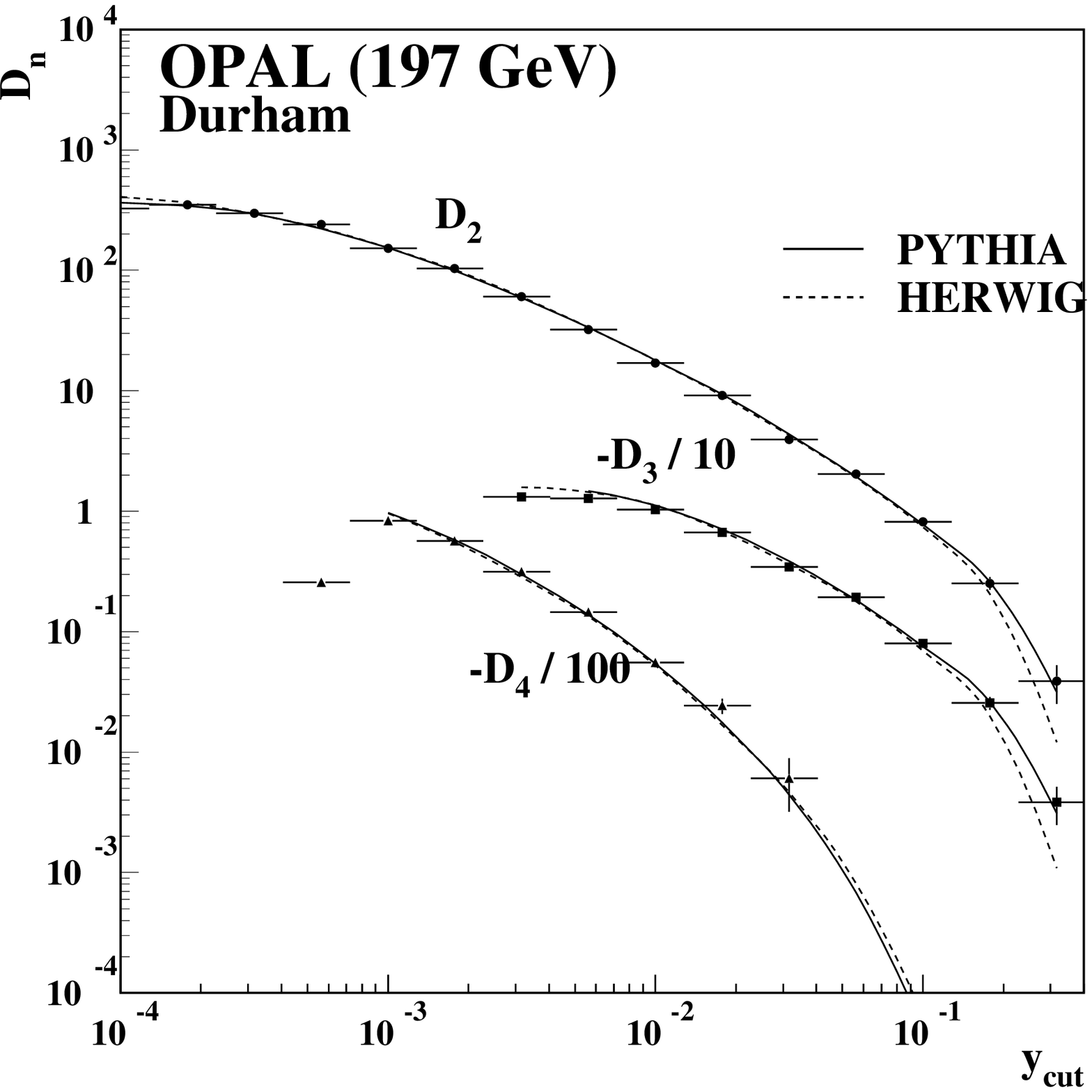}
      \hspace*{-.9cm}
      \includegraphics[width=.59\textwidth]{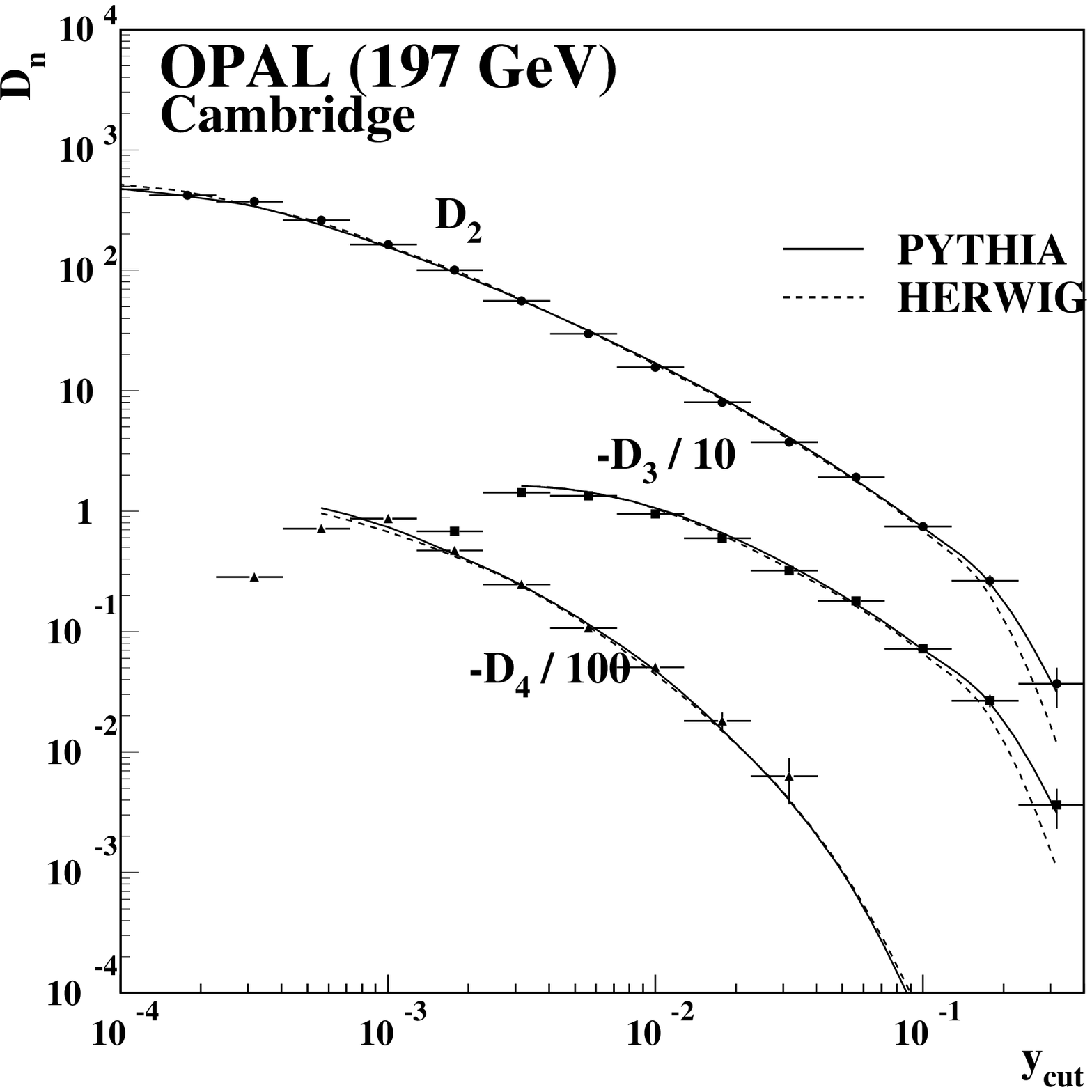}}
\vspace{-.7cm}
\caption[The Cambridge and Durham differential jet rates a function of
$y_{cut}$ for at $\sqrt{s}=103$~GeV]
{The differential 2-, 3- and 4-jet rates at the hadron
   level as a function of $y_{cut}$ for 
   the Cambridge and Durham algorithms for two combined LEP2 datasets with
   $\sqrt{s}=177$~GeV (top) and $\sqrt{s}=197$~GeV (bottom).  
   PYTHIA and HERWIG Monte Carlo expectations
   are represented by the curves.  The differential
   3- and 4-jet rates have a negative slope in the region of large \yc\ 
   and therefore the negative values of $D_{3}$ and $D_{4}$ are plotted
(the positive values are not seen on the curves).  Note
that $D_{3}$ and
   $D_{4}$ are scaled down by one and two orders of magnitude,
   respectively, for clarity.  Error bars indicate total
   (statistical + systematic) errors.} 
\label{fig:dn186}
\end{figure}
\begin{figure}[H]
\centering
\includegraphics[width=1.1\textwidth]{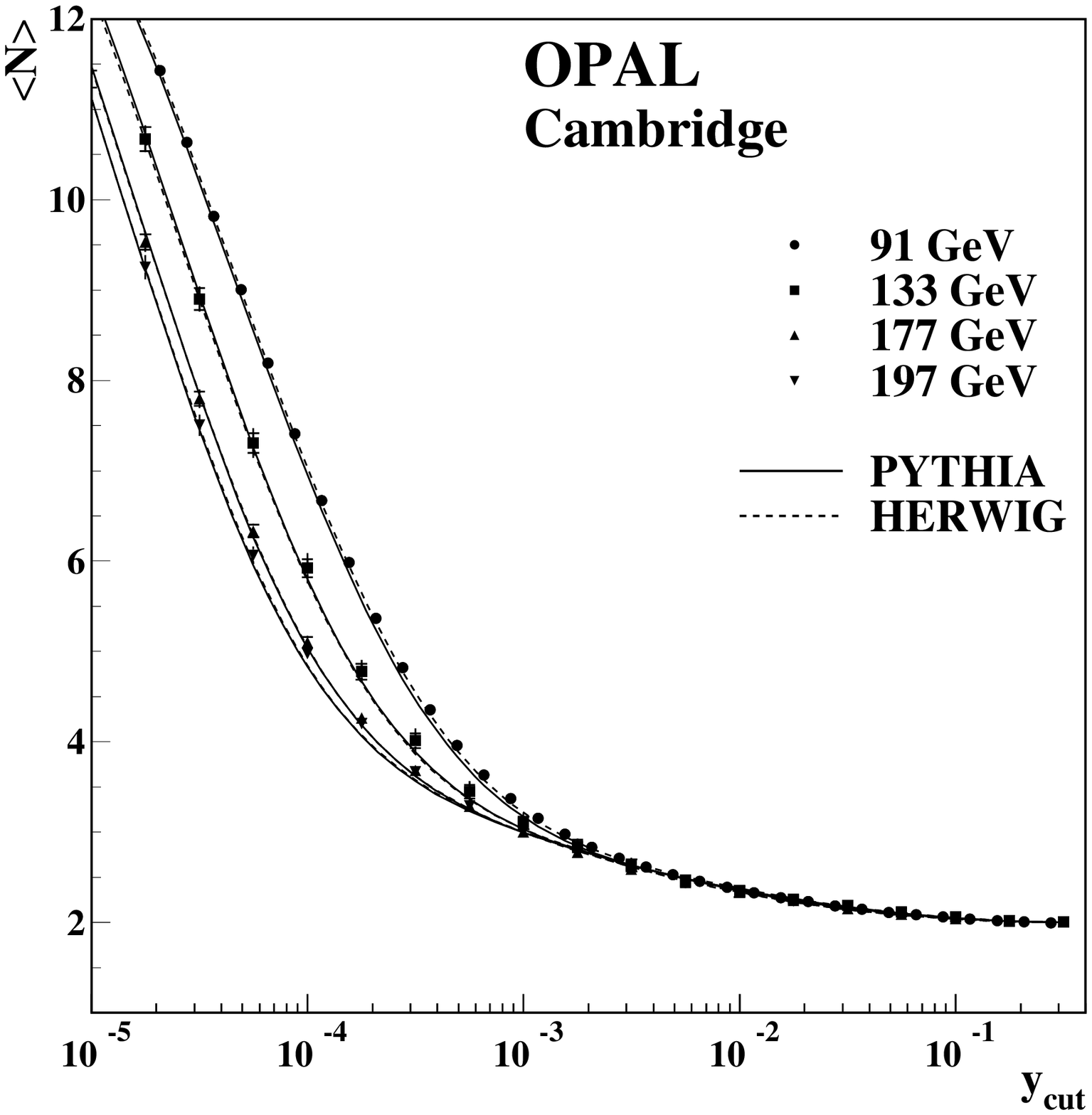}
\caption{The average jet rates at the hadron level as a function of $y_{cut}$
for 
   the Cambridge algorithm for the all \cfm energies. PYTHIA and HERWIG Monte
Carlo expectations
   are represented by the curves.  Error bars indicate total (statistical +
systematic) errors.} 
\label{fig:avn91}
\end{figure}
\begin{figure}[H]
\centering
\includegraphics[width=1.1\textwidth]{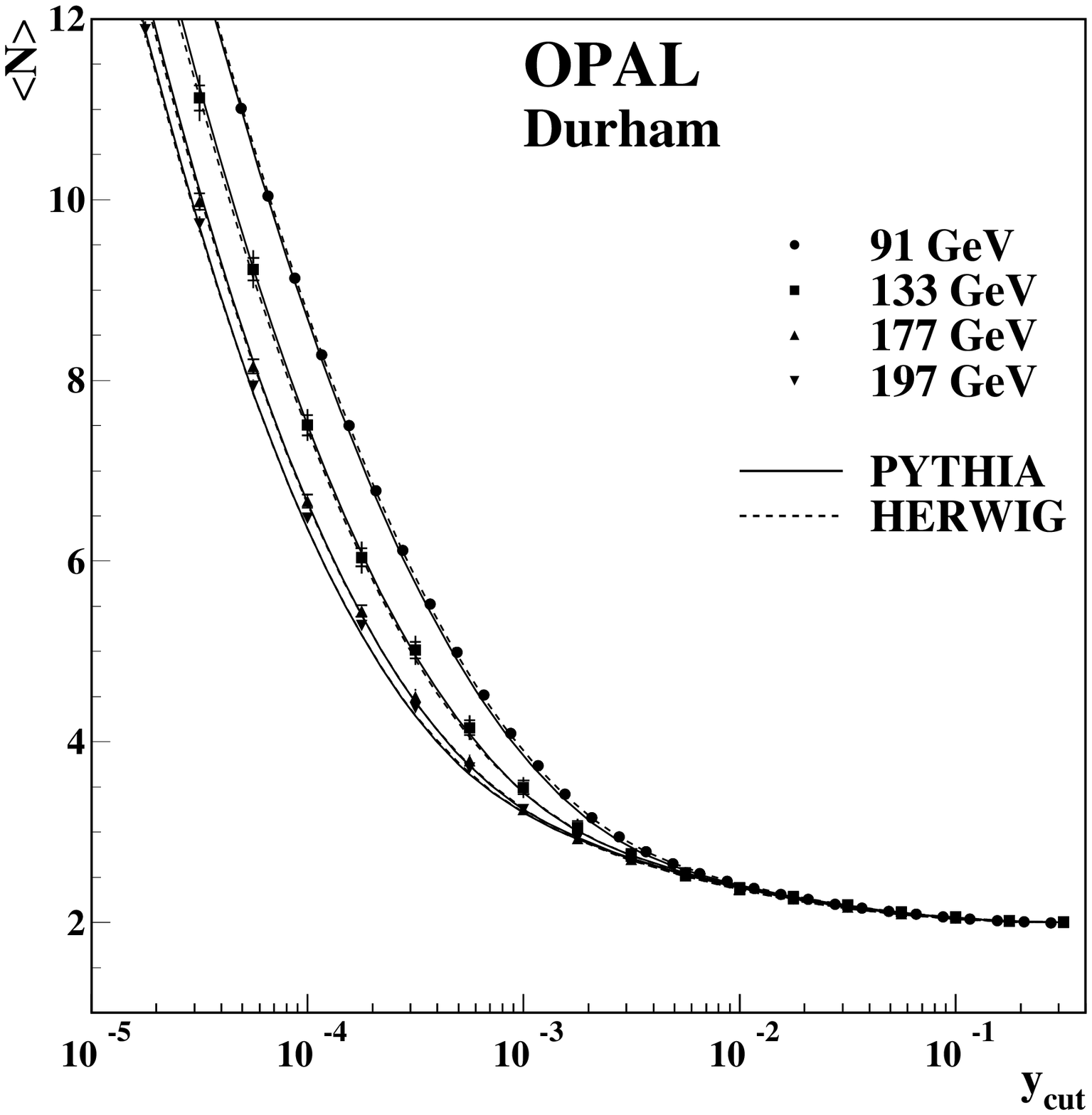}
\caption{The average jet rates at the hadron level as a function of $y_{cut}$
for 
   the Durham algorithm for the all \cfm energies. PYTHIA and HERWIG Monte Carlo
expectations
   are represented by the curves.  Error bars indicate total (statistical +
systematic) errors.} 
\label{fig:avn186}
\end{figure}

\begin{figure}[H]
   \centering
   \hspace*{-1cm}\mbox{
      \includegraphics[width=.55\textwidth]{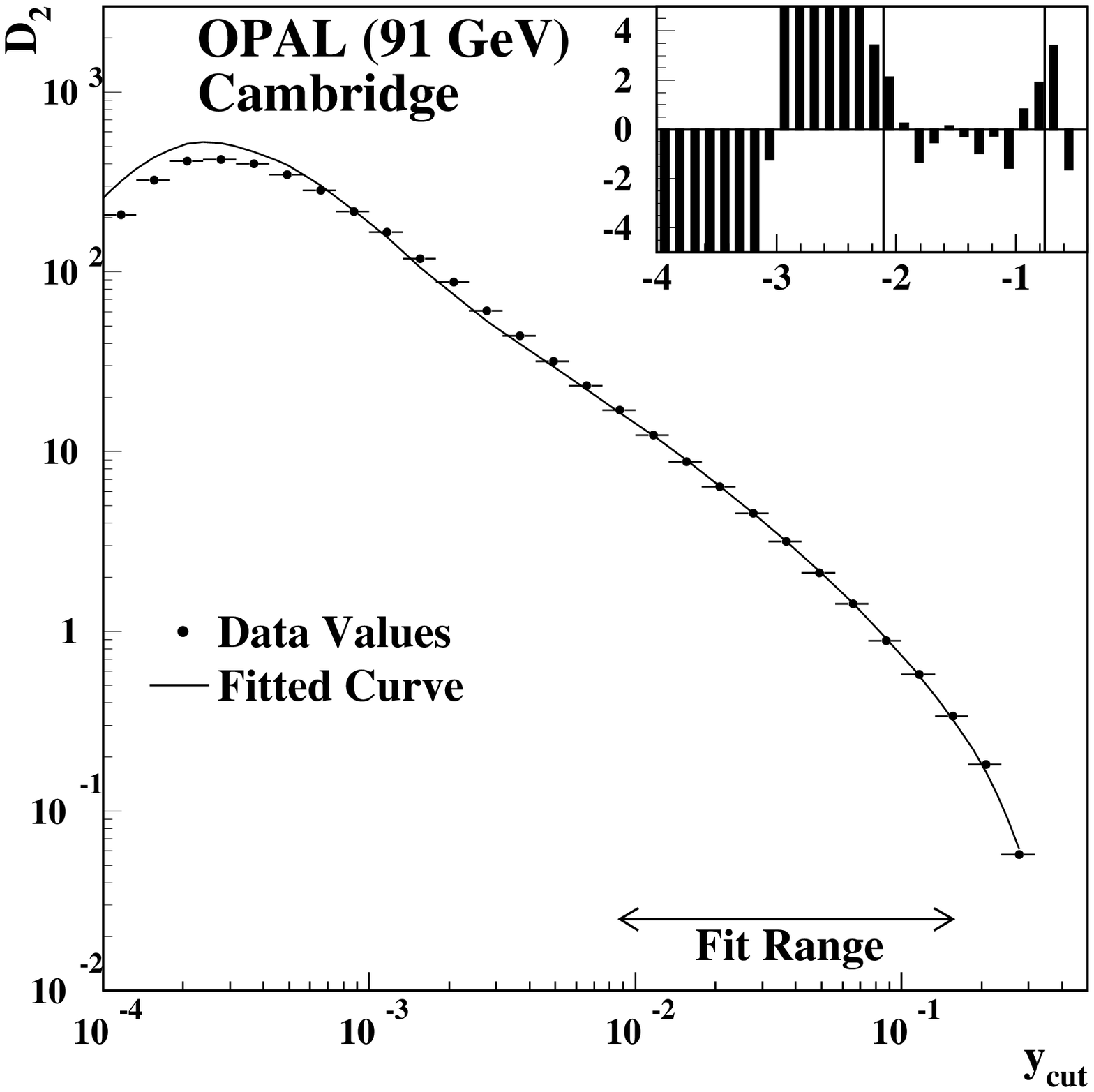}         
      \hspace*{-.9cm}
      \includegraphics[width=.55\textwidth]{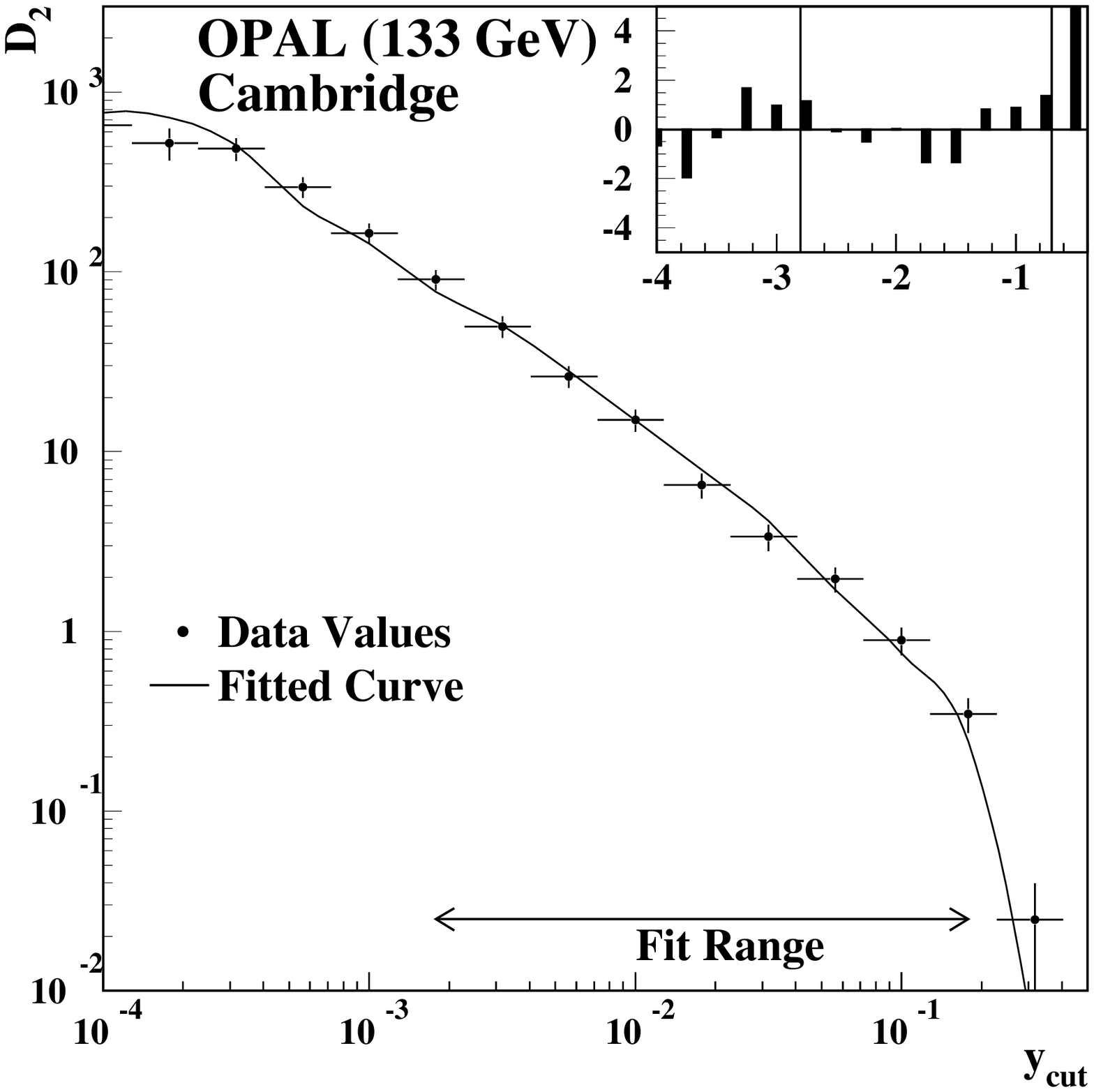}}         
   \hspace*{-1cm}\mbox{
      \includegraphics[width=.55\textwidth]{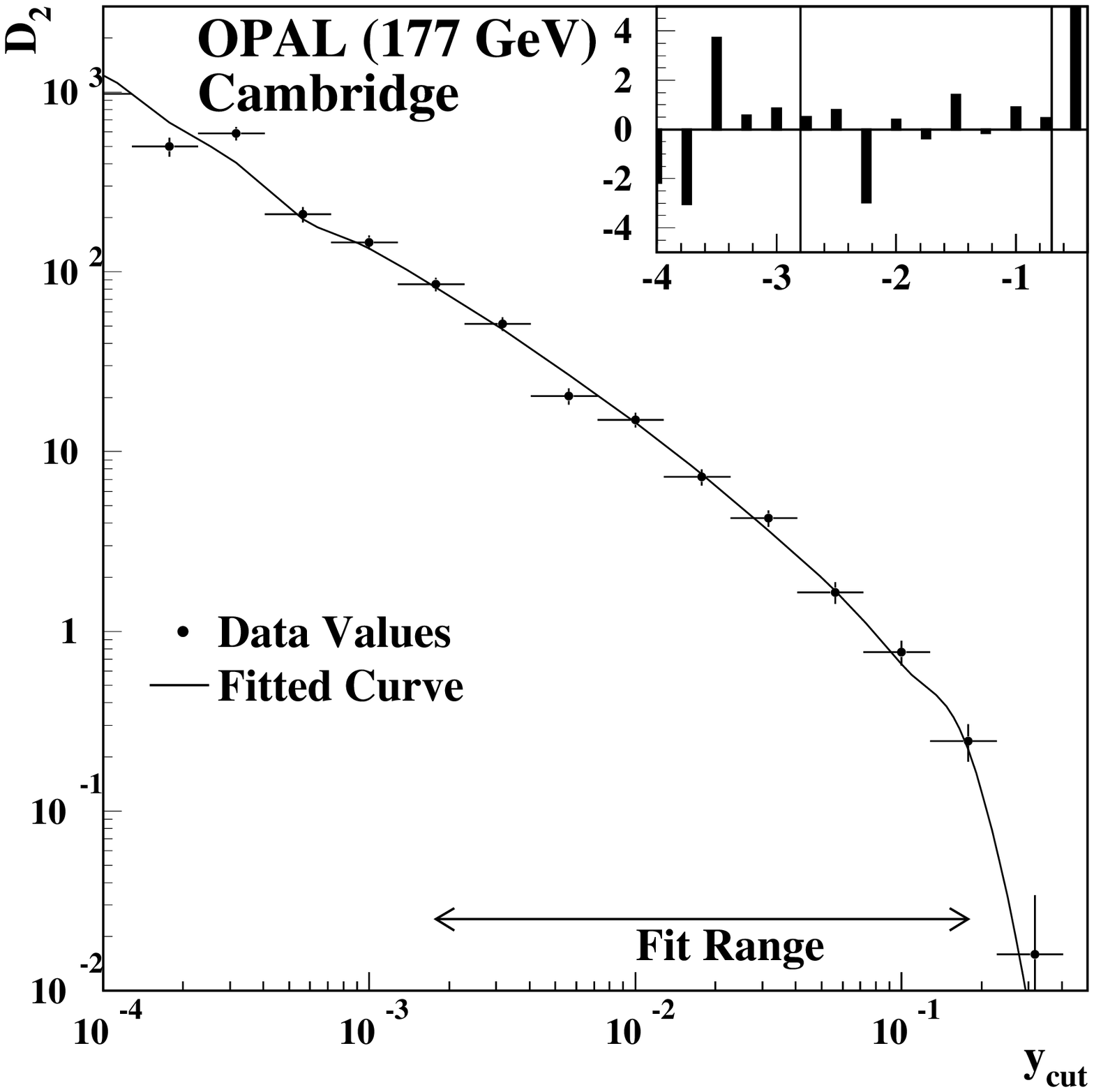}         
      \hspace*{-.9cm}
      \includegraphics[width=.55\textwidth]{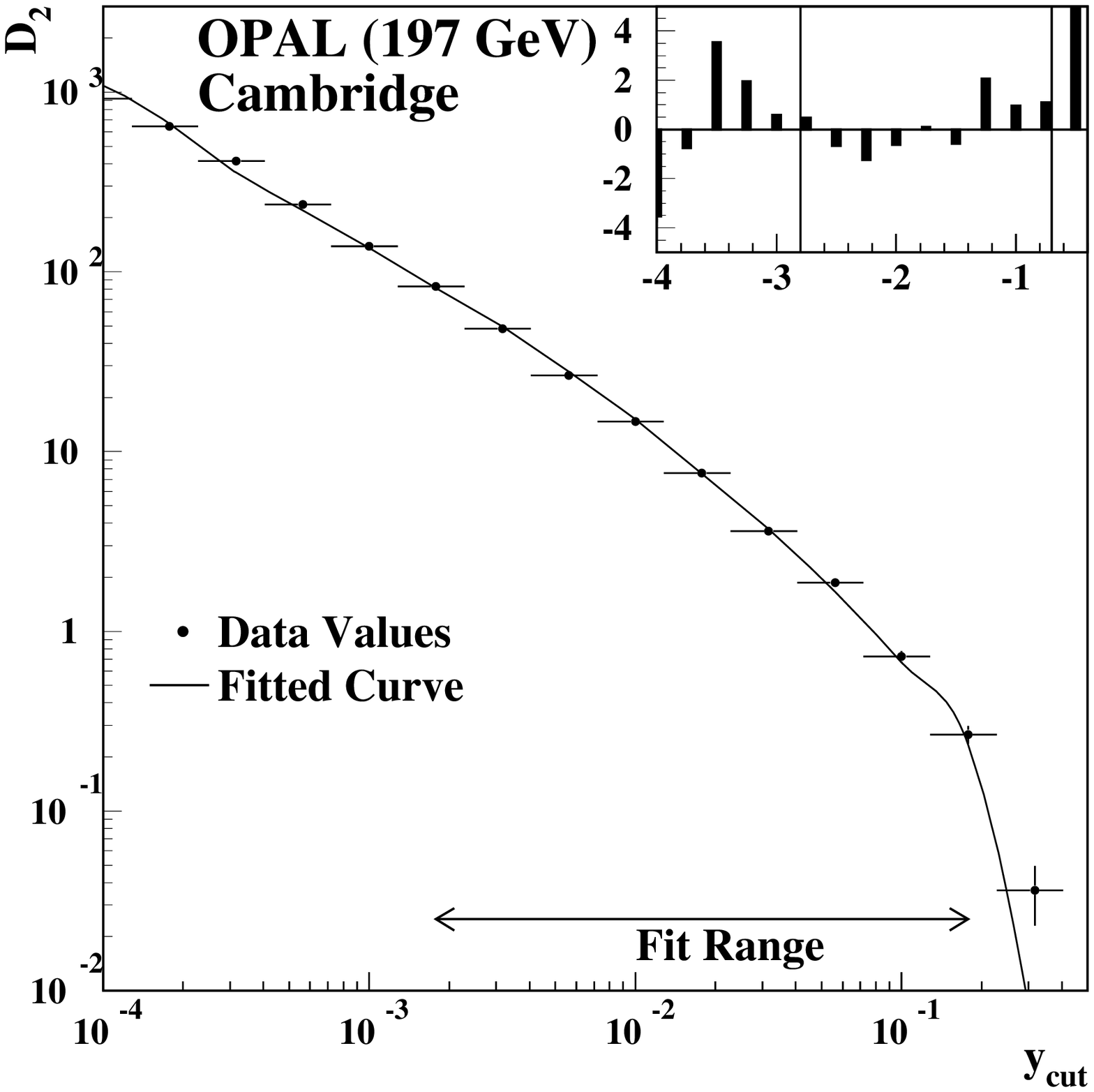}}
   \caption{Fits of the parton level $D_{2}$ distribution using the Cambridge
algorithm 
      as a function of $\yc$ to the
      \mbox{ln $R$} prediction for the 91~GeV (top left), 133~GeV (top right),
179~GeV (bottom
      left) and 198~GeV (bottom right) datasets.  The inset plots show the pull
of each point
      from the line of best fit vs $\log_{10}\yc$.
      The two vertical lines delineate the fit
range.  Vertical error
      bars
      represent statistical errors only.} 
   \label{d2cfits}
\end{figure} 

\begin{figure}[H]
   \centering
   \hspace*{-1cm}\mbox{
      \includegraphics[width=.55\textwidth]{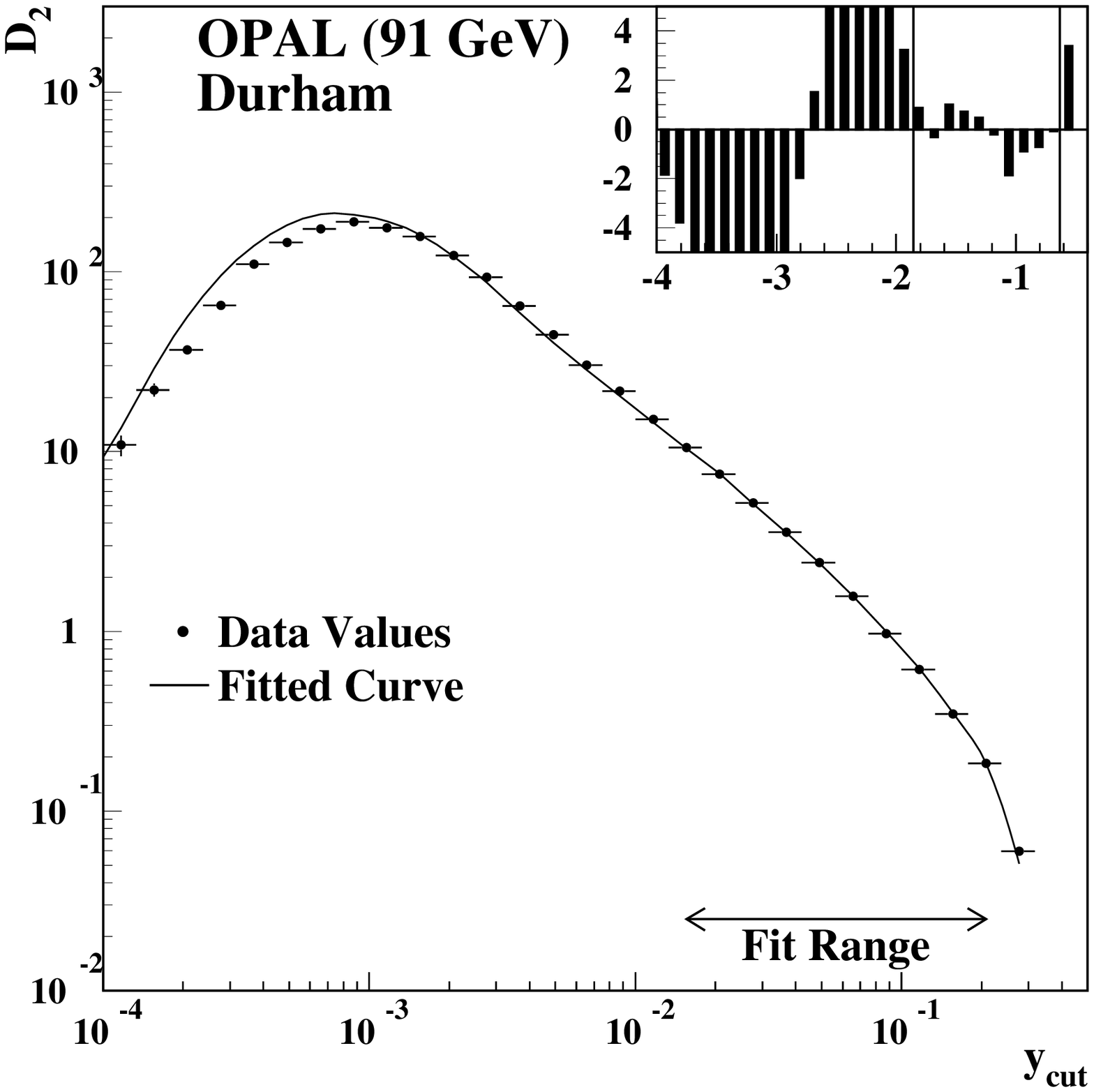}         
      \hspace*{-.9cm}
      \includegraphics[width=.55\textwidth]{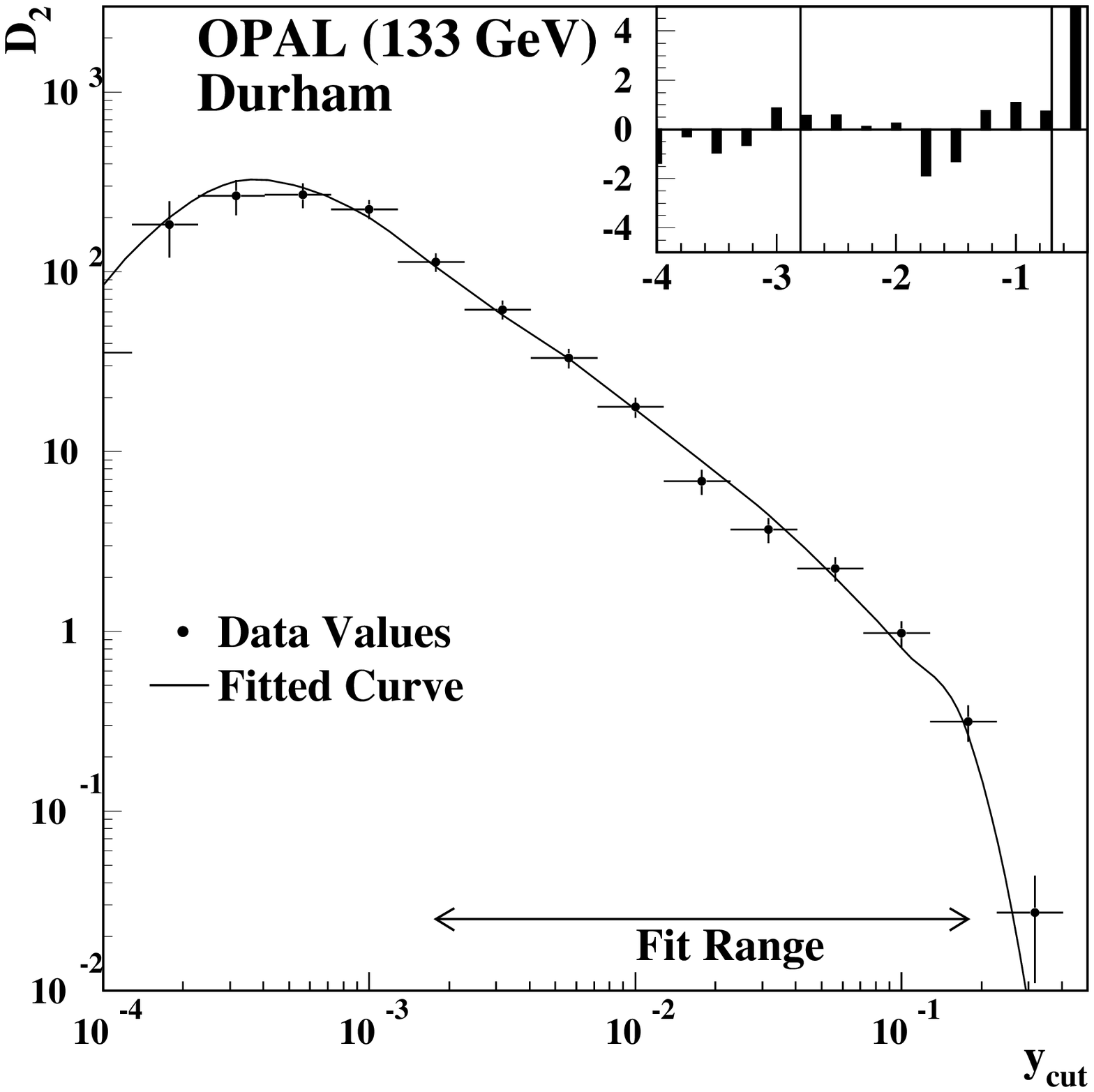}}         
   \hspace*{-1cm}\mbox{
      \includegraphics[width=.55\textwidth]{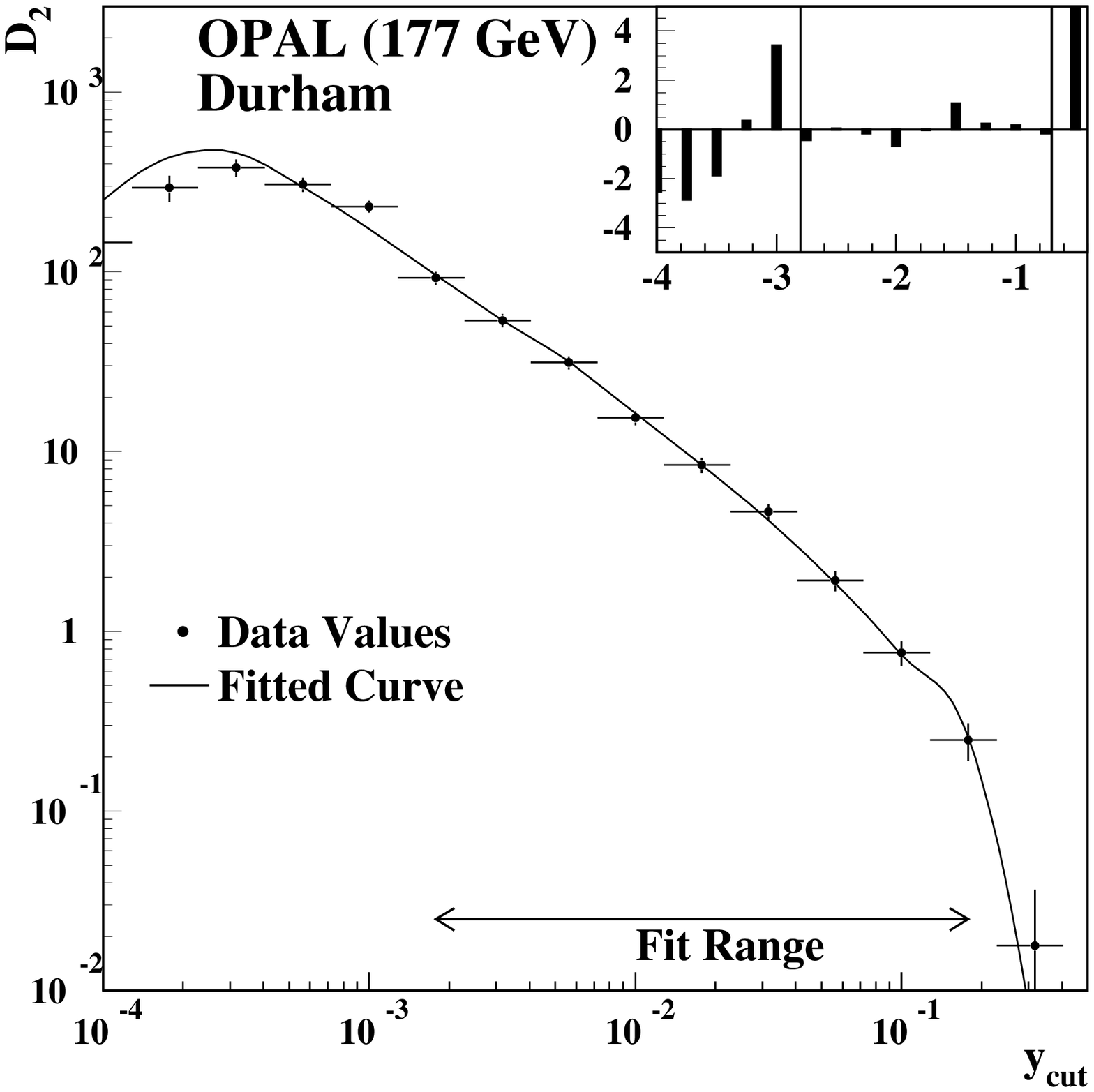}         
      \hspace*{-.9cm}
      \includegraphics[width=.55\textwidth]{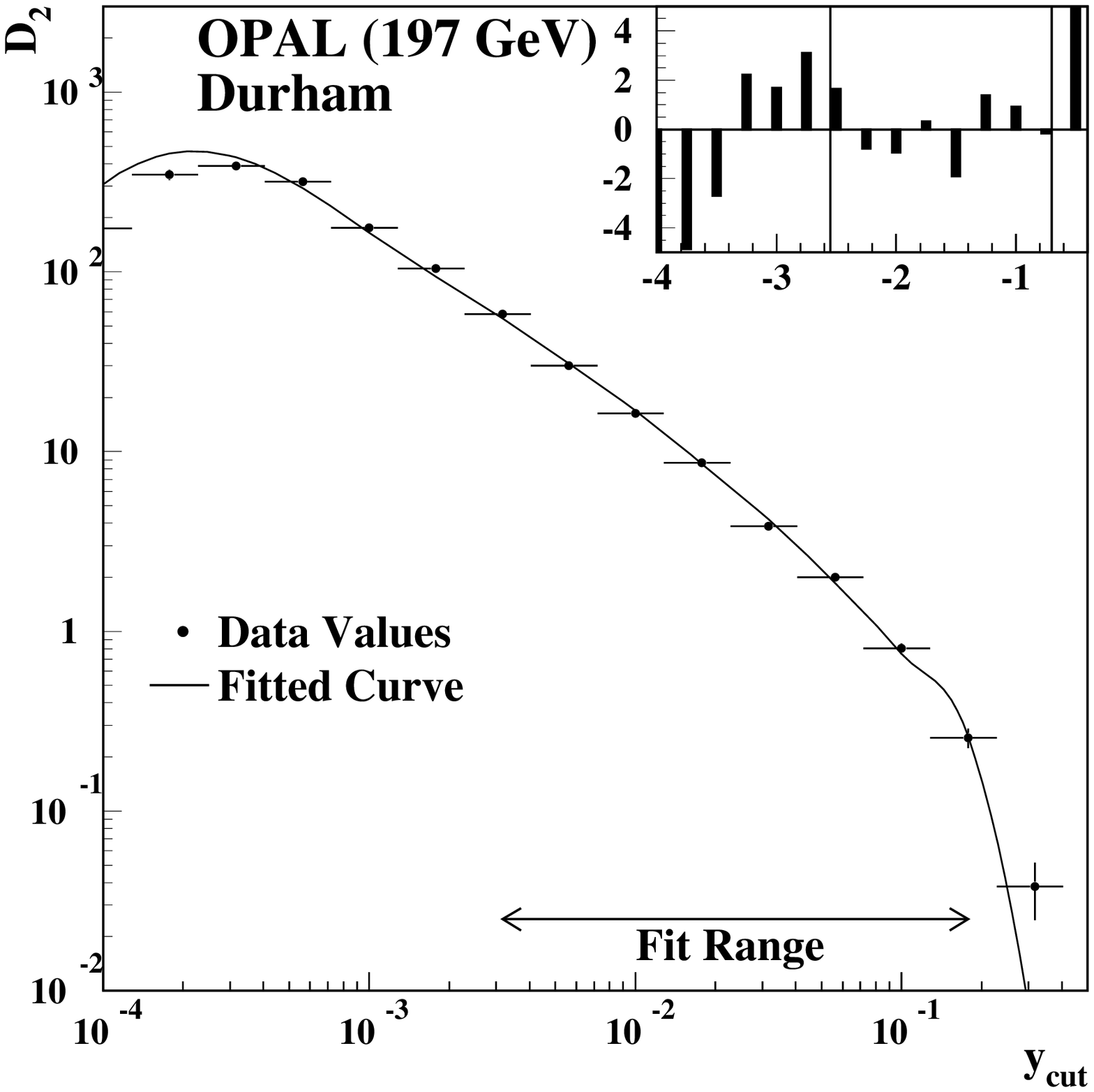}}
   \caption{Fits of the parton level $D_{2}$ distribution using the Durham
algorithm 
      as a function of $\yc$ to the
      \mbox{ln $R$} prediction for the 91~GeV (top left), 133~GeV (top right),
179~GeV (bottom
      left) and 198~GeV (bottom right) datasets.  The inset plots show the pull
of each point
      from the line of best fit vs $\log_{10}\yc$. 
      The two vertical lines delineate the fit range.  Vertical Error
      bars
      represent statistical errors only.} 
   \label{d2dfits}
\end{figure} 

\begin{figure}[H]
   \centering
   \hspace*{-1cm}\mbox{
      \includegraphics[width=.58\textwidth]{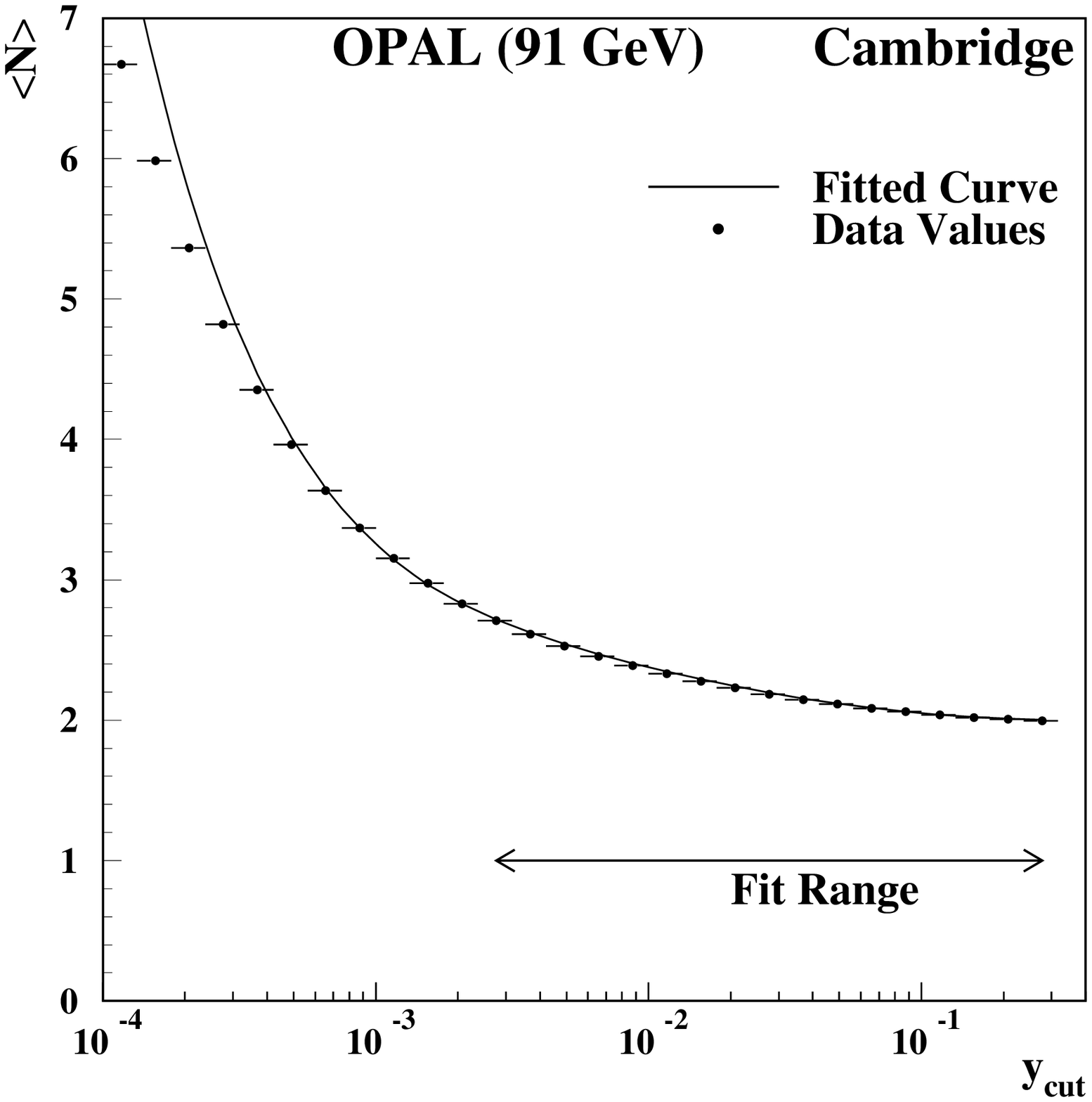}         
      \hspace*{-.9cm}
      \includegraphics[width=.58\textwidth]{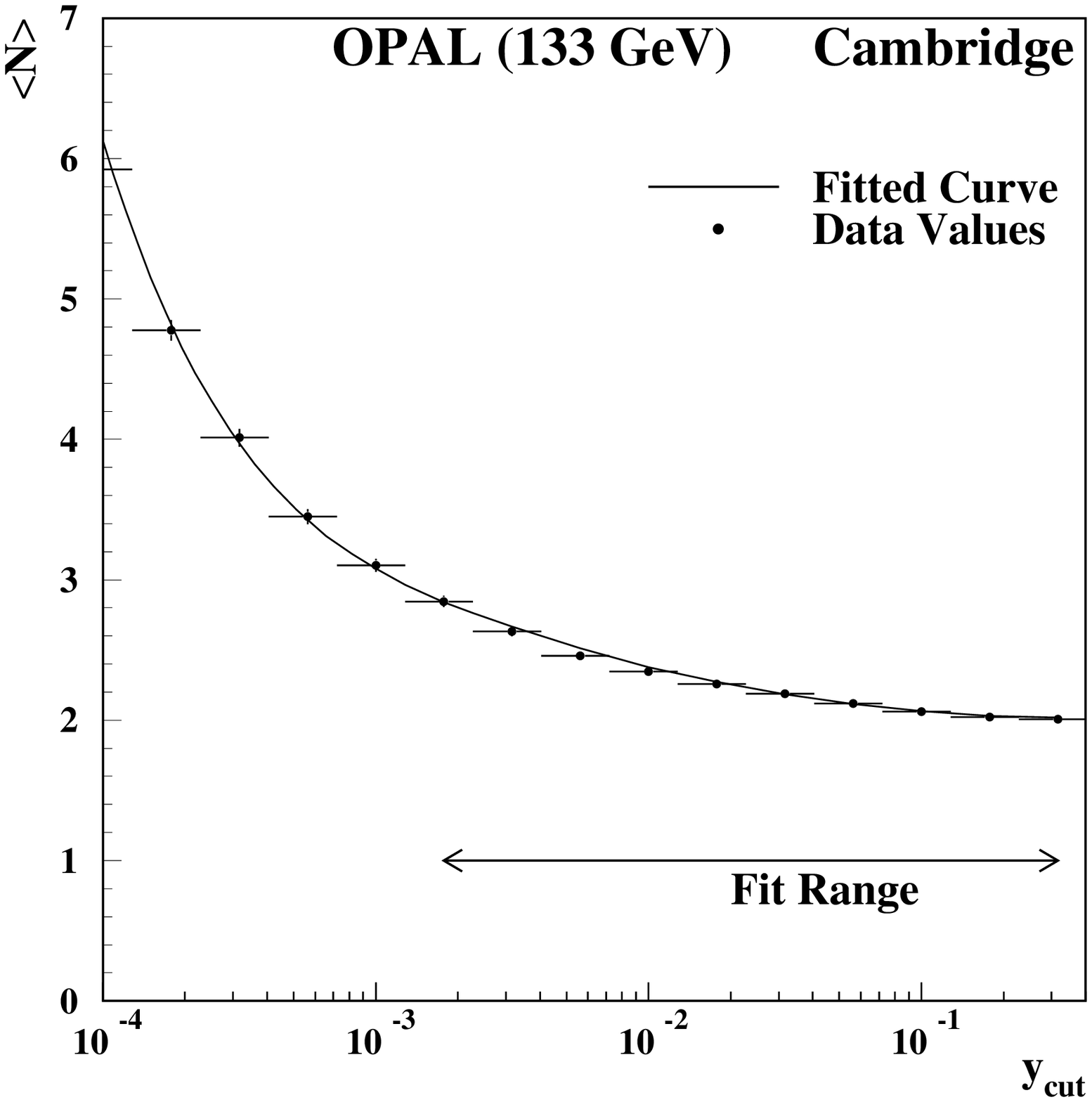}}         
   \hspace*{-1cm}\mbox{
      \includegraphics[width=.58\textwidth]{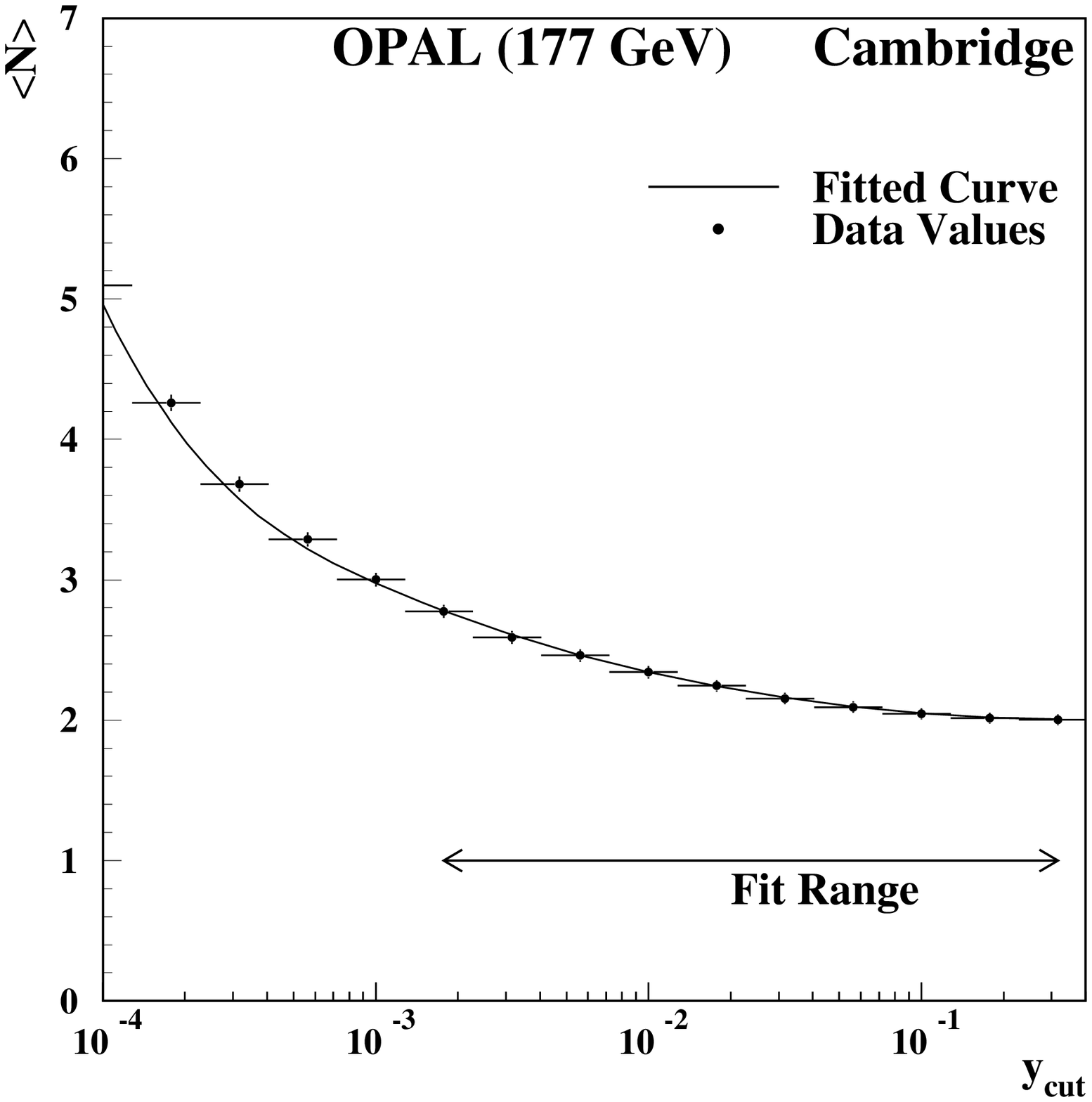}         
      \hspace*{-.9cm}
      \includegraphics[width=.58\textwidth]{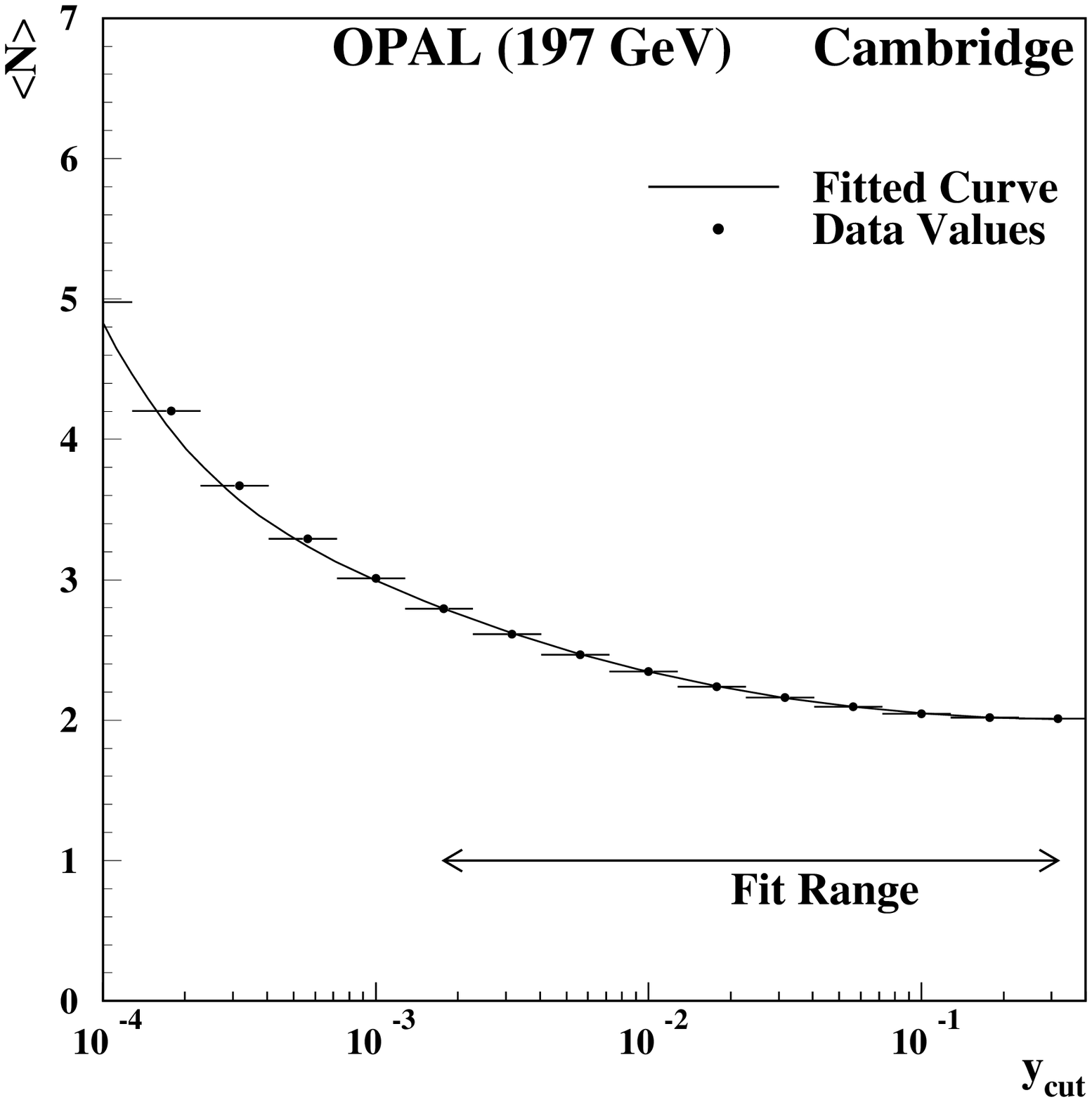}}
   \caption{Fits of the parton level $\avn$ distribution using the Cambridge
algorithm 
      as a function of $\yc$ to the
      \mbox{ln $R$} prediction for the 91~GeV (top left), 133~GeV (top right),
179~GeV (bottom
      left) and 198~GeV (bottom right) datasets.  Vertical error bars represent
statistical errors
      only.  Note: pulls are not shown as there are very large bin-to-bin
correlations.} 
   \label{ncfits}
\end{figure} 

\begin{figure}[H]
   \centering
   \hspace*{-1cm}\mbox{
      \includegraphics[width=.58\textwidth]{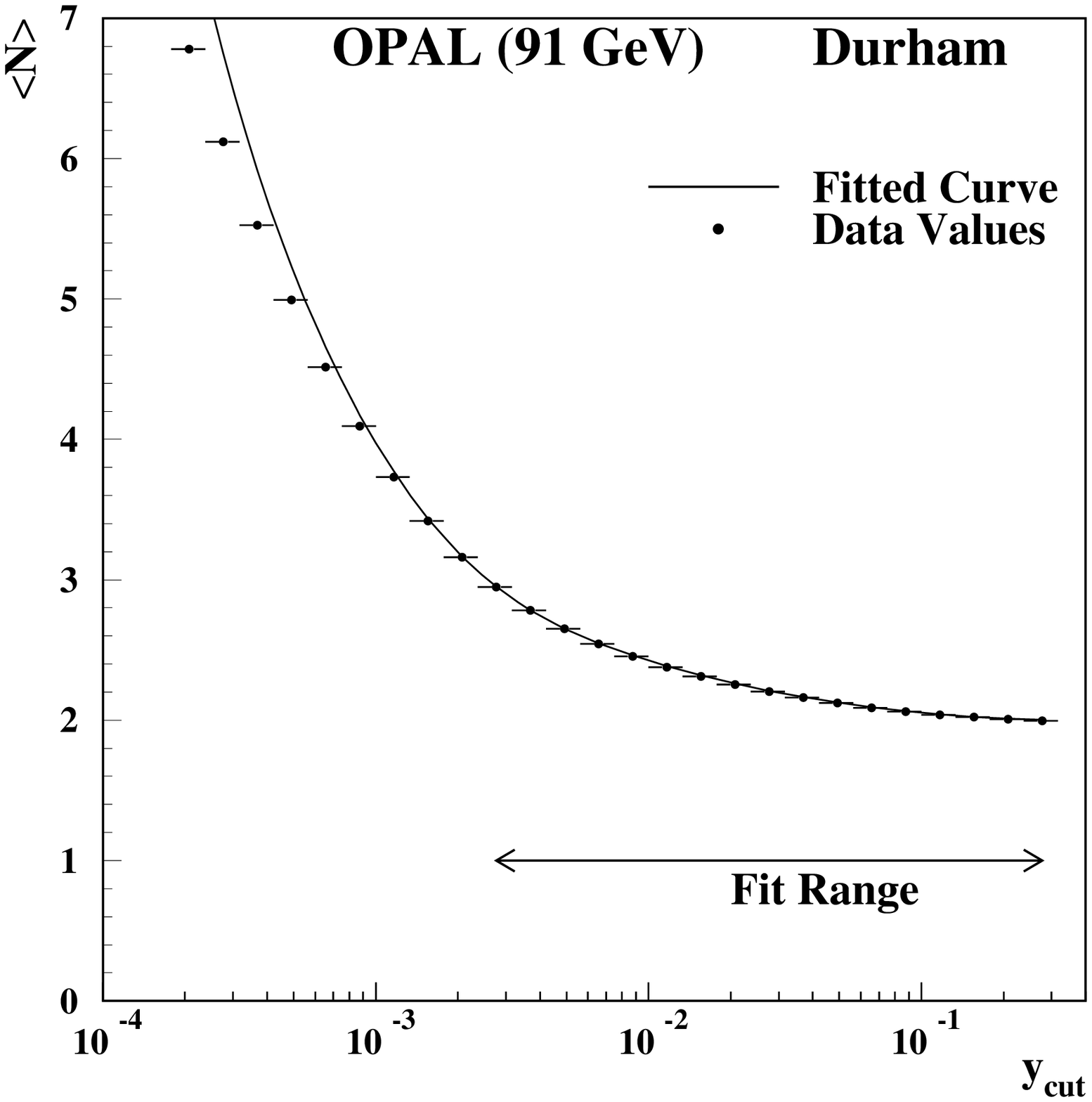}         
      \hspace*{-.9cm}
      \includegraphics[width=.58\textwidth]{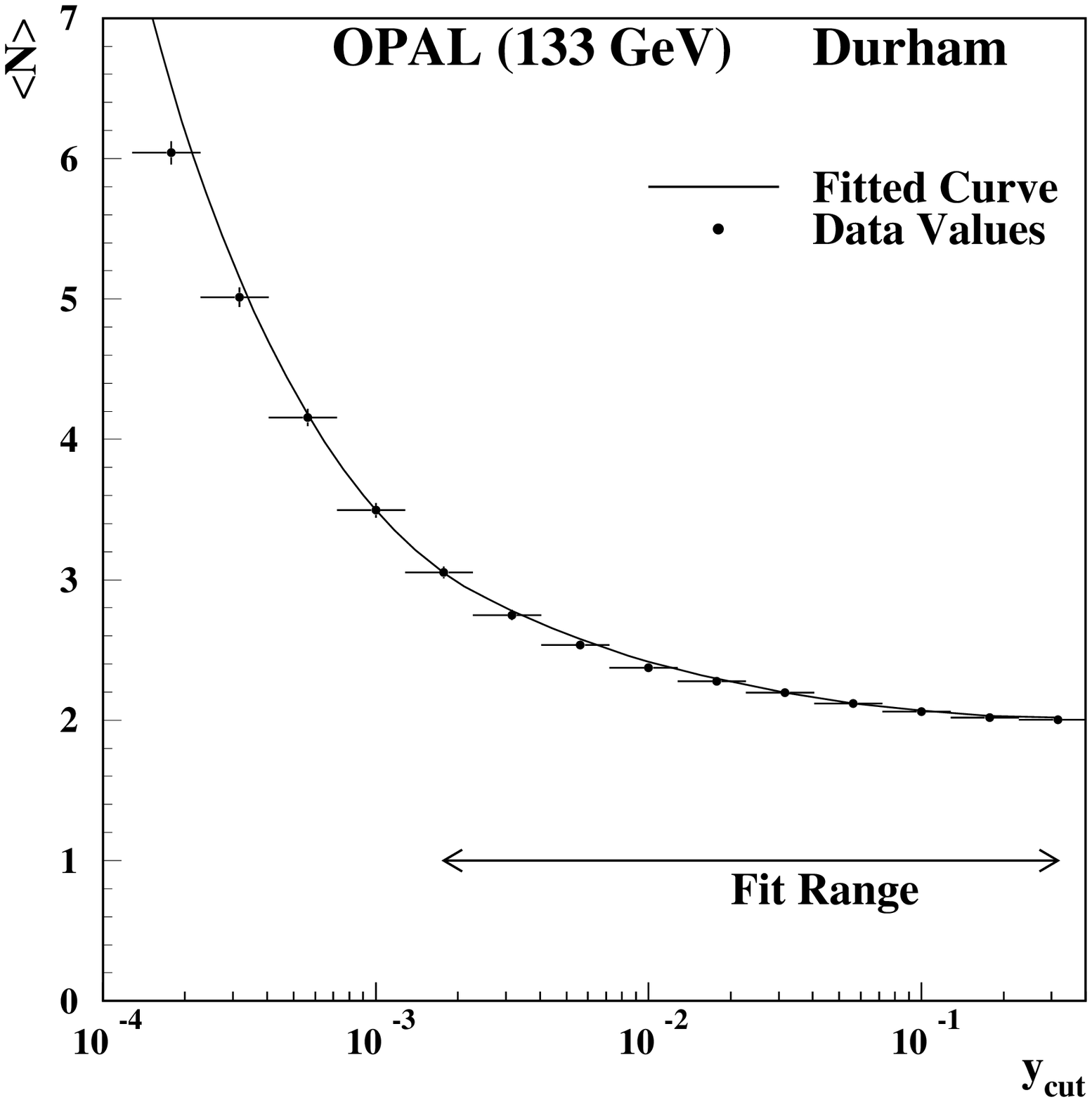}}         
   \hspace*{-1cm}\mbox{
      \includegraphics[width=.58\textwidth]{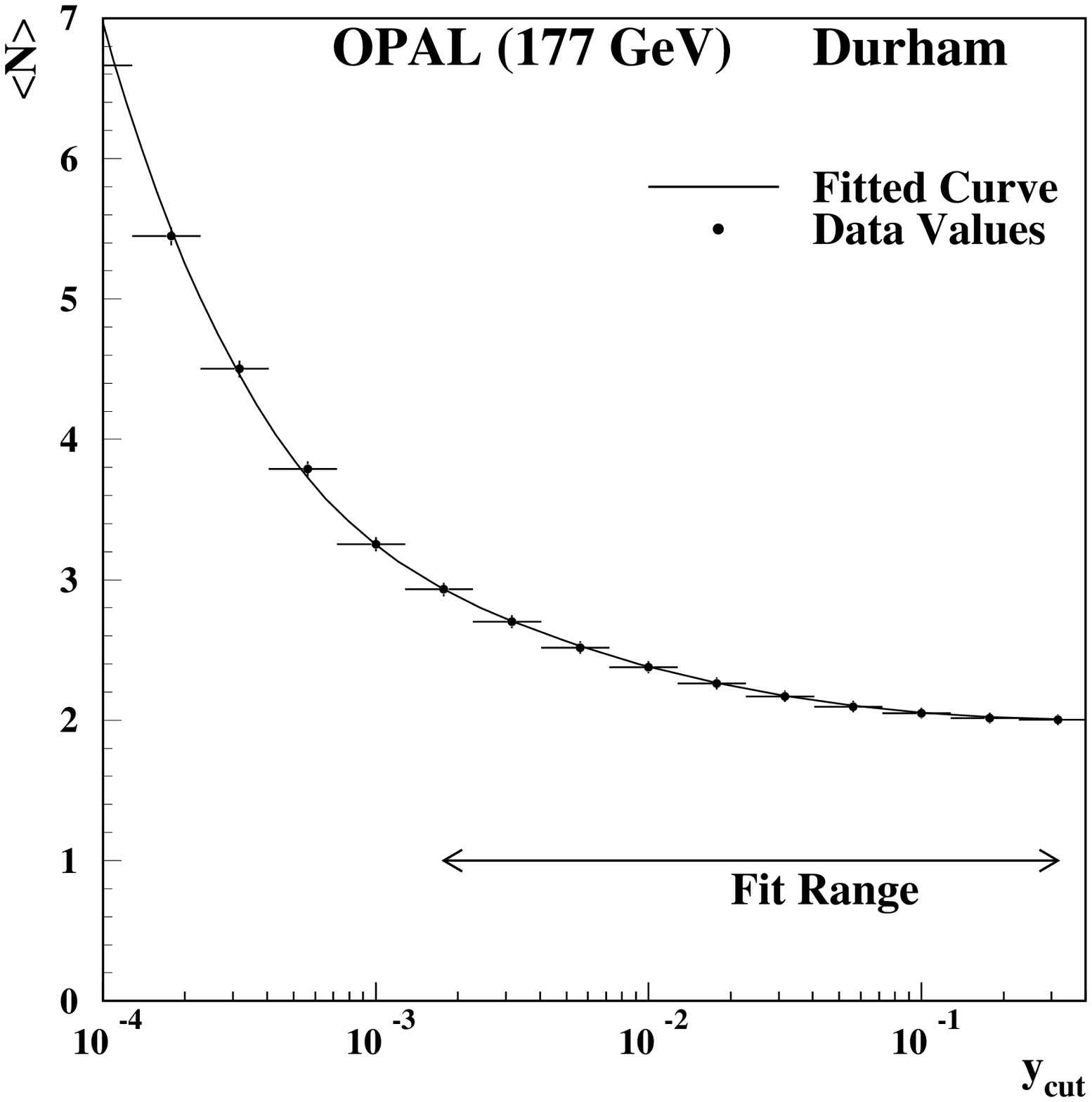}         
      \hspace*{-.9cm}
      \includegraphics[width=.58\textwidth]{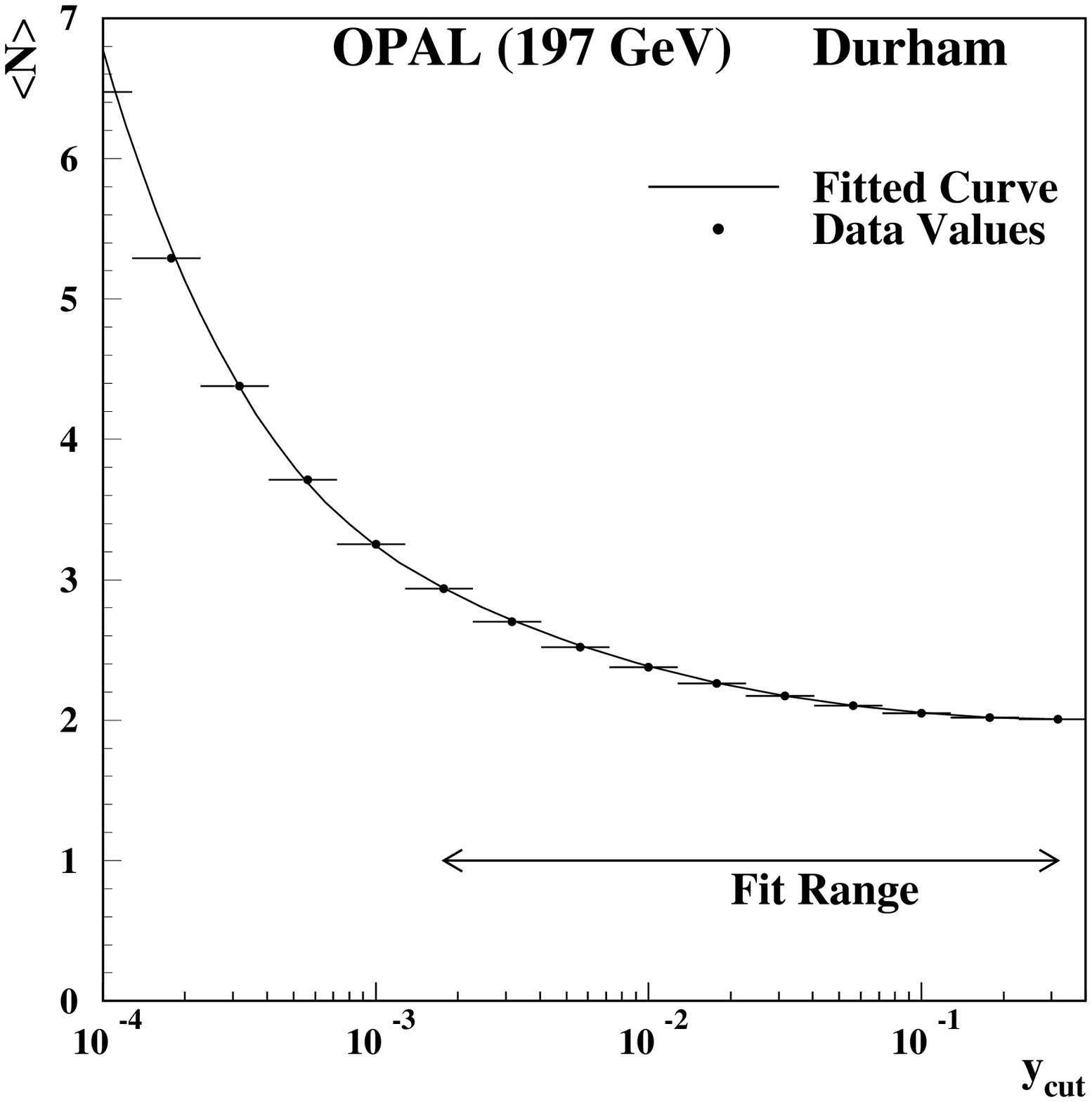}}
   \caption{Fits of the parton level $\avn$ distribution using the Durham
algorithm 
      as a function of $\yc$ to the
      \mbox{ln $R$} prediction for the 91~GeV (top left), 130~GeV (top right),
189~GeV (bottom
      left) and 207~GeV (bottom right) datasets.  Vertical error bars represent
statistical errors
      only.  Note: pulls are not shown as there are very large bin-to-bin
correlations.}    
   \label{ndfits}
\end{figure} 
\newpage
\begin{figure}[H]
\centering
   \hspace*{-.5cm}\includegraphics[width=1.15\textwidth]{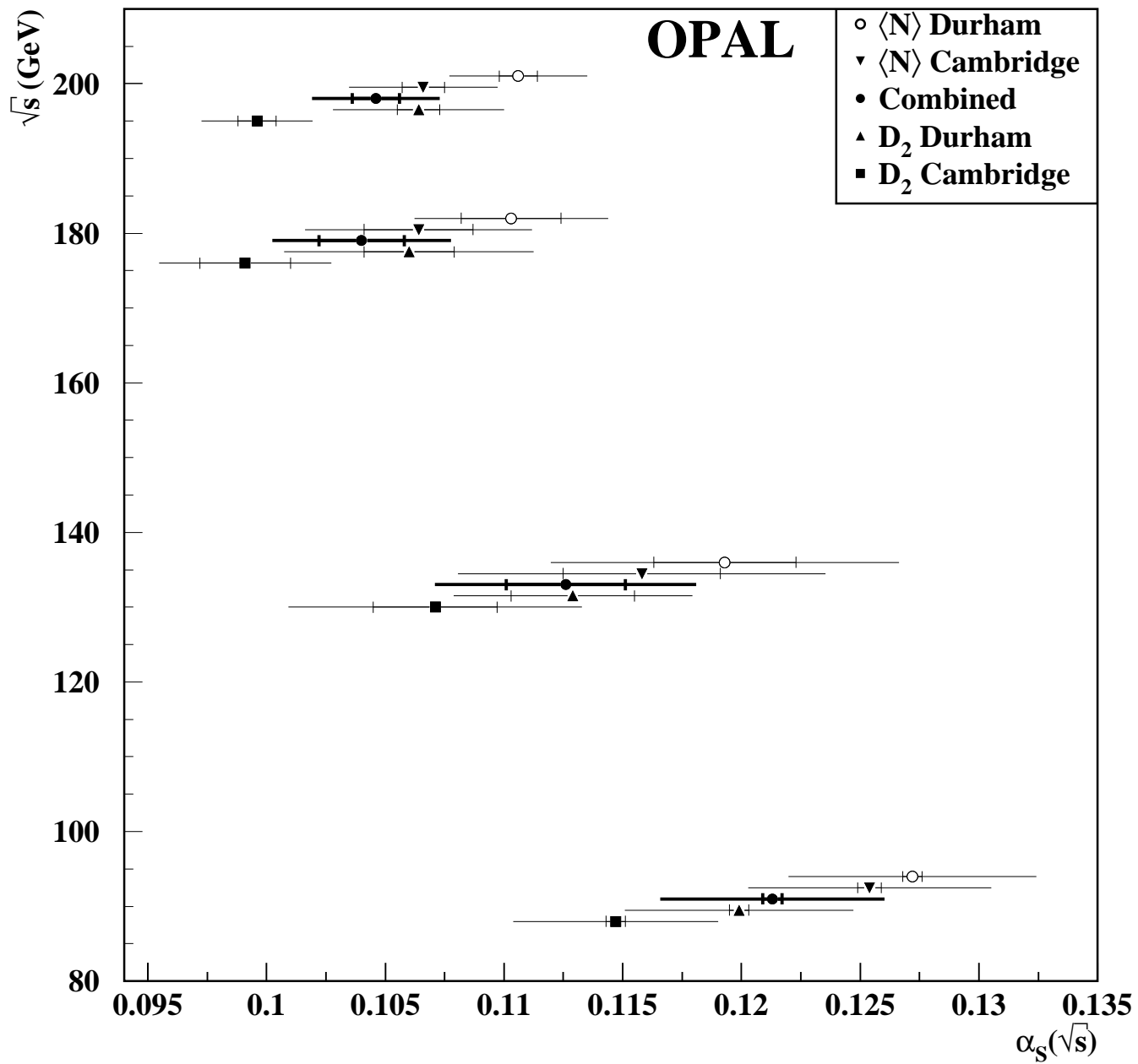}
   \caption{Comparison of combined \alsx values with those determined separately
from
      the $D_{2}^{D}$, $D_{2}^{C}$, $\avn^{C}$ and $\avn^{D}$ distributions for
each \cfm energy
      dataset.  Outer error bars indicate the size of the total errors while
inner bars indicate the
      size of the statistical errors.}
   \label{asrun_comp}
\end{figure} 

\begin{figure}[H]
\centering
   \includegraphics[width=\textwidth]{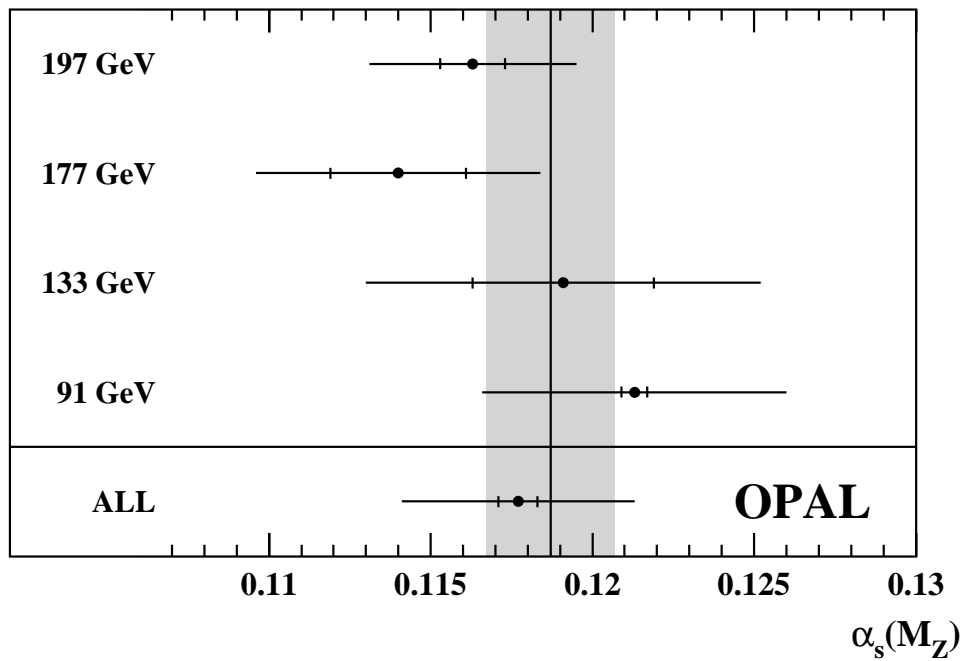}
   \caption{Comparison of the \alsx values after having been run back to the
     \PZo pole for each
      of the datasets.  The point labelled by 'ALL' 
      represents the value determined from the weighted mean of
      the four combined \als(\PMZ) determinations.  
      The shaded band corresponds to the one standard
      deviation range of the world average value of \als(\PMZ)~\cite{pdg}.  The
      inner error bars represent the
      statistical errors while the outer error bars represent the total error.}
   \label{as_comp_mz}
\end{figure}

\begin{figure}[H]
   \centering
   \hspace*{-.7cm}\includegraphics[width=1.1\textwidth]{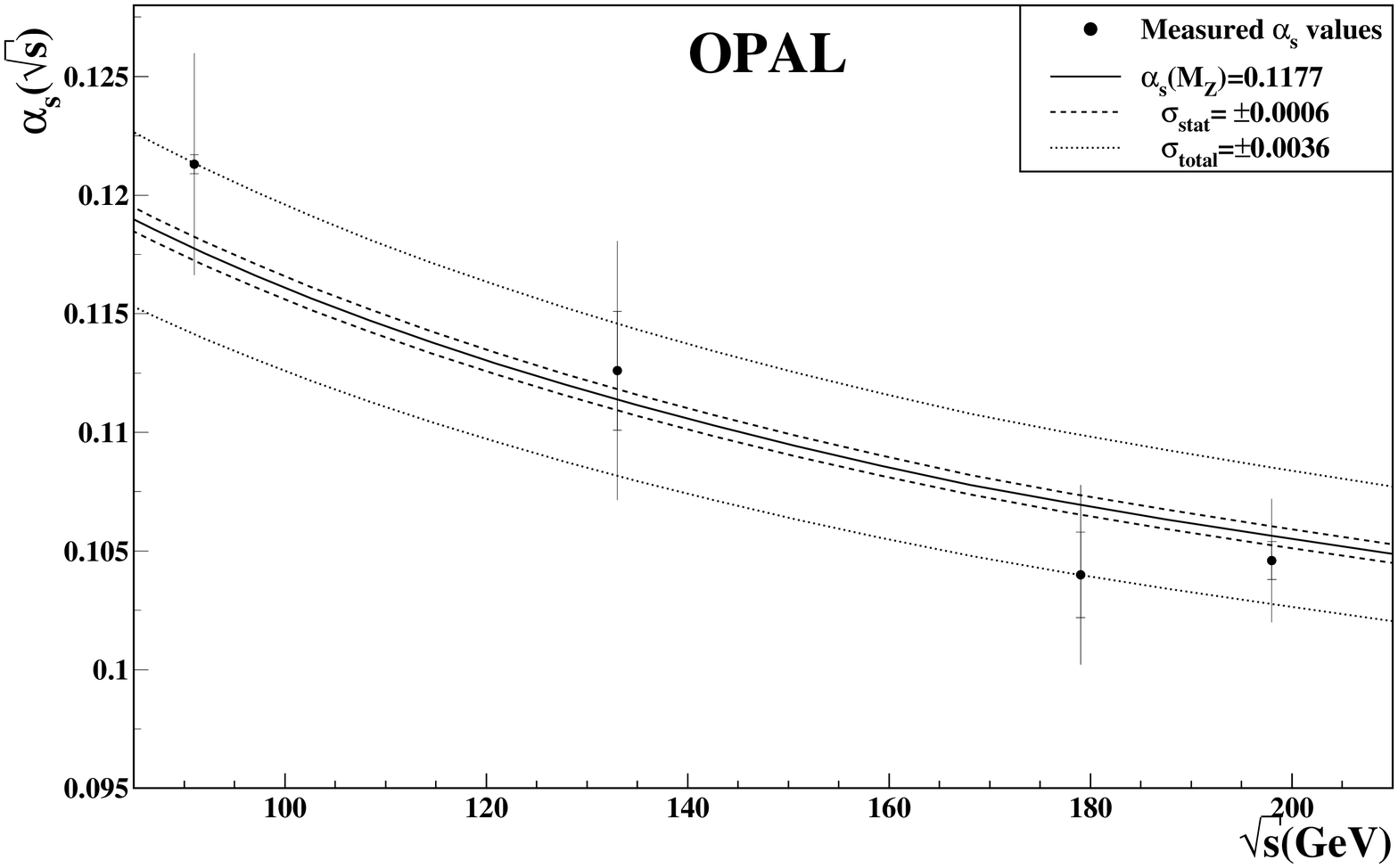}
   \caption{The running of \alsx as a function of \cfm energy.  The points
correspond to the
   \PZo-calibration dataset, the combined LEP1.5 data and the two LEP2 regions
161--183 and
   189--209, respectively.  The solid line corresponds to the expectation based
on the weighted mean
   of the four combined \als(\PMZ) determinations, and the
   outer dashed line to the total (statistical + systematic) uncertainty.}
   \label{fit:asrun}
\end{figure} 
\end{document}